\documentclass[11pt,nofootinbib]{article}
\pdfoutput=1

\usepackage{url}
\usepackage{cite}
\usepackage{multirow}
\usepackage{footmisc}
\usepackage{jheppub,fancyhdr,graphicx,epstopdf,color,amsmath,cases,slashed,hyperref}
\usepackage{makecell}
\usepackage{comment}
\usepackage{caption, subcaption}  
\usepackage{dcolumn}   
\usepackage{amsmath, amssymb}        
\usepackage{color,latexsym, array,multirow,  verbatim, enumerate,cancel}
\usepackage[normalem]{ulem} 
\usepackage{slashed, soul}
\usepackage{bigcenter}
\usepackage{soul}
\usepackage{footnote}
\makesavenoteenv{tabular}
\makesavenoteenv{table}
\usepackage[titletoc]{appendix}
\usepackage{float}

\def\issue(#1,#2,#3){{\bf #1}, #2 (#3)}

\catcode`\@=11
\def\lsim{\mathrel{\mathpalette\@versim<}}
\def\gsim{\mathrel{\mathpalette\@versim>}}
\def\@versim#1#2{\vcenter{\offinterlineskip
\ialign{$\m@th#1\hfil##\hfil$\crcr#2\crcr\sim\crcr } }}
\catcode`\@=12

\catcode`@=12

\newcommand{\newc}{\newcommand}
\newc{\wt}{\widetilde}
\newc{\ra}{\rightarrow}

\def\beq {\begin{equation}}
\def\eeq {\end{equation}}
\def\bi {\begin{itemize}}
\def\ei {\end{itemize}}
\def\bea {\begin{eqnarray}}
\def\eea {\end{eqnarray}}

\newcommand{\br}{\begin{eqnarray}}
\newcommand{\er}{\end{eqnarray}}
\newcommand{\be}{\begin{equation}}
\newcommand{\ee}{\end{equation}}

\newcommand{\ch}{\widetilde \chi^\pm}

\def \ch2p {{\wt\chi_2^+}}

\def \ch2m {{\wt\chi_2^-}}

\newc{\dmchi}{\Delta m_{\wt\chi}}

\def\issue(#1,#2,#3){{\bf #1}, #2 (#3)}

\hypersetup{
   colorlinks=true,       
   linkcolor=blue,        
   citecolor=blue,         
   filecolor=magenta,      
   }

\title{Tagging ultra-boosted jets at FCC-hh using machine learning techniques}

\author[a]{Sanchari Bhattacharyya,}
\author[a]{Biplob Bhattacherjee,}
\author[a]{Camellia Bose,}
\author[b]{Debtosh Chowdhury,}
\author[b]{Swagata Mukherjee}

\affiliation[a]{Centre for High Energy Physics, Indian Institute of Science, Bangalore 560012, India}
\affiliation[b]{Department of Physics, Indian Institute of Technology, Kanpur 208016, India}

\emailAdd{biplob@iisc.ac.in}
\emailAdd{sancharibhat@iisc.ac.in}
\emailAdd{camelliabose@iisc.ac.in}
\emailAdd{debtoshc@iitk.ac.in}
\emailAdd{swagata@iitk.ac.in}

\date{\today}

\abstract
{The Future Circular Hadron Collider (FCC-hh) will probe unprecedented energy regimes, enabling direct searches for new elementary particles at a scale of tens of TeV. FCC-hh is currently in the planning stage, and one of its primary physics goals is to search for physics beyond the Standard Model by exploring a previously inaccessible kinematic domain. While venturing into uncharted high-energy territories promises excitement, reconstructing objects with enormous transverse momenta will require overcoming major experimental challenges. This work investigates the identification of boosted $W$ bosons and boosted top quarks in the context of three beyond the Standard Model scenarios: heavy vector-like quark ($B'$), heavy neutral gauge boson ($Z'$), and heavy neutral Higgs boson ($H$). We employ machine learning techniques, including eXtreme Gradient Boosting (XGBoost) and convolutional neural networks (CNN), to identify these ultra-boosted objects in the collider from their SM background counterpart. We evaluate the performance of these techniques in distinguishing $W$ jets and top jets from QCD jets at extremely high transverse momenta ($p_{T}$) values, demonstrating their potential for future FCC-hh analyses.}

\begin{document}
\maketitle

\section{Introduction}

Experiments at the Large Hadron Collider (LHC) have contributed significantly to our understanding of the Standard Model (SM) of particle physics, including the discovery of the Higgs boson and precision measurements of its properties \cite{higgsdiscatlas, higgsdisccms, higgsprecision1, higgsprecision4, higgsprecision2,  higgsprecision3, pdg}.  While the SM has been rigorously tested with great precision, unanswered puzzles remain, such as the hierarchy problem, the observed phenomena of dark matter, neutrino masses, and matter-antimatter asymmetry in the universe. Although these gaps in our understanding of the universe suggest the existence of Beyond the Standard Model (BSM) physics, there is no consensus about the typical scale at which the new physics should appear. 

The lack of conclusive evidence for new physics at the LHC suggests that BSM phenomena may occur at energy scales inaccessible to current collider experiments. Various BSM scenarios predict new particles with large masses. For example, a simple U(1) extension of the SM in the gauge sector predicts the existence of a new gauge boson $Z'$ \cite{RevModPhys.81.1199, Zpmodel, Zpmodel1}. Similarly, BSM extensions of the Higgs sector in theories such as the Two Higgs Doublet Model \cite{PhysRevD.8.1226, THDM, 2hdm6, 2hdm7, 2hdm5, 2hdm4, 2hdm3, 2hdm2} contain charged and neutral heavy Higgs bosons. Future collider initiatives are essential for exploring such BSM phenomena at energy scales beyond the LHC's capabilities. The production of such heavy BSM particles at future colliders results in highly energetic and heavily boosted decay products, complicating event reconstruction and object identification. This study focuses on efficiently identifying these heavy states, common to many prominent BSM physics models, in a typical general-purpose experiment of the FCC-hh collider.

The FCC-hh \cite{FCC_CDR1, FCC_CDR3} program at CERN, Geneva, boasts an exceptionally broad physics program, spanning Standard Model precision studies to direct searches for BSM. It will explore uncharted territories in high-energy physics, enabling searches for new physics at unprecedented mass scales. However, this endeavor also poses unique challenges in object reconstruction, particularly for objects with extremely large transverse momenta ($p_T$) \cite{top_FCC_4, top_FCC_7, Bandyopadhyay:2024glc}. In the case of FCC-hh, the large $p_T$ of objects resulting from the decay of a highly massive particle leads to highly collimated decay products. Accurate reconstruction of these events depends on high-resolution detectors to resolve the individual signatures within these ultra-boosted objects \cite{top_FCC_5}. 
 
In BSM physics analyses featuring final states that contain the $W$ boson or the top quark, high selection efficiencies can generally be attained by leveraging hadronic decay channels. At sufficiently large Lorentz boost (typically $p_{T} > 200$~GeV or so), the final-state hadrons from the decay of $W$ bosons or top quarks merge into a single large-radius (i.e., fat) jet. In such cases, jet substructure techniques are normally used to identify the jets arising from decays of $W$ bosons or top quarks, exploiting the multiple, experimentally separable subjets within the fat jet~\cite{Thaler:2010tr, Larkoski:2013eya, Larkoski:2014gra, Larkoski:2014zma, Larkoski:2015kga, Moult:2016cvt}. However, in FCC-hh, where $p_T$ of top quark or $W$ boson can be in the multi-TeV range, the subjets will overlap and may not be experimentally separable anymore. Then, a key question that arises is how robust or useful the jet substructure observables (like N-Subjettiness \cite{Thaler:2010tr}) will remain at such high energy scales. The highly collimated nature of jets at FCC-hh will make it very challenging to identify their subjet information in the same way as it is still possible at the LHC. The proposed solution in this direction has been to use observables constructed on track jets, as the tracking devices have the best resolution, much better than the calorimeters. Track-based observables offer improved discrimination properties of jet substructure variables compared to calorimeter-based methods, owing to the better resolution of the tracking system at FCC-hh \cite{top_FCC_4, top_FCC_7}. In this context, we present studies on the robustness of the improved discriminating power under various conditions of energy resolution of calorimeters, as well as the granularity of calorimeters and trackers. Moreover, previous studies have highlighted the importance of choosing an optimum jet radius to avoid contamination from excess radiation \cite{top_FCC_4}. In this work, we focus on optimizing the jet radius from parton-level information of the top quark and $W$ boson decay.

Furthermore, the performance of jet tagging at the LHC has steadily been improving and heavily relies on various Machine Learning (ML) techniques \cite{Erdmann:2013rxa,Almeida:2015jua,Kasieczka:2017nvn,Erdmann:2017hra,Butter:2017cot,Macaluso:2018tck,Moore:2018lsr,Qu:2019gqs,Kasieczka:2019dbj,Roy:2019jae,Diefenbacher:2019ezd,Chakraborty:2020yfc,Bhattacharya:2020aid,Shmakov:2021qdz,Lim:2020igi,Dreyer:2020brq,Aguilar-Saavedra:2021rjk,Andrews:2021ejw,Qu:2022mxj,Dreyer:2022yom,Ahmed:2022hct,Munoz:2022gjq,Choi:2023slq,He:2023cfc,Bogatskiy:2023nnw,Shen:2023ofd,Isildak:2023dnf,Sahu:2023uwb,Baron:2023yhw,Bogatskiy:2023fug,Liu:2023dio,Batson:2023ohn,Furuichi:2023vdx,Ngairangbam:2023cps, ATLAS:2020lks,ATL-PHYS-PUB-2020-017,ATL-PHYS-PUB-2022-039,CMS:2012bti,CMS:2017ucf,CMS:2021beq,Keicher:2023mer, top_shap, ml_review, Baruah:2024wrn, Barman:2024wfx, Fraser:2024dhp,Mondal:2024nsa,Chowdhury:2023jof,Choudhury:2023eje,Chakraborty:2023dhw,Chakraborty:2023hrk,Alvarez:2022qoz,Alvestad:2021sje}, which also inspires jet tagging algorithms for the proposed FCC-hh detectors. The focus of this work is mainly to provide a basic understanding of boosted top and $W$ jet identification in FCC-hh detectors using Extreme Gradient Boosting (XGBoost) \cite{DBLP:journals/corr/ChenG16} and Convolutional Neural Networks (CNN) \cite{Fukushima1980NeocognitronAS, FUKUSHIMA2013103,Lecun}. These two relatively simple ML models, based on high-level jet features and low-level jet constituents, respectively, are used to establish a baseline for preliminary feasibility studies at 100 TeV. The goal is to understand how well these taggers perform in the FCC-hh environment compared to their performance in the LHC environment, which can then serve as a reference point for further improvements involving more complex and state-of-the-art ML taggers.
We thus investigate the feasibility of efficiently distinguishing multi-prong $W$ jets and top quark jets from single-prong QCD jets, where these objects have extremely high $p_T$, far exceeding those encountered at the LHC. 
Previous work in Ref.~\cite{Coleman:2017fiq} employed boosted decision trees (BDTs) in the context of 100~TeV $pp$ collisions, focusing on jets with $p_T \sim 5$~TeV. Our study focuses on multiple jet $p_T$ bins ranging from 2~TeV to 16~TeV.
The models in our work are trained as multi-class classifiers. We use the score from the classifiers to tag the final state top and $W$ jets, thus allowing us to reconstruct the massive BSM particle. The study is based on the fast simulation of the reference FCC-hh detector \cite{Selvaggi:2717698,FCC_CDR1, FCC_CDR3}. 

Earlier works have studied the discovery potential of the FCC-hh for heavy resonances using BDTs~\cite{top_FCC_3, top_FCC_6}. In our study, we explore the variation of jet kinematics with $p_T$ and make use of the information to build jet identification strategies spanning a broad range of transverse momenta. The reason is that any heavy resonance decays to jets with $p_T$ ranging from, say, $\mathcal{O}(1)$~TeV to as high as half of its mass. The capability of FCC-hh to directly produce such heavy resonances necessitates the development of $p_T$-specific taggers, as demonstrated in our study. We perform a detailed analysis of three benchmark BSM scenarios: a high-mass $Z'$ boson decaying to boosted top jets, a heavy neutral CP-even Higgs boson decaying to boosted $W$ jets, and a high-mass vector-like quark ($B'$) decaying to a boosted top quark jet and a boosted $W$ jet.

This paper explores the feasibility of efficiently distinguishing between boosted top jets, $W$ jets, and QCD jets at extremely high transverse momenta (multi-TeV range). We focus on developing and validating simple machine learning-based top and $W$ taggers rather than optimizing the analysis flow for setting cross-section limits on BSM scenarios. The example analyses presented in this paper are illustrative, and the analysis strategy can be refined further to improve the cross-section limits presented here. Our goal is to lay the groundwork for future studies rather than striving for the most stringent limits. 

Another caveat is that we have not included pileup in our study. The FCC-hh is expected to have a mean pileup of 500-1000 at an instantaneous luminosity of $30\times 10^{34}~\text{cm}^{-2}\text{s}^{-1}$~\cite{FCC_CDR3}. However, pileup mitigation strategies (pile-up subtraction algorithms like PUPPI~\cite{Bertolini:2014bba}) for FCC-hh are still under development and not yet standardised. Moreover, the final states analysed in this study involve highly energetic jets in the multi-TeV range. In this regime, the relative effect of pileup is expected to be significantly reduced. Therefore, while the omission of pileup represents a limitation, its impact can be neglected to first approximation for the purposes of this exploratory study.

The outline of this paper is as follows: In Section~\ref{sec:bsm_theory}, the physics models we consider are described. A brief description of the proposed FCC facility is provided in Section~\ref{sec:fcc}. Details of event simulation are discussed in Section~\ref{sec:simu}. 
The challenges that occur while reconstructing boosted jets at the $\sqrt{s} = 100$~TeV collider are highlighted in Section~\ref{sec:challenge}. 
It presents a detailed discussion about jet radius, jet substructure observables, and boosted $W$ boson jets that may appear in the final states of various BSM scenarios. The section also addresses the effect of finite calorimeter and tracker resolution on the jet observables, along with the impact of jet grooming. Details of the ML models used in this study are provided in Section~\ref{sec:ML}. Finally, the analyses of the three benchmark BSM scenarios are presented: a heavy neutral $Z'$ boson in Section~\ref{sec:Zp_analysis}, a heavy neutral CP-even  Higgs boson in Section~\ref{sec:H_analysis}, and a high-mass vector-like bottom-partner ($B'$) in Section~\ref{sec:Bp_analysis}. The conclusions and outlook are summarised in Section~\ref{sec:conclusion}.

\section{Theoretical Scenarios}
\label{sec:bsm_theory}
At the LHC, the top quarks and $W$ bosons are produced via various SM interactions, like the strong and electroweak interactions. They can also originate from BSM interactions, such as the decay of heavy new particles predicted by BSM. There are several BSM scenarios where heavy resonances are produced, which then decay to top quarks or $W$ bosons. 
The top quark eventually decays to a $W$ boson and a $b$ quark. Depending on the decay channel of the $W$ boson, i.e., whether $W$ decays to a lepton and a neutrino (leptonic decay) or to a quark-antiquark pair (hadronic decay), the final states will be different, leading to different signatures at the collider.
In this analysis, we take into account three such BSM scenarios, which act as an additional source of top quarks and $W$ bosons in addition to the SM. 

\begin{figure}[hbt!]
\centering
    \begin{subfigure}[b]{0.4\linewidth}
    \includegraphics[width=\linewidth]{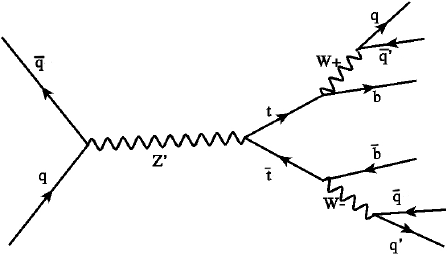}
    \caption{}
    \label{fig:feyn_Zp}
    \end{subfigure}\hfill
    \begin{subfigure}[b]{0.5\linewidth}
    \includegraphics[width=\linewidth]{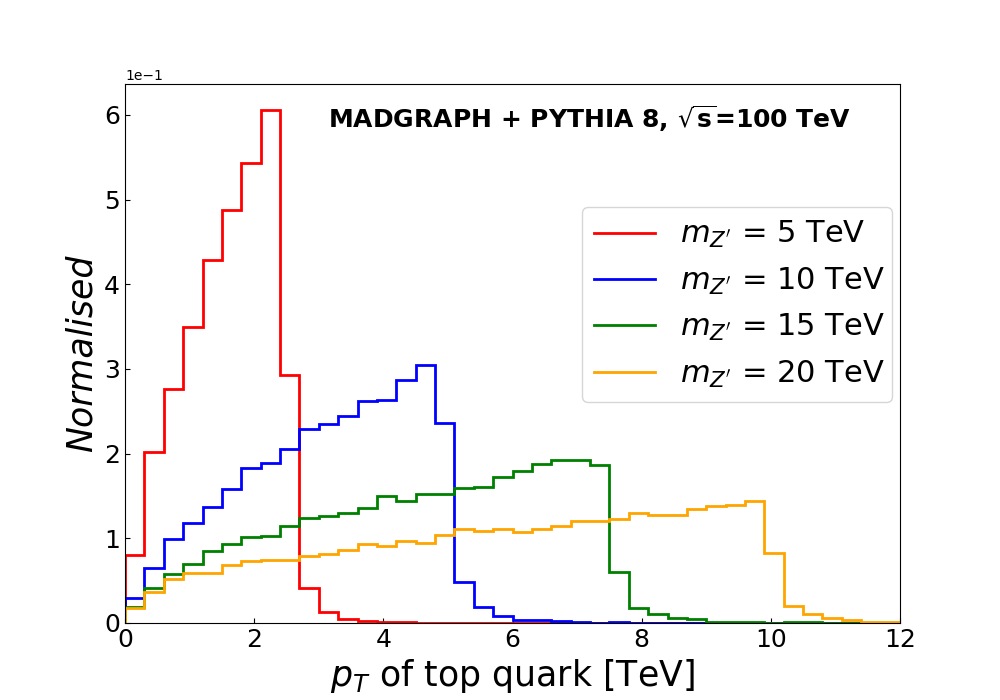}
    \caption{}
    \label{fig:pT_Zp_top}
    \end{subfigure}
    \caption{(a) Feynman diagram of $s$-channel production of a $Z'$, with the $Z'\to t\bar{t}$ decay where $t$ and $\bar{t}$ decay hadronically. (b) Transverse momentum distribution of a top quark coming from the decay of a $Z'$ boson produced at the FCC-hh.}
\label{fig:feyn_pt_Zp}	
\end{figure}

To begin with, we consider a BSM scenario that involves a heavy neutral gauge boson, $Z'$, in addition to the SM spectrum. Such a heavy gauge boson is generic to many BSM scenarios, e.g., in the U(1) extended models \cite{Zpmodel, Zpmodel1}, Left-Right symmetric models (LRSM) \cite{Zplrsm1, Zplrsm2, zplrsm3}, 3-3-1 model \cite{Zp331}, Vector-leptoquark model \cite{Zpleptoquark}, top-phillic $Z'$ model \cite{Zptopphil1, Zptopphil2}, etc. $Z'$ may have a kinetic mixing with the SM $Z$ boson, and depending on the symmetries of the corresponding models, it can decay into SM fermions. If the mass of $Z'$ is quite high, the decay products are very boosted \cite{zprime_atlas_1, zprime_atlas_2, zprime_atlas_3, zprime_atlas_4}. 
If kinematically allowed, a $Z'$ can decay into a pair of top quarks. The results of an ATLAS study, where they have searched for a $Z'$ boson decaying into a pair of hadronically decaying top quarks at the 13~TeV run of LHC with 139 fb$^{-1}$ integrated luminosity~\cite{ATLAS:2020lks}, have excluded masses of $Z'$ in the topcolor-assisted-technicolor model up to 3.9 TeV and 4.7 TeV, for the decay widths of 1\% and 3\%, respectively. In the context of FCC-hh, where the center of mass energy is 100~TeV, the FCC-hh can directly produce such a gauge boson with a very high mass (around tens of TeV). In this analysis, we have generated a multi-TeV scale $Z'$ boson, and depending on its mass, we consider four benchmark points where $m_{Z'} = 5$, 10, 15, and 20~TeV with a width to mass ratio ($\Gamma/m_{Z'}$) of 10\%.
We generate 10,000 events via the process $p \, p \rightarrow Z' \rightarrow t\, \bar{t}$ at 100~TeV run of FCC-hh using \texttt{PYTHIA 8} (refer to the Feynman diagram in Figure \ref{fig:feyn_Zp}.
The top quarks decaying from such a heavy $Z'$ will be highly boosted.
Figure \ref{fig:pT_Zp_top} represents the transverse momenta of such boosted top quarks for four different masses of the $Z'$. Similar final states ($t\bar{t}$) also arise in the SM due to QCD interactions, e.g., $ pp \to t\bar{t}$ production, which is an important background for $Z' \to t\bar{t}$ search. The SM dijet production ($pp \to jj$) also acts as a major background process in the search for $Z'$. In section \ref{sec:Zp_analysis}, we will discuss the signal-background analysis for $Z'$ in detail.

Next, we consider another BSM scenario that involves a heavy CP-even Higgs boson. Theories involving one or more heavy neutral and charged scalars have been extensively studied in the literature. One popular example is the Two Higgs Doublet Model (THDM) \cite{THDM, 2hdm2, 2hdm3, 2hdm4, 2hdm5, 2hdm6, 2hdm7}. In addition to that, many of the well-studied BSM theories like Minimal Supersymmetric Standard Models (MSSM) \cite{susy, susy1, susy2, susy3, susy4, susy5, susy6}, Left-Right Symmetric Models (LRSM) \cite{Zplrsm1, Zplrsm2, zplrsm3, lrsmhiggs}, Extra-Dimensional models \cite{ed1, ed2}, Grand Unified Theories (GUTs) \cite{e6gut, so10gut}, multi-Higgs Doublet models \cite{3hdm, multihiggs1, multihiggs2, multihiggs3, multihiggs4, multihiggs5}, Georgi-Machacek model (GM) \cite{gm1, gm2, gm3, gm4, gm5, gm6, gm7} have at least one heavy CP-even scalar other than the SM-like light Higgs boson in its mass spectrum. Both ATLAS and CMS collaborations have searched for such heavy scalars, neutral as well as charged \cite{heavyHiggs, higgscms1, higgscms2, higgscms3, higgscms4, higgsatlas1, higgsatlas2, higgsatlas3, higgsatlas4, higgsatlas5}.
 
The ATLAS collaboration performed a search for a heavy neutral resonance in the $W^+W^- \to \mu^{+}\bar\nu\mu^{-}\bar\nu$ channel using $13$~TeV data corresponding to an integrated luminosity of $139~\text{fb}^{-1}$. For a narrow-width Higgs-like scalar, 95\% CL upper limits are set on the production cross-section times branching ratio. At $m_{H} = 3.8$~TeV, values above $0.0048$~pb and $0.0045$~pb are excluded in the ggF and VBF production modes, respectively~\cite{ATLAS:2022qlc}. The FCC-hh shall be able to directly produce a much heavier Higgs boson. Just as in the case of $Z'$, here too we choose four benchmark points by varying the mass of $H$, i.e., $m_H = 5$, 10, 15, 20~TeV, with the decay mode $H\to W^+W^-$. In our study, we consider the fully hadronic decay of the boosted $W$'s as we aim to study the boosted $W$ jet tagging scenario at FCC-hh.

\begin{figure}[hbt!]
\centering
    \begin{subfigure}[b]{0.4\linewidth}
    \includegraphics[width=\linewidth]{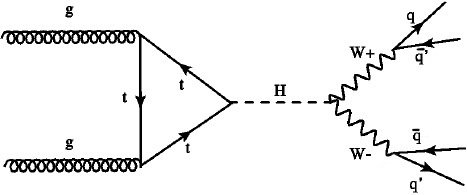}
    \caption{}
    \label{fig:feyn_H}
    \end{subfigure}\hfill
    \begin{subfigure}[b]{0.5\linewidth}
    \includegraphics[width=\linewidth]{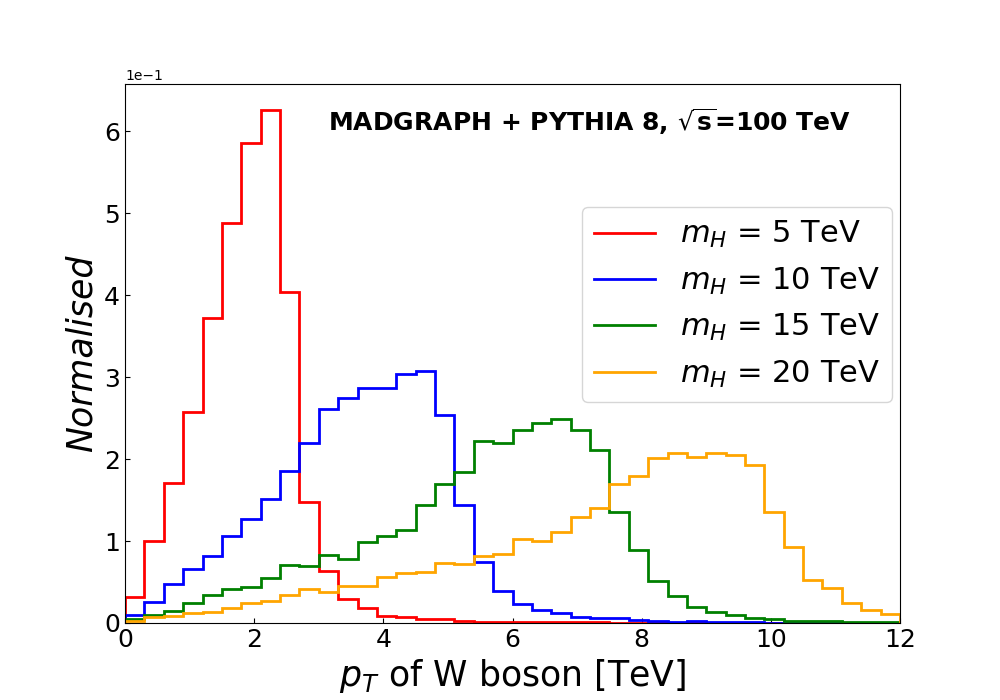}
    \caption{}
    \label{fig:pT_H0_w}
    \end{subfigure}
	\caption{(a) Feynman diagram of gluon-gluon fusion production of $H$ decaying via $H\to W^+W^-$ where the $W$'s decay hadronically. (b) Transverse momentum distribution of $W$ boson produced from the decay of the heavy Higgs at the FCC-hh.}
	\label{fig:feyn_pt_H}	
\end{figure}

We generate 10,000 $pp\to H\to WW$ events using \texttt{PYTHIA 8} with $\Gamma/m_{H}=10\%$, where the $W$ bosons decay hadronically to jets (refer to the Feynman diagram in Figure \ref{fig:feyn_H}).
The decay of such massive Higgs bosons results in ultra-boosted $W$ bosons. Figure \ref{fig:pT_H0_w} depicts the transverse momentum ($p_T$) of the boosted $ W$'s considering four different heavy Higgs boson masses. As observed from the figure, $W$ bosons from the decay of the massive Higgses are extremely boosted (very large $p_T$). In this case, the major SM background processes include $W^+W^-$ pair production, $t\bar{t}$ production, and dijet production, the analysis of which is presented in detail in Section \ref{sec:H_analysis}.

In our study, the $W$ bosons are treated as unpolarised. While the results may be sensitive to the polarisation of the $W$ bosons, as discussed in Ref.~\cite{CMS:2013uea}, a detailed analysis of polarisation effects is beyond the scope of this work.

The last model that we consider in our current analysis is a BSM scenario involving a vector-like quark (VLQ). A vector-like fermion is a Dirac fermion whose left- and right-chiral components have similar transformations under the SM gauge group. Thus, their representations are free from chiral anomaly. The minimal model for a VLQ analysis could be a VLQ singlet or doublet added to the SM Lagrangian. Theories with one or more VLQs have been considered in the literature where different multiplets of VLQs have been added to the SM Lagrangian and their exotic decay modes, and related phenomenology has been studied \cite{MLVLQ5, MLVLQ4, vlq5, vlq4, vlqavik3, MLVLQ3, MLVLQ2, MLVLQ1, vlq1, vlqavik, vlq3, vlqavik2, vlq2}.

A generic VLQ model may be categorised into models having bottom-like ($B'$) or top-like ($T'$) partners, or models where the VLQ has fractional charges other than $-1/3$ or $2/3$ \cite{vlq1, vlqavik, vlqavik2}. ATLAS and CMS collaborations have extensively searched for $B'$ and $T'$ quarks in the 13~TeV run of LHC \cite{vlfreview, vlqref1, vlqref2, vlqref3, vlqref4, vlqref5, vlqref6, vlqref7, vlqref8} via its various decay modes.
In many BSM theories, $B'$ or $T'$ quarks mainly interact with third-generation SM quarks. Depending on the charge of the VLQ, it can decay to a top (or bottom) quark and a $W$ boson (or SM/BSM particle, for example, SM Higgs/Heavy Higgs/$Z$ boson, etc.) \cite{VLQmodel, VLQmodel2, VLQmodel3}. 
If the VLQ is very heavy, then the top quark, coming from its decay, will be highly boosted. In such cases, the decay products of the top quark become highly collimated, and separating them from the usual background of jets becomes challenging. The search for VLQs with boosted top signatures has been performed in multiple studies \cite{single_vlq_cms, vlq_atlas_1, vlq_atlas_2, vlq_atlas_3, vlq_atlas_4, vlq_atlas_5}.
The ATLAS collaboration has searched for a vector-like $T'$ quark at 13~TeV LHC run, where $T'$ decays to a top quark and a $Z$ boson, followed by $t\to bW,~W\to jets$ and $Z\to \nu\bar\nu$\cite{vlq_atlas_5}. 
They have carried out the analysis with boosted top topology and have quoted the lower mass limit of the VLQ to be 1.8~TeV. The CMS collaboration has also searched for vector-like $B'$ quark decaying to a $W$ boson and a top quark at 13~TeV run with 138 fb$^{-1}$ integrated luminosity~\cite{single_vlq_cms}. This analysis excludes the mass of the $B'$ quark up to 1.56~TeV.
 
At FCC-hh, VLQs can be produced with masses of tens of TeV. We consider a BSM scenario where we have added one SU(2) singlet bottom-like VLQ, $B'$ having $SU(3)_C \otimes SU(2)_L \otimes U(1)_Y$ gauge quantum numbers as $(3,1,-1/3)$, to the  SM Lagrangian which interacts with the first and third generation SM quarks \cite{VLQmodel, VLQmodel2, VLQmodel3}. The interactions relevant for our study reads as \cite{VLQmodel3},
\begin{equation}
\mathcal{L} \supset \dfrac{g \tilde{g}}{2} \left[ \sqrt{\dfrac{R_L}{1+R_L}}\; \bar{B}^{\prime}_L W_{\mu}^{-} \gamma^{\mu} u_L + \sqrt{\dfrac{1}{1+R_L}}\; \bar{B}^{\prime}_L W_{\mu}^{-} \gamma^{\mu} t_L \right] + \mathrm{h.c.}
\end{equation}
where, $g$ stands for $\text{SU(2)}_L$ gauge coupling constant of SM, and $\tilde{g}$ and $R_L$ are real-valued parameters. This $B'$ can be singly-produced or pair-produced at the LHC or FCC-hh via electroweak or strong processes. In the present analysis, we produce a single $B'$ in association with a light jet via the process $pp\to jB'\to jtW$. This production mode is chosen due to its relatively enhanced cross-section compared to pair production, especially at higher $B'$ masses.
We use the model file described in Ref.~\cite{VLQmodel} and set the coupling parameter values fixed at $\tilde{g}=0.1$ and $R_L=0.5$. 

\begin{figure}[hbt!]
\centering
    \begin{subfigure}[b]{0.4\linewidth}
			\includegraphics[width=\linewidth]{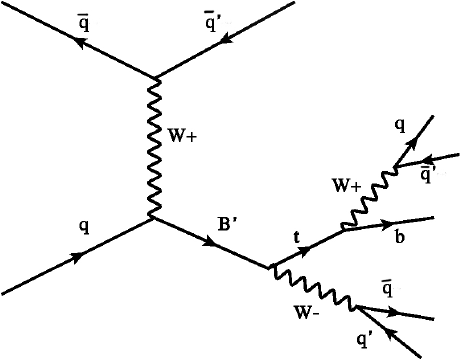}
		\caption{}
        \label{fig:feyn_Bp}
	\end{subfigure}\hfill
	\begin{subfigure}[b]{0.5\linewidth}
		\includegraphics[width=\linewidth]{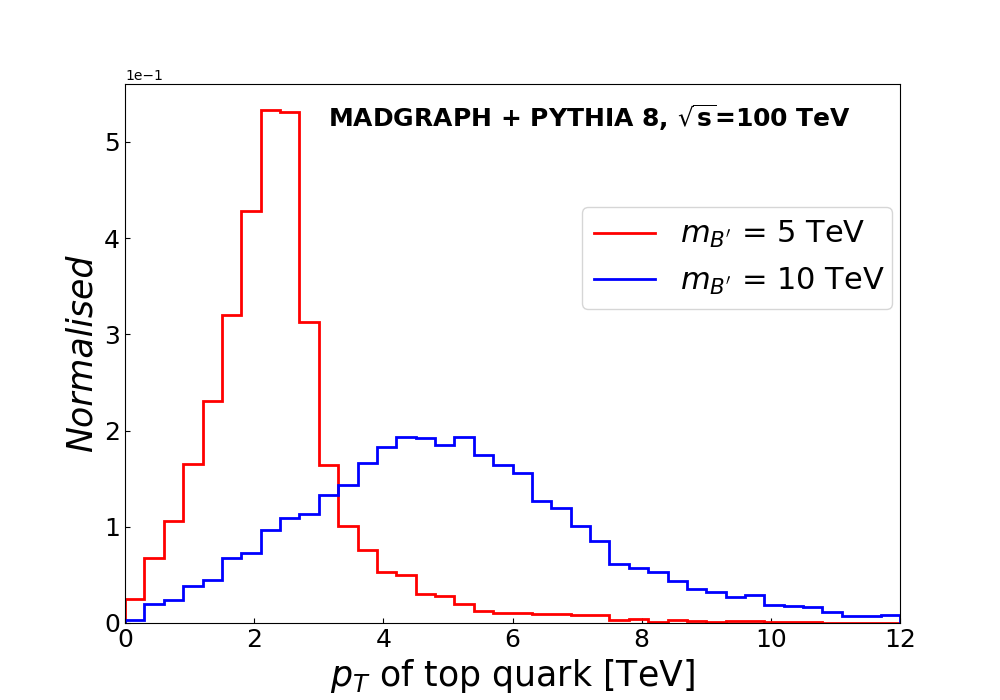}
		\caption{}
        \label{fig:pT_Bp_top}
	\end{subfigure}
	\begin{subfigure}[b]{0.5\linewidth}
		\includegraphics[width=\linewidth]{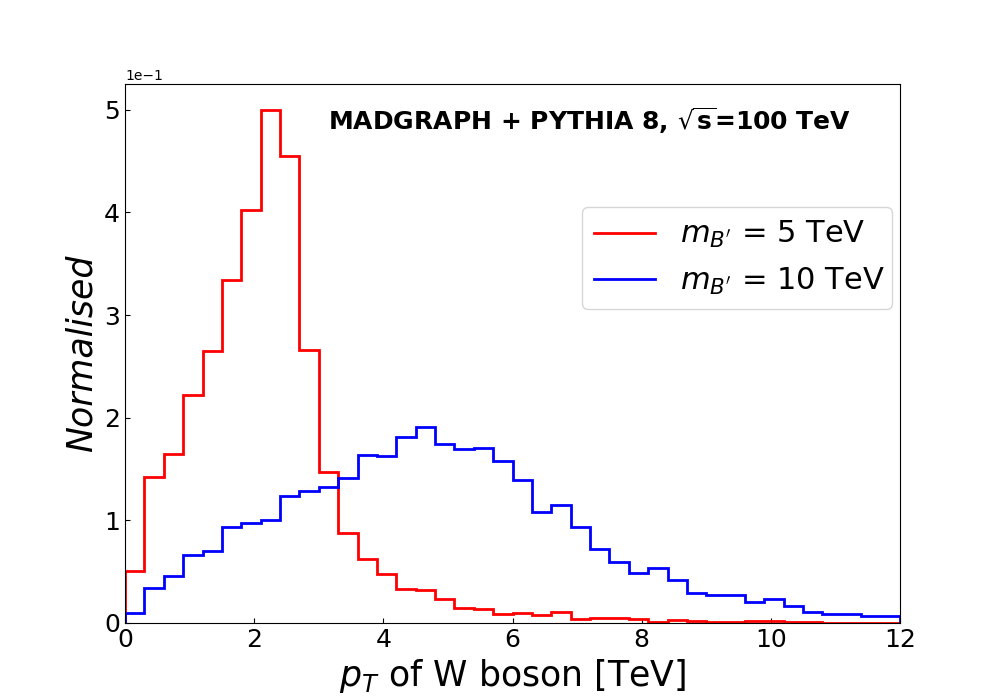}
		\caption{}
        \label{fig:pT_Bp_w}
	\end{subfigure}\hfill
    \begin{subfigure}[b]{0.5\linewidth}
			\includegraphics[width=\linewidth]{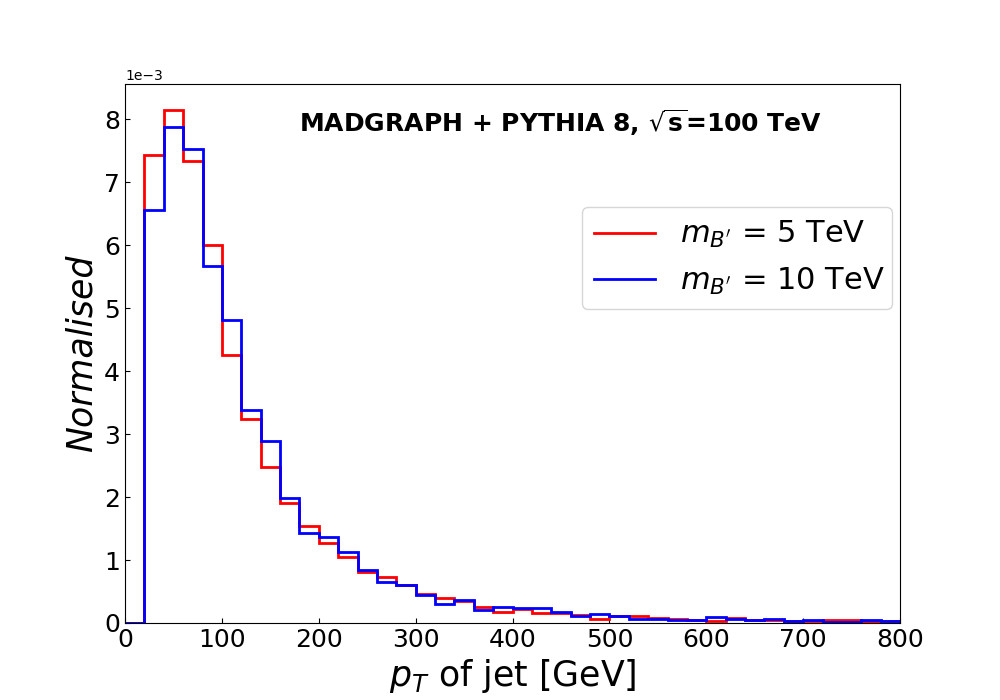}
		\caption{}
        \label{fig:pT_Bp_j}
	\end{subfigure}
	\caption{(a) Feynman diagram of $t$-channel production of $B'$ with an associated jet and $W$ exchange, where $B'$ decays as $B'\to t W$ followed by hadronic decay of $W$. Distribution of transverse momentum of (b) the top quark coming from the decay of $B'$, (c) the $W$ boson coming from the decay of $B'$, and (d) the light jet produced in association with the $B'$, for two different mass values of the singlet $B'$ produced at the FCC-hh. These plots are made at the parton level.}
	\label{fig:feyn_pT_Bp}	
\end{figure}

We generate 50,000 events featuring $B'$, with two benchmark masses of 5~TeV and 10~TeV, produced in association with a light jet, were generated using the event generator \texttt{MADGRAPH v2.9.18} \cite{Alwall:2011uj} and the UFO file from Ref.~\cite{feynrules}. The production cross-sections of $pp\to jB'\to jtW$ for $m_{B'}=5~\text{TeV}$ and $m_{B'}=10~\text{TeV}$ are 287 fb and 33 fb, respectively. After being produced, the $B'$ decays to $tW$ with a maximum branching ratio (BR) compared to its other decay modes \cite{VLQmodel, VLQmodel2}. The top quark then further decays to $Wb$ with a nearly 100\% branching ratio (refer to the Feynman diagram in Figure \ref{fig:feyn_Bp}). 

In Figures \ref{fig:pT_Bp_top}, \ref{fig:pT_Bp_w}, and \ref{fig:pT_Bp_j}, we present the transverse momenta ($p_T$) of the boosted top quark and the boosted $W$ boson produced in the $B'$ decay for two different masses of $B'$ at the FCC-hh using \texttt{PYTHIA 8}. We observe that the $p_T$ of the light jet produced in association with $B'$ is very small compared to the $p_T$ of the top quark or the $W$ boson. The top quark and the $W$ boson have almost similar $p_T$ distributions. Therein arises an issue where a $W$ jet can be mistagged as a top jet or a QCD jet. Similar SM processes, as discussed earlier, will act as major backgrounds for this signal. The signal-background analysis considering this benchmark model will be described in section \ref{sec:Bp_analysis}.

In the following sections, we first discuss the FCC detector and the event simulation elaborately, followed by the challenges and possible solutions while dealing with boosted top quarks and boosted $W$ bosons. Then, we present our results in detail, considering the above-mentioned theoretical scenarios.

\section{CERN FCC facility} 
\label{sec:fcc}

The Future Circular Collider (FCC)~\cite{FCC_CDR1} is a proposed particle accelerator designed to push the boundaries of the energy frontier and significantly surpass the center-of-mass energies of previous colliders like the LHC. The FCC project includes three main planned scenarios: FCC-ee for electron-positron collisions, FCC-eh for electron-hadron collisions, and FCC-hh for hadron-hadron collisions. The FCC-hh aims for a center-of-mass collision energy of 100~TeV, which is nearly seven times higher than the LHC. It will offer a very exciting opportunity to explore physics at the multi-TeV scale. The collider will be housed in a 100 km tunnel, making it the largest particle accelerator ever built. The main motivation for FCC-hh is the exploration of the energy frontier as best as possible. With an unprecedented center-of-mass energy of 100~TeV, it will significantly extend the mass reach for discovering new BSM particles. The collider will enable precision measurements of the Higgs boson, top quark, and electroweak symmetry-breaking properties, surpassing the capabilities of the LHC. The high energy and luminosity of FCC-hh will allow for detailed studies of rare processes and interactions, providing insights into the fundamental forces of nature. 

One of the key technological challenges is the development of high-field superconducting magnets, which are essential for bending the proton beams along the collider's circular path. The FCC-hh will also require advancements in cryogenics and vacuum systems to maintain the necessary conditions for particle collisions. The magnetic dipole system shall offer a field of up to 16 Tesla using $\text{Nb}_3\text{Sn}$ superconducting magnets. The FCC-hh is planned to have two high luminosity and two low luminosity interaction points. The detectors for the FCC-hh are designed to handle extreme conditions and highly energetic particles originating from 100~TeV proton-proton collisions. 

The FCC-hh detector is designed to have a diameter of $20~\text{m}$ and length of $50~\text{m}$~\cite{FCC_CDR3}. The innermost layer of the detector consists of the tracking system, which tracks the trajectories of charged particles in the presence of a strong magnetic field. Surrounding this is the calorimeter, which is divided into electromagnetic and hadronic components to measure the energy of electrons, photons, and hadrons. The outermost layer contains the muon detector, designed to identify and measure the momentum of muons, which penetrate through the calorimeter. The detector design has been simulated using the FCC-hh configuration card in ~\texttt{DELPHES}~\cite{deFavereau:2013fsa, Selvaggi:2717698}.

The detectors of FCC-hh will feature highly granular tracking and calorimetry systems. The innermost tracking layer is designed to have three layers consisting of $25–33.3\mu\text{m} \times 50–400\mu\text{m}$ pixels within $r < 200~\text{mm}$, $33.3\mu\text{m}\times 400\mu\text{m}$ macro-pixels in the $200 < r < 900~\text{mm}$ region, and $33.3\mu\text{m} \times 2-50~\text{mm}$ macro-pixels in the  $900 < r < 1600~\text{mm}$ region \cite{FCC_CDR3}. The combined tracker resolution per detector layer is $7.5-9.5~\mu~\text{m}$. The total number of readout channels combining the three layers is $16\times 10^9$, which is significantly larger than that of the Phase II ATLAS and CMS tracking systems ($6\times 10^9$ and $2.2\times 10^9$ respectively) \cite{FCC_CDR3}.

At large $p_T$, the momentum resolution ($\sigma(p_T)$) is largely dependent on the detector resolution, which can be estimated using the approximation:

\begin{equation}
\frac{\sigma(p_T)}{p_T} \simeq \frac{\sigma_{\theta} p_T}{B R^2}
\end{equation}

where $\sigma_\theta$ is the tracker resolution per layer, B is the uniform magnetic field, and R is the tracker radius.  
in the most optimistic scenario, the momentum resolution $\sigma(p_T)/p_T$ with the proposed granularity is $\sim$20\% for $p_T = 10~\text{TeV}$, and $\sim$2\% for $p_T = 1~\text{TeV}$.

In the case of FCC-hh, the large $p_T$ of objects resulting from the decay of a highly massive particle leads to highly collimated decay products. Accurate reconstruction of these events depends on high-resolution detectors to resolve the individual signatures within these ultra-boosted objects \cite{top_FCC_5}.

The proposal for the electromagnetic calorimeter (ECAL) and the hadron calorimeter (HCAL) of the FCC-hh collider assumes a ($\Delta \eta,\Delta \phi$) segmentation of (0.01, 0.012) and (0.025, 0.025), respectively in the central region, i.e., $|\eta|<2.5$. The energy deposits of the particles in a tower are termed \textit{hits}. The energy deposits are smeared by a log-normal distribution with a standard deviation $\sigma(\eta,E)$, which is also the resolution, given by:
\begin{equation}
    \left(\frac{\sigma}{E}\right)^2 = \left(\frac{S(\eta)}{\sqrt{E}}\right)^2 + C(\eta)^2
    \label{eq:energy_res}
\end{equation}

where $E$ is the energy of the particle, $\eta$ is the pseudorapidity, $S$ and $C$ are the \textit{stochastic} term and the \textit{constant} term respectively. We have neglected the \textit{noise} term: $(N(\eta)/E)^2$, as it does not contribute much at high energies. In fact, for highly energetic particles, $C(\eta)$ contributes dominantly to the energy resolution of the calorimeters. For FCC-hh, the terms are listed in Table \ref{tab:cal_res}.
\begin{table}[hbt!]
\centering
\begin{tabular}{|c|c|c|c|}
\hline
 & $|\eta| < 1.7$ & $1.7 < |\eta| < 4.0$ & $4.0 < |\eta| < 6.0$ \\ \hline
$S_{\text{ECAL}}$ & 10$\%$ & 10$\%$ & 30$\%$ \\
$C_{\text{ECAL}}$ & 0.7$\%$ & 0.7$\%$ & 3.5$\%$ \\ \hline
$S_{\text{HCAL}}$ & 50$\%$ & 60$\%$ & 100$\%$ \\
$C_{\text{HCAL}}$ & 3$\%$ & 3$\%$ & 10$\%$ \\ \hline
\end{tabular}
\caption{The stochastic and constant terms of the energy resolution of ECAL and HCAL as a function of pseudorapidity in the conceptual detector design of FCC-hh \cite{FCC_CDR3}.}
\label{tab:cal_res}
\end{table}

One of the significant challenges is managing the high radiation levels and particle fluxes, which require robust radiation-hard materials and advanced cooling systems. The detectors must also handle high levels of pile-up, where multiple proton-proton collisions occur in the same bunch crossing, complicating data analysis. In the ultimate scenario when FCC-hh runs with an instantaneous luminosity of $30\times 10^{34}~\text{cm}^{-2}\text{s}^{-1}$, the average pile-up is projected to reach a maximum value of 1000. In our study of highly boosted jets, we have neglected the impact of pile-up. Over 25 years of operation, it is expected to collect 30~$\text{ab}^{-1}$ of data.

In the following section, we provide an overview of the simulation details employed for generating signal and background events.

\section{Simulation of events}
\label{sec:simu}
We generated the following standard model (SM) processes: QCD dijet production ($pp\to jj$), pair-production of top and anti-top quarks ($pp\to t\bar{t}$), pair production of $W$ boson ($pp\to W^+W^-$), as the source of boosted QCD jets, top jets, and $W$ jets, respectively. 
These samples are generated using \texttt{PYTHIA 8} \cite{Sjostrand:2007gs} with \texttt{NNPDF 2.3 LO} PDF \cite{NNPDF:2014otw}. We generate the samples in five bins according to $p_T$ of the boosted jet that we wish to tag. The $p_T$ bins considered in this study are 2--4~TeV, 4--6~TeV, 6--8~TeV, 8--10~TeV, 10--12~TeV, 12--14~TeV, and 14--16~TeV. 

\begin{figure}[hbt!]
    \centering
    \includegraphics[width=0.5\textwidth]{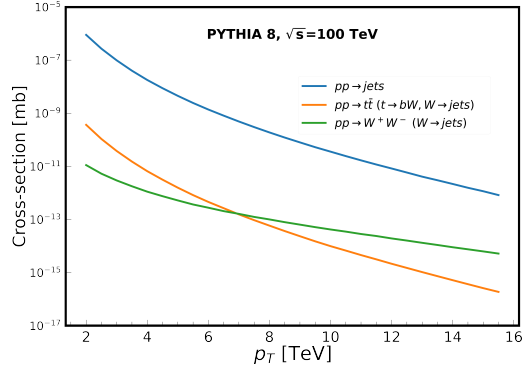}
    \caption{Cross-section as a function of $p_T$ at FCC-hh for the three major SM background processes considered in this study: $pp\to jj$, $pp\to t\bar{t}$, $pp\to W^+W^-$.}
    \label{fig:cs}
\end{figure}

Figure \ref{fig:cs} shows the cross-section of the processes as calculated by \texttt{PYTHIA 8}. The QCD di-jet events have the largest cross-section. At lower energies, the SM $t\bar{t}$ is the next dominant process, while at higher energies, the SM $WW$ process has a higher cross-section.

As discussed in Section \ref{sec:bsm_theory}, we consider three BSM processes - the $Z'$ production with $f\bar{f}\to Z' \to t\bar{t}$, which results in two top jets in the final state, the neutral heavy Higgs boson ($H$) production that decays to two $W$ jets using the $p p \to H \to W^+W^-$ process, and the 
$pp\to j B'$ production, with the singlet $B'$ decaying to a $t$ and a $W$. The generation procedure of the signal benchmarks has been discussed previously.

The backgrounds of the above BSM processes also include many other SM processes, which have been simulated for the analysis. These include SM $pp\to WZ$, and $pp\to ZZ$, which have been generated using \texttt{PYTHIA 8}. Other processes like SM $pp\to W+\text{jet}$ and SM $pp\to Z+\text{jet}$ have been generated using \texttt{MADGRAPH 5}, followed by showering and hadronisation via \texttt{PYTHIA 8}. In all the cases, we only consider the hadronic decay of the $W$ boson. These backgrounds have not been included in the training samples, but are only considered during the BSM analysis later.

For our analysis, we have multiplied the cross-sections of QCD, $t\bar{t}$, $VV$, and $V+jets$ ($V=W^\pm/Z$) processes with a k-factor of 2 to approximately account for the higher-order corrections. For the signal, we have not used any k-factor.

All the above-mentioned samples are passed through \texttt{DELPHES 3.5.0} \cite{deFavereau:2013fsa} simulation of a typical FCC-hh detector.  Table \ref{tab:delphes_res} shows the resolution of the tracking system and the $\eta-\phi$ segmentation of the Electromagnetic Calorimeter (ECAL) and Hadronic Calorimeter (HCAL) of FCC-hh as parametrised in \cite{Selvaggi:2717698}.

\begin{table}[hbt!]
\centering
\begin{tabular}{|ccc|}
\hline
\multicolumn{3}{|c|}{Tracker} \\ \hline
\multicolumn{1}{|c|}{$|\eta| < 6.0$} & \multicolumn{2}{c|}{$\sigma_\theta = 0.001$ } \\ \hline
\multicolumn{3}{|c|}{Calorimeters} \\ \hline
\multicolumn{1}{|c|}{} & \multicolumn{1}{c|}{ECAL} & HCAL \\ \hline
\multicolumn{1}{|c|}{$|\eta| < 2.5$} & \multicolumn{1}{c|}{$(\Delta\eta,\Delta\phi)=(0.01,0.012)$} & $(\Delta\eta,\Delta\phi)=(0.025,0.025)$ \\ \hline
\multicolumn{1}{|c|}{$2.5 < |\eta| < 6.0$} & \multicolumn{1}{c|}{$(\Delta\eta,\Delta\phi)=(0.025,0.025)$} & $(\Delta\eta,\Delta\phi)=(0.05,0.05)$ \\ \hline
\end{tabular}
\caption{Angular resolution of the tracking system and segmentation in the Electromagnetic and Hadronic calorimeters of FCC-hh as parametrised in \cite{Selvaggi:2717698}. }
\label{tab:delphes_res}
\end{table}

The tracking efficiency for charged hadrons and the identification efficiency used by \texttt{DEPLHES} for electrons, muons, and photons are listed in Table~\ref{tab:delphes_eff} in Appendix~\ref{app:delphes}. In case of two or more tracks lying within the angular resolution $\sigma_\theta$, only the track with the highest momentum is reconstructed.

We relied on the particle-flow (PF) event reconstruction in \texttt{DELPHES}, which is based on a simplified approach that combines momentum and energy measurement from the tracking systems as well as calorimeter towers for event reconstruction. The algorithm is based on the complementarity between track momentum resolution and calorimeter energy resolution. For electrons and charged hadrons, the FCC-hh tracker provides excellent resolution up to $E=500~\text{GeV}$ and $E=2~\text{TeV}$ respectively, beyond which the calorimeter measurement dominates \cite{Selvaggi:2717698}. \textit{Particle-flow tracks} are reconstructed for each charged object with an associated track. Neutral objects like photons, neutral hadrons, and charged hadrons with no associated tracks have significant energy deposits in the calorimeters. If the energy deposited in the calorimeters exceeds that measured from tracks, the final measurement is calculated as the difference between the two. These are reconstructed as \textit{particle-flow photons} and \textit{particle-flow neutral hadrons}.

\texttt{FASTJET 3.4.2} \cite{Cacciari:2005hq, Cacciari:2011ma} is used on the PF candidates to construct jets of radius $R$ = 0.8. We choose $R = 0.8$ because it is a typical choice of jet radius for large-radius jets in the LHC experiments. In the next section, we study the disadvantages of using fixed-radius jets. To mitigate the disadvantages, we also discuss the usage of dynamic radius jets. We select jets that satisfy the pseudorapidity cut of $|\eta| < 2.5$. These jets are then passed through the default flavour tagger modules of \texttt{DELPHES}. In this paper, we have not considered pile-up, which is left as an open topic for future studies. 

The ML taggers are originally trained on 1M QCD jets, 1M top jets, and 1M jets distributed over seven $p_T$ bins. The dataset is split in an 80:20 ratio between training and validation. Model performance is evaluated on an equivalent, independent test set. Finally, for the BSM analysis, we generate 560K $jj$, 560K $t\bar{t}$, 560K $tW$, 560K $VV$, and 560K $Vj$ events, where $V=W,~Z$, over seven different $p_T$ bins.

\section{Challenges and mitigation strategies for jets at $\sqrt{s} = 100$~TeV}
\label{sec:challenge}
At the 100~TeV collider, the high center-of-mass energy of $pp$ collisions produces heavy BSM particles with masses up to tens of TeV. When such particles decay into jets, the jets are highly boosted and thus highly collimated. In our study, we consider top and $W$ jets with $p_T$ ranging from 2~TeV to 16~TeV. 
As a consequence of poor jet energy resolution at very high $p_T$, we need to construct a tagger that caters to multiple $p_T$ ranges because the characteristics of jets vary widely across the entire $p_T$ range (see Appendix~\ref{app:mig_jet}). Reconstructing boosted particles from their final state requires addressing the unique challenges posed by these extreme energies. This section explores these challenges and discusses potential techniques to overcome them.

\subsection{Effect of fixed jet radius on top and QCD jet mass}
\begin{figure}[hbt!]
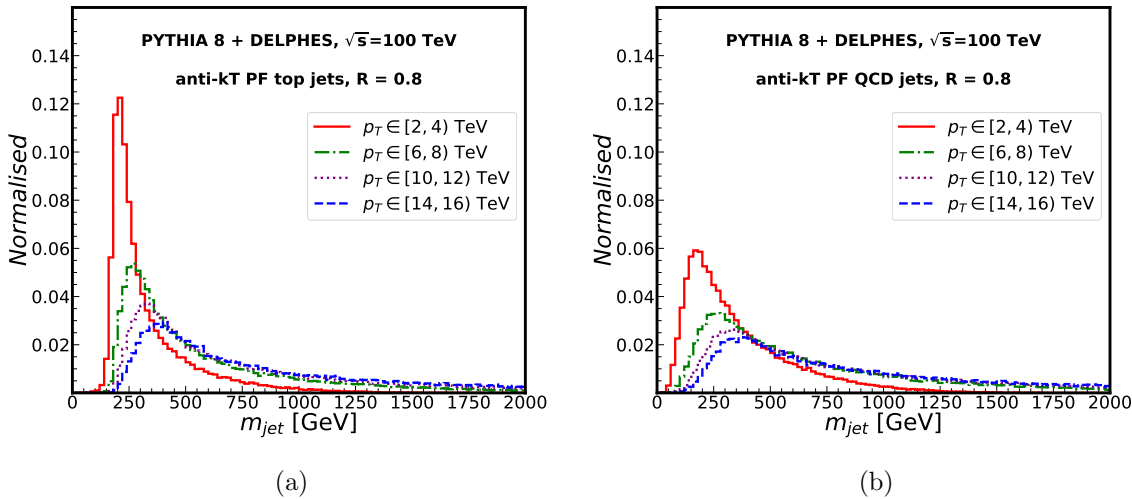

\centering
\begin{subfigure}[c]{0.5\linewidth}
\includegraphics[width=\linewidth]{mjet_pT_top.pdf} 
\caption{}
\label{fig:2a}
\end{subfigure}\hfill    
\begin{subfigure}[c]{0.5\linewidth}
\includegraphics[width=\linewidth]{mjet_pT_qcd.pdf}
\caption{}
\label{fig:2b}
\end{subfigure}
\caption{Normalised $m_{\text{jet}}$ distribution of (a) top jets and (b) QCD jets with transverse momenta ($p_T$) of 2--4~TeV, 6--8~TeV, 8--10~TeV, and 14--16~TeV. Jets have been constructed with a fixed jet radius $R=0.8$.}
\label{fig:mjet_pT}
\end{figure}

At sufficiently large boosts, the final-state hadrons from decays of resonances (like $W$ boson or top quark) merge into a
single jet. The jet mass is a potent discriminator between resonance jets and background QCD jets (i.e., jets originating from the light-flavour quarks or gluons).
The distribution of jet mass ($m_{\text{jet}}$) for the top and QCD jets constructed with the radius of the jet, $R = 0.8$, is shown in Figure \ref{fig:mjet_pT}. The jets carry $p_T$ in the range of 2--4~TeV, 6--8~TeV, 10--12~TeV, and 14--16~TeV.

 The mean squared invariant mass of a QCD jet in the collinear approximation is given by \cite{Ellis_2008, Salam_2010, Larkoski:2015yqa},

\begin{equation}
    \langle m^2 \rangle \simeq  C\cdot\frac{\alpha_S}{\pi}p_T^2 R^2
\end{equation}
where $C$ is a coefficient that depends on the jet algorithm. At high $p_T$ and with a fixed jet radius $R$, the mass increases, as shown in Figure \ref{fig:mjet_pT}. In addition, the jet captures contamination from the initial-state radiation (ISR), final-state radiation (FSR), and underlying events (UE). As described in \cite{Larkoski:2015yqa}, in the case of highly boosted top jets, the FSR from the top quark is not dead-cone suppressed any more, as the dead cone radius shrinks as $\simeq m_{\text{top}}/p_T$ for very high $p_T$. On the other hand, a single emission at the jet boundary with transverse momenta $p_{T,ISR}$ can substantially change the jet mass, as
\begin{equation}
m^2 \simeq m_{\text{top}}^2 + p_T \cdot p_{T,ISR} \cdot R^2
\end{equation}
The authors in \cite{Larkoski:2015yqa} showed that an ISR jet with transverse momenta $p_{T,ISR} \sim m_{\text{top}}^2/p_TR^2$ can change the jet mass by $\mathcal{O}$(1). At the FCC, where $p_T$ can reach tens of TeV, radiative emissions of a few GeV can alter and increase the jet mass. The large tail in the jet mass distribution manifests in QCD jets, too, as seen in Figure \ref{fig:2b}.

Jet grooming algorithms are applied to jets to remove contaminating radiation from uncorrelated sources, with the aim of improving the jet mass scale and resolution. However, they are not optimised for high $p_T$ jets~\cite{Larkoski:2014bia}. As a result, when grooming algorithms such as the soft drop technique are applied to fixed radius high-$p_T$ jets, observables like the jet mass still possess the problem of a long tail in distributions (see Figure~\ref{fig:groom} in Appendix~\ref{app:soft drop}).

To mitigate the above issues and reduce the contamination, we make use of a dynamic jet radius. Dynamic jet radius has already been used at the LHC for various studies~\cite{Mukhopadhyaya:2023rsb, Jain:2023rln, Lapsien:2016zor}. The choice of the jet radius $R$ is crucial because not only should the jet algorithm reconstruct the top quark well, but it also should be able to mitigate the contamination from underlying events. In the following section, we construct the dynamic radius using parton decay information.

\subsubsection{Jet radius estimation at the parton level}
To choose the radius that captures the correct jet substructure, we go back to the parton-level information of top jets. The top quark ($t$) decays to a $b$ quark and a $W$ boson, which in turn decays hadronically to a light quark and an anti-quark, which we shall refer to as $u$ and $d$. 

\begin{figure}[hbt!]
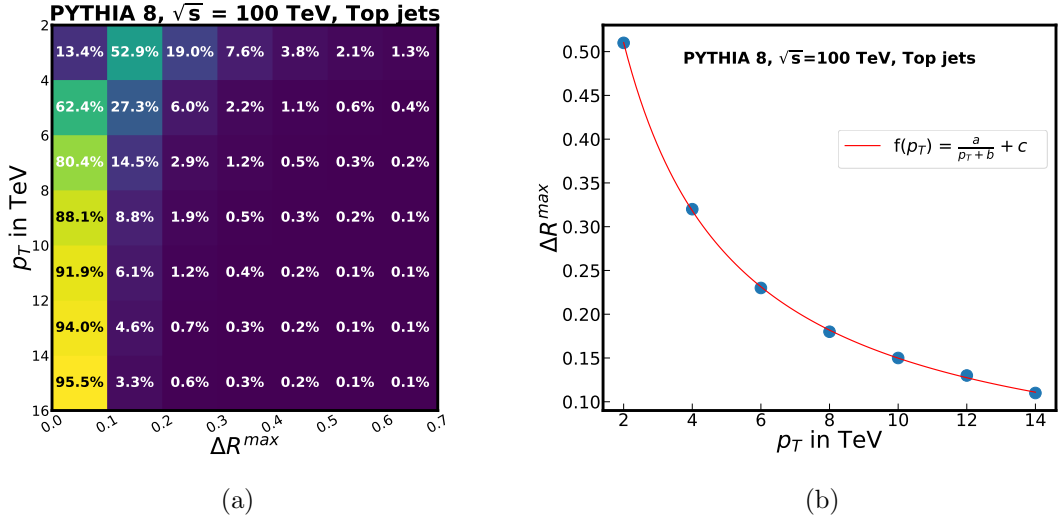

\centering
\begin{subfigure}[c]{0.5\linewidth}
\includegraphics[width=\linewidth]{radius.pdf} 
\caption{}
\label{fig:1a}
\end{subfigure}\hfill    
\begin{subfigure}[c]{0.5\linewidth}
\includegraphics[width=\linewidth]{fit.pdf}
\caption{}
\label{fig:1b}
\end{subfigure}
\caption{(a) A two-dimensional matrix showing the percentage of $t\to bW (W\to ud)$ events that have all the three daughter partons within radius $\Delta R^{max}$ from the mother top quark. (b) The variation of the $\Delta R^{max}$ value with 95$\%$ success rate of containing the full top decay, as a function of the $p_T$ of the top quark. Here, $f(p_T)$ is the fitting function.}
\label{fig:parton_DR}
\end{figure}

The top decay products lie within a cone, whose size depends on the initial $p_T$. For events in each bin, we calculate the distance $\Delta R$ between the parent top quark $t$ and its three daughter quarks, $b$, $u$, and $d$. We then find out the maximum of the three distances, $\Delta R^{max} = \text{max}(\Delta R_{tb}, \Delta R_{tu}, \Delta R_{td})$. $\Delta R^{max}$ serves as an approximate measurement of the cone size. We expect that this number decreases with increasing $p_T$. Figure \ref{fig:1a} displays a two-dimensional matrix of the $p_T$ ranges and the $\Delta R^{max}$ values. Each square shows the percentage of top jets with $p_T$ in the $i^{th}$ bin and $j^{th}$ $\Delta R^{max}$. 

From Figure \ref{fig:1a}, we see that as we move on to higher $p_T$, the top jets have smaller decay cone size, as denoted by the smaller $\Delta R^{max}$. Only 13.4$\%$ of tops generated with $p_T$ of 2--4~TeV have the $b$, $u$, and $d$ contained fully within $\Delta R = 0.1$, while in the case of 14--16~TeV tops, the number is 95.5$\%$. Using this information, we can estimate the cone radius that should be used so that a reconstructed jet contains the full information of the top.

We determine the $\Delta R^{max}$ value such that the top quark decay products, $b$, $u$, and $d$, are fully contained within a cone of this radius 95$\%$ of the time. We assign this $\Delta R^{max}$ value to the lower value of the corresponding $p_T$ bin. For instance, if the top quarks in the 2--4~TeV have a $\Delta R^{max}$ of 0.5, we use 0.5 for 2~TeV. Similarly, if the 4--6~TeV bin has a $\Delta R^{max}$ of 0.3, we use 0.3 for 4~TeV, and we follow this for all the remaining bins. 
In Figure \ref{fig:1b}, we have plotted the $\Delta R^{max}$ corresponding to each $p_T$ bin.  We fit the data points using a function of the form,
\begin{equation}
f(p_T) = \frac{a}{p_T+b} + c
\label{eq:radius}
\end{equation}
where  $a$,  $b$, and $c$ are fitting parameters, which are set to 1.64, 1.3, and 0.001, respectively. The dependence of R on the inverse $p_T$ is inspired by the approach in \cite{Larkoski:2015yqa}. As the $p_T$ reaches $\approx$ 15.7~TeV, $\Delta R^{max}$ goes below 0.1.

In the later part of our work, we use the function to calculate the jet radius dynamically instead of using a fixed radius.

\begin{figure}[hbt!]
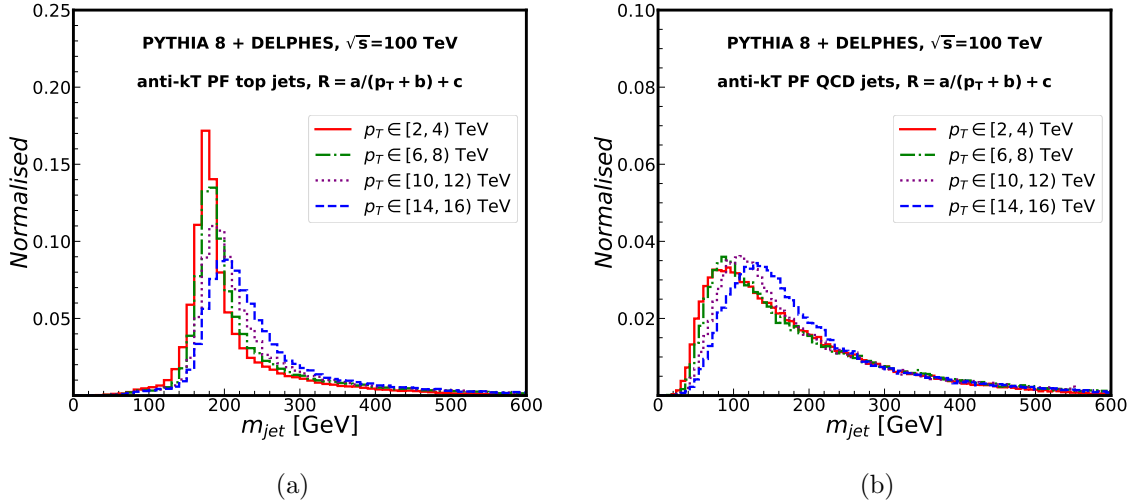

\centering
\begin{subfigure}[c]{0.5\linewidth}
\includegraphics[width=\linewidth]{recl_mjet_pT_top.pdf} 
\caption{}
\label{fig:3a}
\end{subfigure}\hfill    
\begin{subfigure}[c]{0.5\linewidth}
\includegraphics[width=\linewidth]{recl_mjet_pT_qcd.pdf}
\caption{}
\label{fig:3b}
\end{subfigure}
\caption{Normalized $m_{\text{jet}}$ distribution of (a) top jets and (b) QCD jets with transverse momenta ($p_T$) of 2--4~TeV, 6--8~TeV, 8--10~TeV, and 14--16~TeV. The jet constituents have been reclustered using radius $R(p_T)=a/(p_T+b) +c$. }
\label{fig:mjet_dynR}
\end{figure}

As stated earlier, we construct anti-$k_T$ jets using \texttt{FASTJET} on particle-flow (PF) objects, using a fixed radius of 0.8. Now we recluster the jet constituents using the $p_T$ dependent function in Equation \ref{eq:radius}. Figure \ref{fig:mjet_dynR} shows the distributions of jet mass for top jets and QCD jets after reclustering. 

As seen in Figure \ref{fig:3a} and \ref{fig:3b}, using a dynamic jet radius that is a function of the jet's $p_T$ corrects the jet mass. The jet mass falls off faster in this case. However, the peak of the jet mass and the width of the distribution still increase with increasing $p_T$. This indicates that even with the new dynamic radius, the contamination could not be completely eliminated.

\subsection{Effects of finite detector resolution on top and QCD jet mass}

To further investigate the source of the high jet mass, we look into the detector parametrisation of the 100~TeV collider because the energy-momenta of the jet constituents depend to a large extent on the resolution of the detector elements.
As stated in Section \ref{sec:simu}, we use the \texttt{DELPHES} modeling of the FCC-hh detector for simulation. The resolution used in the calorimeters and the tracking system is listed in Table \ref{tab:delphes_res} in Section \ref{sec:fcc}. Since we are using particle-flow (PF) objects, consisting of tracks and calorimeter towers, we focus on the resolution effects of these two detector elements.

\subsubsection{Calorimeters}
The \texttt{DELPHES} algorithm iterates over a collection of tower hits and processes them to fill a tower by adding the total energy deposited in the tower. The tower's $\eta-\phi$ is then calculated as the $\eta-\phi$ of the cell's geometric centre, smeared by a Gaussian distribution. The four-momentum of the tower is recalculated from the $\eta, \phi, E$ information. 

\begin{figure}[hbt!]
\centering
\begin{subfigure}[c]{0.5\linewidth}
\includegraphics[width=\linewidth]{rec_mjet_2_4TeV_sigma.pdf} 
\caption{}
\label{fig:4a}
\end{subfigure}\hfill    
\begin{subfigure}[c]{0.5\linewidth}
\includegraphics[width=\linewidth]{rec_mjet_2_4TeV_hcal.pdf}
\caption{}
\label{fig:4b}
\end{subfigure}
\begin{subfigure}[c]{0.5\linewidth}
\includegraphics[width=\linewidth]{rec_mjet_14_16TeV_sigma.pdf} 
\caption{}
\label{fig:4c}
\end{subfigure}\hfill    
\begin{subfigure}[c]{0.5\linewidth}
\includegraphics[width=\linewidth]{rec_mjet_14_16TeV_hcal.pdf}
\caption{}
\label{fig:4d}
\end{subfigure}
\caption{Normalised $m_{\text{jet}}$ distribution of 2--4~TeV ((a), (b)) and 14--16~TeV ((c), (d)) top jets and QCD jets using different calorimeter configurations. Subfigures (a), (c) show the change in $m_{\text{jet}}$ when the energy resolution is varied. Subfigures (b), (d) show the change in $m_{\text{jet}}$ when the angular resolution is varied.}
\label{fig:mjet_det}
\end{figure}

With a poorer energy resolution and granularity than the values that are currently under consideration, the jet mass resolution worsens (see Figure~\ref{fig:worse_mjet} in Appendix~\ref{app:mig_jet}). To examine the opposite effect, we construct two hypothetical scenarios in which the energy resolution and granularity are improved relative to the present values. First, we set the stochastic and constant terms in the ECAL and HCAL resolution parameter $\sigma$ to a very low value of 0.1$\%$, i.e., $S$ = $C$ = 0.1\%. This reduces the extent to which the energy can be smeared. Then, we use a more granular segmentation of the HCAL by setting the cell size of HCAL towers to that of the ECAL towers. Reducing the HCAL cell size increases the number of cells within the jet clustering radius. Figure \ref{fig:mjet_det} shows the jet mass distribution for 2--4~TeV jets (Figures \ref{fig:4a}, \ref{fig:4b}) and for 14--16~TeV jets (Figures \ref{fig:4c}, \ref{fig:4d}).

Improving the energy resolution did not affect the jet mass reconstruction. On the other hand, increasing the $\eta-\phi$ segmentation of the HCAL towers by a factor of two changes the reconstructed jet mass substantially for the 14--16~TeV jets, although it does not affect the 2--4~TeV jets. The 14--16~TeV top jets have a sharper peak, and the overlap between the top and the QCD jets increases as the QCD distribution peaks at lower values.  The improved angular resolution of the HCAL allows the highly collimated decay products of the 14--16~TeV to deposit energy in an increased number of highly granular tower cells, causing the PF reconstruction to be more efficient when matching the charged tracks to towers. For 2--4~TeV jets, the default segmentation already offers sufficient granularity to resolve the particles, hence increasing the segmentation does not provide much improvement. 

To conclude, we find that in the study of ultra-boosted objects, granularity plays a more important role than energy resolution, as seen in Figure \ref{fig:4d}. If such a fine granularity, where HCAL and ECAL cell sizes are of the same size, is indeed feasible to construct at the FCC-hh experiment, it will improve the mass reconstruction of highly boosted objects. However, we continue our study with the currently proposed granularity of ECAL and HCAL. We can also obtain a high angular resolution by using charged particle tracks, which we study in the next section.

\subsubsection{Tracking system}
\begin{figure}[hbt!]
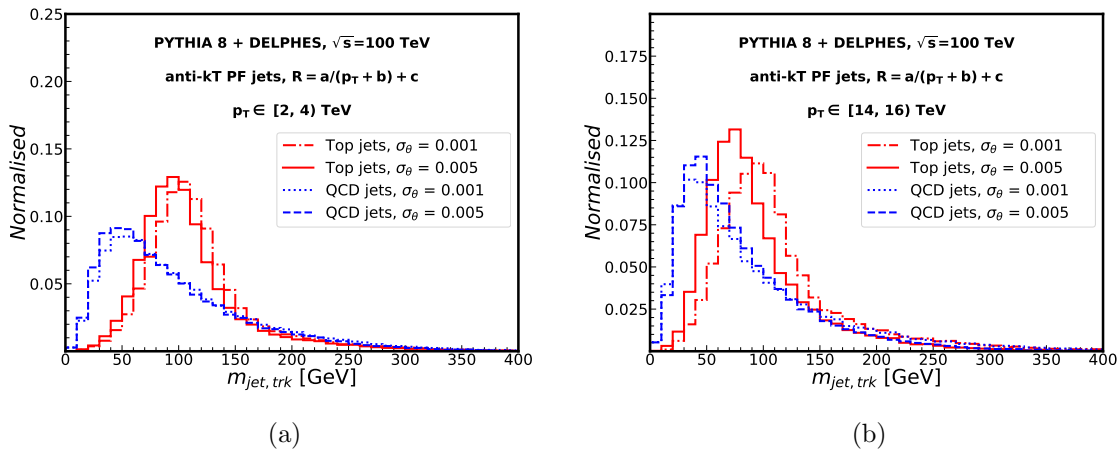

\centering
\begin{subfigure}[c]{0.5\linewidth}
\includegraphics[width=\linewidth]{track_rec_mjet_2_4TeV.pdf} 
\caption{}
\label{fig:5a}
\end{subfigure}\hfill    
\begin{subfigure}[c]{0.5\linewidth}
\includegraphics[width=\linewidth]{track_rec_mjet_14_16TeV.pdf} 
\caption{}
\label{fig:5b}
\end{subfigure}\hfill    
\caption{Normalised track-based $m_{\text{jet}}$ distribution of (a) 2--4~TeV and (b) 14--16~TeV top jets and QCD jets with varying angular resolution of the tracker.}
\label{fig:mjet_trk}
\end{figure}

The proposed angular resolution for tracking in the FCC detector design is $\sigma_\theta = 0.001$, which is an order of magnitude less than that of the ECAL. When two or more tracks are separated with a distance less than $\sigma_\theta$, the one with the maximum $p_T$ is considered by the dense track filtering algorithm in \texttt{DELPHES}. We calculate the mass on the constituent tracks of the PF jet with reclustered radius ($m_{\text{trk}}$). We also compute $m_{\text{jet}}^{\text{trk}}$ with a degraded tracking resolution, $\sigma_\theta$ = 0.005. Figure \ref{fig:mjet_trk} shows the distribution of mass calculated on tracks for 2--4~TeV (\ref{fig:5a}) and 14--16~TeV jets (\ref{fig:5b}), with two values of the $\sigma_\theta$. 

The mass distribution for tracks assumes lower values because they do not capture the full energy of a jet. Additionally, as we see in Figure \ref{fig:mjet_trk}, $m_{\text{jet}}^{\text{trk}}$, especially for top jets, is very sensitive to the angular resolution. The track-based mass needs to be rescaled with a ratio of the original jet $p_T$ to the track $p_T$ \cite{Larkoski:2015yqa} if we want to obtain the correct estimation of the jet mass. In the next section, we perform a similar study and check the detector effects on the jet substructure.

\subsection{Effects of finite detector resolution on N-Subjettiness observables}
\label{sec:sub_str}
For the purpose of identifying the top jets and QCD jets based on their substructure properties, a popular method is the use of the N-Subjettiness \cite{Thaler:2010tr} ($\tau_N$) observable, which measures the \textit{prongness} of a jet relative to $N$ subjet directions $\hat{n}_{j}$, 

\begin{equation}
\tau_N^{(\beta)} = \frac{1}{\sum_{\alpha\in\text{jet}} p_{T,\alpha}R_0^{\beta}} \sum_{\alpha\in\text{jet}}p_{T,\alpha} \min_{k=1,...,N}(\Delta R_{k,\alpha})^{\beta}
\label{eq:Nsub}
\end{equation}
\noindent
where $\Delta R_{,\alpha}$ is the separation between the $\alpha$-th jet constituent and the $k$-th axis, $\beta$ (which should be greater than zero) is an arbitrary weighting exponent to guarantee infrared safety, and $R_0$ is a radius parameter. 
For a jet having $N$ subjets, $\tau_N<\tau_{N-1}$ and the ratio $\tau_{N}/\tau_{N-1}$ is small. 

\begin{figure}[hbt!]
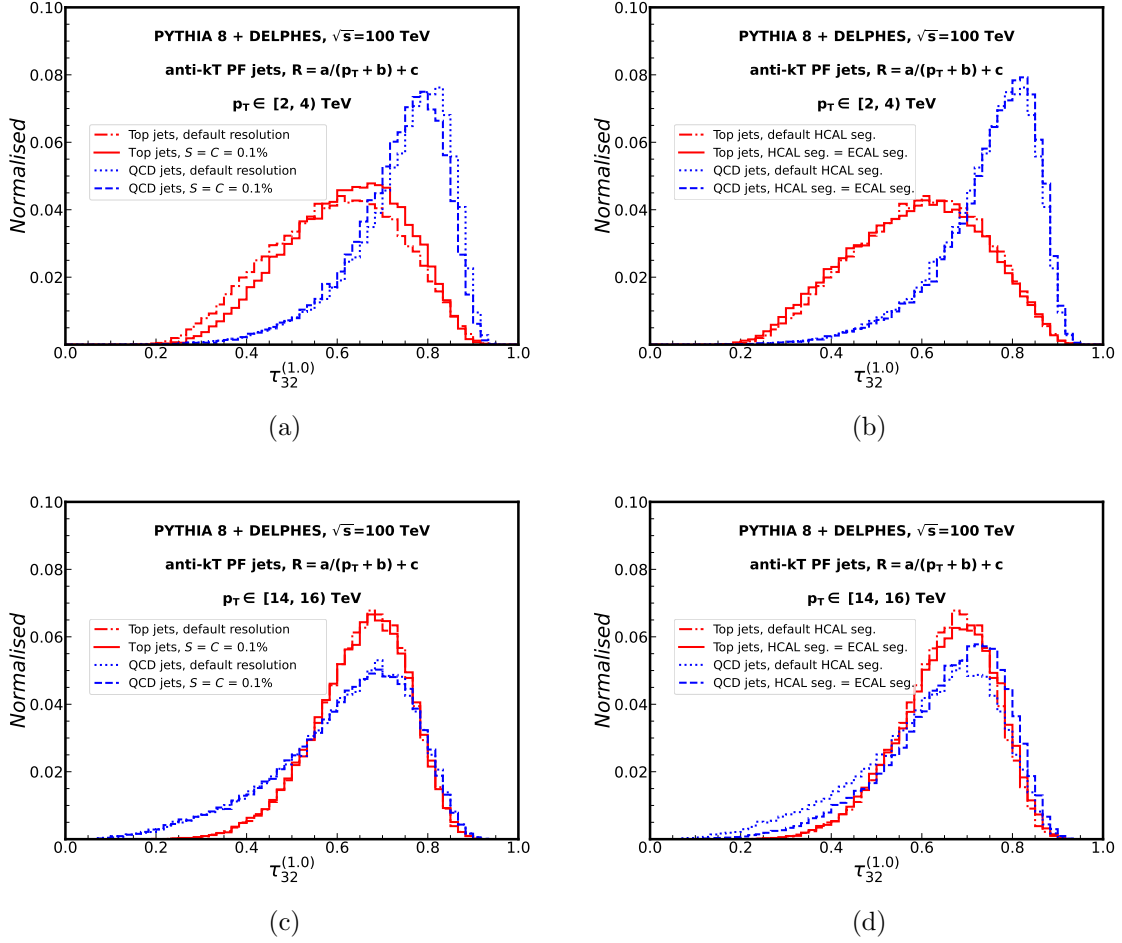

\centering
\begin{subfigure}[c]{0.5\linewidth}
\includegraphics[width=\linewidth]{rec_tau32b1_2_4TeV_sigma.pdf} 
\caption{}
\label{fig:6a}
\end{subfigure}\hfill    
\begin{subfigure}[c]{0.5\linewidth}
\includegraphics[width=\linewidth]{rec_tau32b1_2_4TeV_hcal.pdf}
\caption{}
\label{fig:6b}
\end{subfigure}
\begin{subfigure}[c]{0.5\linewidth}
\includegraphics[width=\linewidth]{rec_tau32b1_14_16TeV_sigma.pdf} 
\caption{}
\label{fig:6c}
\end{subfigure}\hfill    
\begin{subfigure}[c]{0.5\linewidth}
\includegraphics[width=\linewidth]{rec_tau32b1_14_16TeV_hcal.pdf}
\caption{}
\label{fig:6d}
\end{subfigure}
\caption{Normalised $\tau_{32}^{(1.0)}$ distribution of 2--4~TeV ((a), (b)) and 14--16~TeV ((c), (d)) top jets and QCD jets using different calorimeter configurations. Subfigures (a), (c) show the change in $\tau_{32}^{(1.0)}$ when the energy resolution is varied. Subfigures (b), (d) show the change in $\tau_{32}^{(1.0)}$ when the angular resolution is varied.}
\label{fig:tau32b1_det}
\end{figure}

In top quark decays, which yield three distinct subjets, the ratio $\tau_3/\tau_2$ is expected to exhibit a peak at lower values, distinguishing it from the QCD background, which peaks at higher values. Figure \ref{fig:tau32b1_det} shows the distributions of $\tau_{32}^{(1.0)} = \tau_{3}/\tau_{2}$ for 2--4~TeV and 14--16~TeV jets with modifications to the energy and the angular resolution of the calorimeters.

To begin with, we note that the discrimination between top and QCD jets in terms $\tau_{32}^{(1)}$ degrades with an increase in the transverse momenta of the jets. Not only is there a greater overlap between the two for 14--16~TeV jets, but the top and the QCD jets follow reverse trends in $\tau_{32}$. In other words, the N-Subjettiness observable breaks down at high transverse momenta, even with the $(\Delta \eta,\Delta \phi)=(0.01,0.01)$ calorimeter resolution at the FCC-hh. Improving the HCAL resolution corrects the observable to some extent, but it is not sufficient to be advantageous as a tagger. Variable $\tau_{21}$ also exhibits similar behavior. We now calculate the observable on the tracks.

\begin{figure}[hbt!]
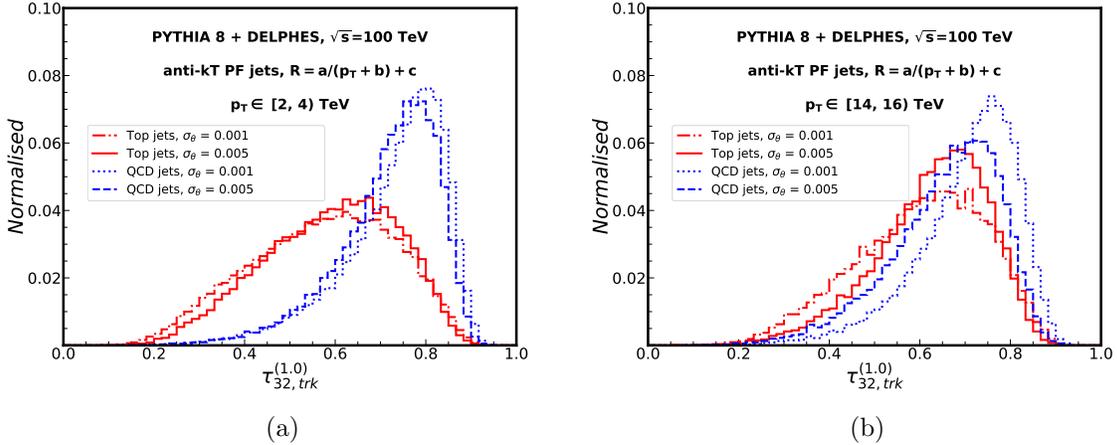

\centering
\begin{subfigure}[c]{0.5\linewidth}
\includegraphics[width=\linewidth]{track_rec_tau32b1_2_4TeV.pdf} 
\caption{}
\label{fig:7a}
\end{subfigure}\hfill    
\begin{subfigure}[c]{0.5\linewidth}
\includegraphics[width=\linewidth]{track_rec_tau32b1_14_16TeV.pdf} 
\caption{}
\label{fig:7b}
\end{subfigure}\hfill    
\caption{Normalised track-based $\tau_{32}^{(1.0)}$ distribution of (a) 2--4~TeV and (b) 14--16~TeV top jets and QCD jets with varying angular resolution of tracker.}
\label{fig:tau32b1_trk}
\end{figure}

Figure \ref{fig:tau32b1_trk} shows that an angular resolution of 0.001 in the tracker is beneficial for the boosted jets. The finer track resolution restores the correct substructure properties of the 14--16~TeV jets. As in the case of jet mass, decreasing $\sigma_\theta$ degrades the N-Subjettines. 

From our studies of the detector effects on the jet mass and $\tau_{32}$, we make the key observation that a full and accurate reconstruction of the momentum and the jet substructure at tens of TeV is limited by finite resolution of the calorimeters. Even the \texttt{DELPHES} particle-flow reconstruction that uses both track and calorimeter information fails to resolve the substructure at high $p_T$. However, particle flow algorithms developed in actual experiments may be built with mechanisms to make up for resolution issues and achieve at least as good a tracking-only reconstruction.

The tracker volume with the proposed angular resolution of $\sigma_\theta$ = 0.001 comes to the rescue. Solely track-based N-Subjettiness captures the \textit{prongness} successfully. However, one must keep in mind that this is the best-case scenario that does not take pile-up into consideration. Further studies with pile-up are required to see if the track-based observables work in that situation.

\subsection{Boosted $W$ jets at $\sqrt{s} = 100$~TeV}
\label{sec:wjet}
As discussed in Section \ref{sec:bsm_theory}, the production of a vector-like quark (VLQ) $B'$ at the FCC-hh that can decay into a top quark and a $W$ boson is not only a source of highly boosted top jets but also highly boosted $W$ jets. Figure \ref{fig:pT_Bp_w} shows the full range of transverse momenta of such $W$ jets coming from the decay of a $B'$, reaching up to 12~TeV for a $B'$ of mass $m_{B'} = 10$~TeV. In the previous section, we discussed the challenges that may arise while reconstructing top jets and the possible ways to mitigate them. First, we cluster the PF candidates with a radius $R$ that is $p_T$ dependent. This reduces the contamination by radiative effects. Next, we saw that the track-based jet substructure observable $\tau_{32}$ acts as a potent discriminant between the top and the QCD jets, even at high $p_T$. Similar challenges shall arise while reconstructing the $W$ boson.

\begin{figure}[hbt!]
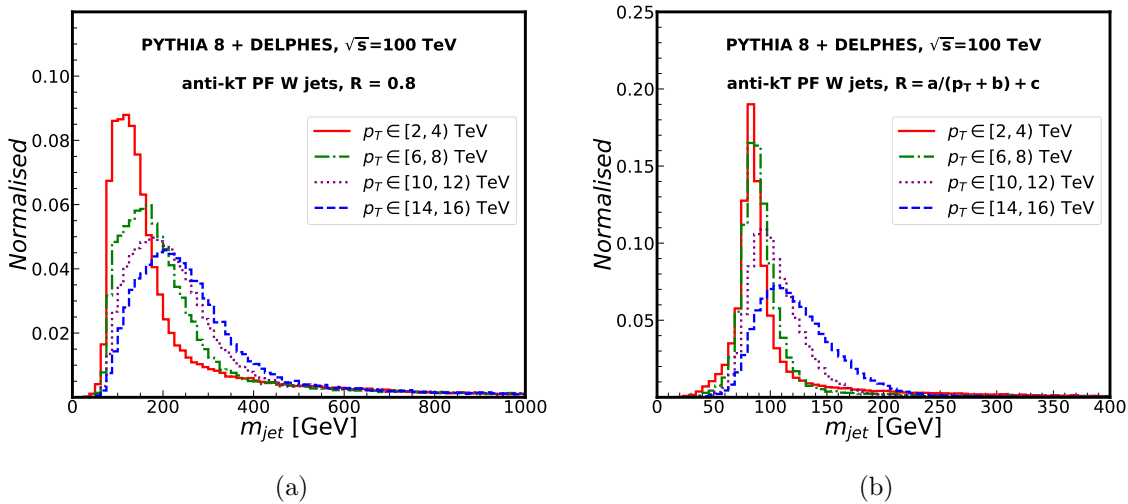

\centering
\begin{subfigure}[c]{0.5\linewidth}
\includegraphics[width=\linewidth]{mjet_pT_ww.pdf} 
\caption{}
\label{fig:8a}
\end{subfigure}\hfill    
\begin{subfigure}[c]{0.5\linewidth}
\includegraphics[width=\linewidth]{recl_mjet_pT_ww.pdf}
\caption{}
\label{fig:8b}
\end{subfigure}
\caption{Normalised $m_{\text{jet}}$ distribution of $W$ jets (a) before and (b) after reclustering with $R(p_T)$.}
\label{fig:mjet_dynR_ww}
\end{figure}

To study boosted $W$ jets, we pair-produce $W$ boson from a pair of quarks at $\sqrt{s} = 100$ TeV, where the $W$ boson further decays hadronically to two light quarks. We follow the same simulation procedure as described in Section \ref{sec:simu}. We use the dynamic radius $R(p_T) = a/(p_T+b)+c$ that was optimized for top jets to cluster the particle-flow objects. Figure \ref{fig:mjet_dynR_ww} shows the jet mass distribution of the boosted $W$ jets for four different $p_T$ ranges, with a fixed radius (Figure \ref{fig:8a}) and a dynamic radius (Figure \ref{fig:8b}). 

As in the case with top and QCD jets, the $W$ jet mass also suffers from radiation at the $\sqrt{s}$ = 100 TeV collider though being an electroweak process, the amount is less than that in the former cases. Using the radius $R(p_T)$, much of the radiative effects could be removed. For $p_T\in$[2,4] TeV jets, the mass is exactly reconstructed at the $W$ boson mass ($m_W \simeq 80$~GeV). With increasing $p_T$, the jet mass peak shifts to higher values due to increasing radiative effects and the worsening resolution of the detector.

\begin{figure}[hbt!]
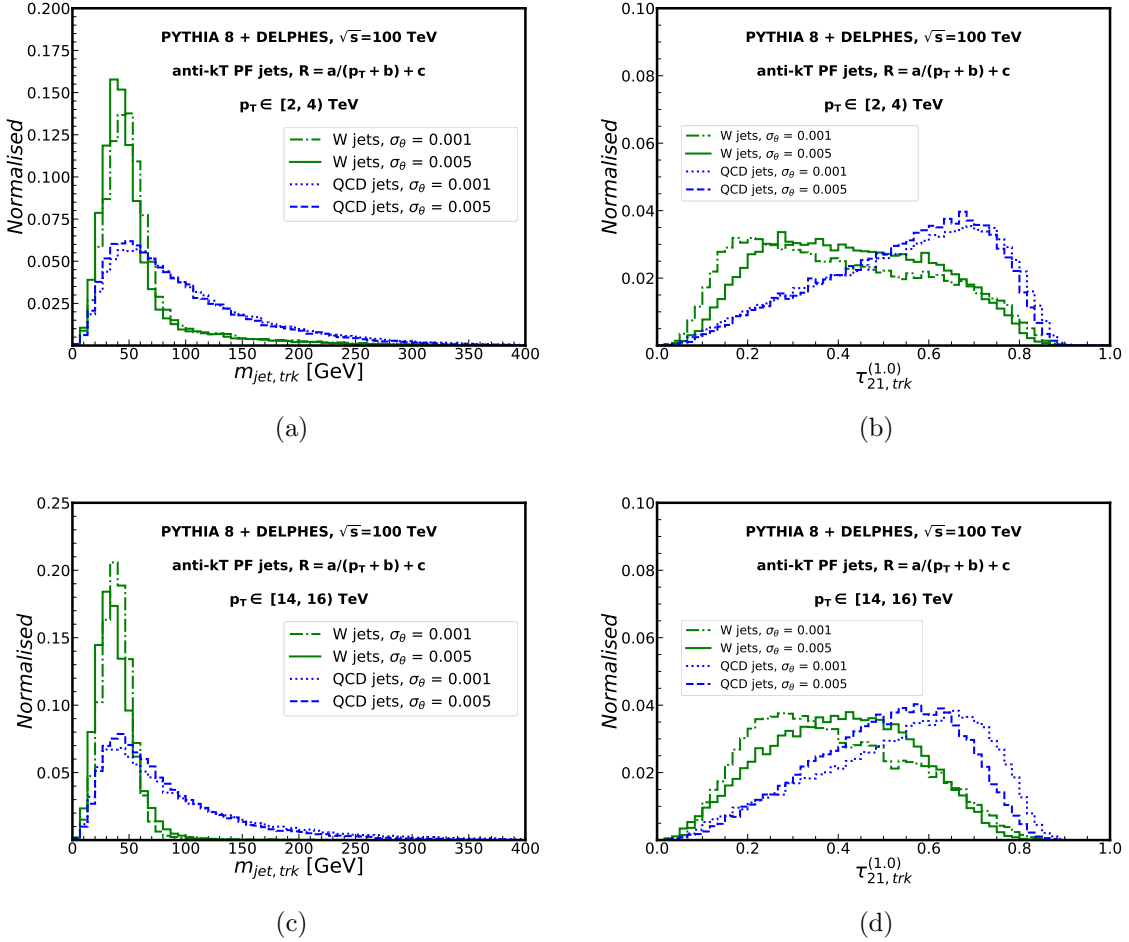

\centering
\begin{subfigure}[c]{0.5\linewidth}
\includegraphics[width=\linewidth]{track_ww_rec_mjet_2_4TeV.pdf} 
\caption{}
\label{fig:9a}
\end{subfigure}\hfill    
\begin{subfigure}[c]{0.5\linewidth}
\includegraphics[width=\linewidth]{track_rec_tau21b1_2_4TeV.pdf}
\caption{}
\label{fig:9b}
\end{subfigure}
\begin{subfigure}[c]{0.5\linewidth}
\includegraphics[width=\linewidth]{track_ww_rec_mjet_14_16TeV.pdf} 
\caption{}
\label{fig:9c}
\end{subfigure}\hfill    
\begin{subfigure}[c]{0.5\linewidth}
\includegraphics[width=\linewidth]{track_rec_tau21b1_14_16TeV.pdf}
\caption{}
\label{fig:9d}
\end{subfigure}
\caption{Normalised track-based $m_{\text{jet}}$ and $\tau_{21}^{(1.0)}$ distributions of 2--4 TeV ((a) and (b) respectively) and 14--16 TeV ((c) and (d) respectively) top jets and QCD jets with varying angular resolution of the tracker.}
\label{fig:mjet_tau21b1_trk_ww}
\end{figure}

Figure \ref{fig:mjet_tau21b1_trk_ww} shows the distributions of the jet mass and the two-prong N-Subjettiness ratio $\tau_{21}$ calculated on tracks for the $W$ jets and the QCD jets from the two extreme ends of the $p_T$ spectrum. Both $W$ and QCD peak at $\sim 50$~GeV, though QCD features a much longer tail. The track-based $\tau_{21}$ shows distinguishing properties between the $W$ and the QCD jets, the former peaking at lower values owing to the two-prong substructure. With the poorer resolution of tracks, the two tend to overlap with each other. 

Using a radius optimized for $W$ jets (using the parton level information of $W\to q \bar{q}$) could eliminate more of the radiation from the jets. We have used the dynamic radius optimized for top jets while reclustering the $W$ decay products because we primarily wanted to study boosted top signatures. In the next section, we discuss the possibility of making further improvements to the reconstructed jets through jet grooming algorithms.

\subsection{Jet grooming techniques}
\label{sec:grm_tech}
Initial-state radiation and the underlying event can substantially alter the measured jet mass. 
Jet grooming algorithms are applied to jets to remove contaminating radiation from uncorrelated sources, with the aim of improving the jet mass scale and resolution. 

While the reclustering technique using a dynamic $p_T$ dependent jet radius prevented ISR, FSR, and multi-parton interactions (MPI) contamination within the jets, we want to see if standard grooming techniques can do better in reducing the residual radiation.

We use the soft drop technique \cite{Larkoski:2014wba, Larkoski:2014bia} on the jets. The original jet with $R=0.8$ is first reclustered within $R(p_T)$ using the Cambridge/Aachen (C/A) algorithm instead of the anti-$k_T$ algorithm to form a structured tree with angular ordering. Then, the tree is declustered using the following algorithm:

\begin{enumerate}
    \item Undo the last step of C/A clustering to break the jet $j$ into two subjets $j_1$ and $j_2$.
    \item Check if $j_1$ and $j_2$ satisfy the condition:
        \begin{equation}
            \frac{min (p_{T_1}, p_{T_2})}{p_{T_1}+p_{T_2}} > z_{\text{cut}}(\frac{\Delta R_{12}}{{R_0}})^{\beta}
        \label{eq:SD}
        \end{equation}
        where the two parameters, $z_{\text{cut}}$, the soft drop threshold, and the angular exponent $\beta$ controls the degree
of jet grooming.
    \item If the above condition is not satisfied, remove the softer subjet, redefine $j$ to be the harder subjet, and reiterate steps 1 and 2 until criteria \ref{eq:SD} is satisfied. 
    \item If the two branchings of jet $j$ pass the condition, $j$ is the final soft dropped jet.
\end{enumerate} 

\begin{figure}[hbt!]
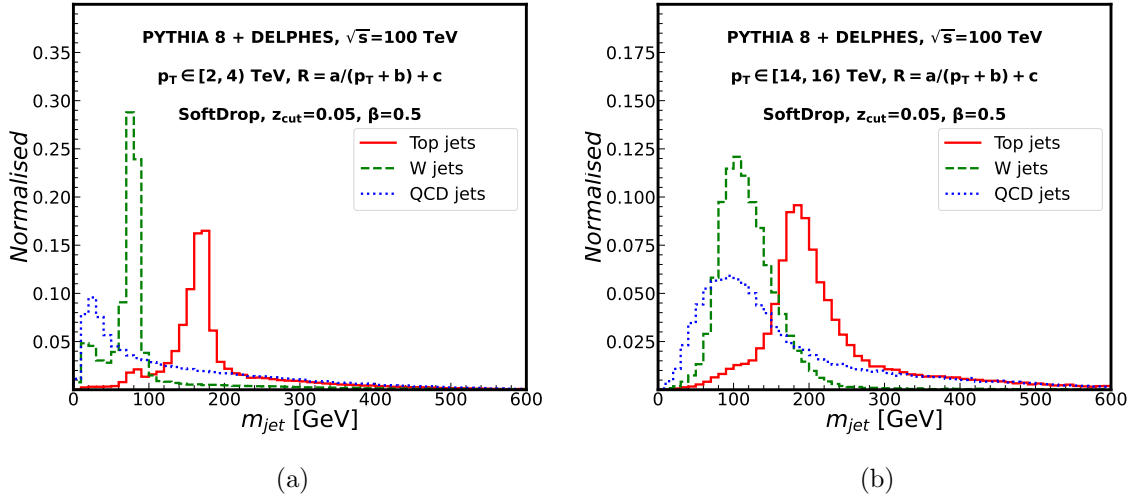

\centering
\begin{subfigure}[c]{0.5\linewidth}
\includegraphics[width=\linewidth]{zcut005_beta05_2_4TeV.pdf} 
\caption{}
\label{fig:10a}
\end{subfigure}\hfill    
\begin{subfigure}[c]{0.5\linewidth}
\includegraphics[width=\linewidth]{zcut005_beta05_14_16TeV.pdf}
\caption{}
\label{fig:10b}
\end{subfigure}
\caption{Normalised $m_{\text{jet}}$ distribution of (a) 2--4 TeV and (b) 14--16 TeV top jets and QCD jets, after applying softdrop grooming with parameters $z = 0.05$, and $\beta = 0.5$.}
\label{fig:mjet_SD}
\end{figure}

We vary the parameters used for soft drop: $z_{\text{cut}} = 0.05$, 0.08, 0.10, 0.20, and $\beta$ = 0.2, 0.5, 1.0, 2.0. Appendix \ref{app:soft drop} shows the $m_{\text{jet}}$ distributions of top, $W$, and QCD jets for all possible combinations of the parameters. We calculate the Wasserstein-1 distance \cite{arjovsky2017wasserstein}, a measure of the separation between two distributions, between the $m_{\text{jet}}$ distributions of top, QCD, and $W$ jets, and choose the parameters $z_{\text{cut}} = 0.05$, $\beta = 0.5$ that maximize this distance for the two extreme ends of our $p_T$ range. Figure \ref{fig:mjet_SD} shows the distribution of the top and the QCD jet mass after using the soft drop algorithm, with $z_{\text{cut}} = 0.05$, $\beta = 0.5$ for $p_T\in[2,4]$ TeV and $p_T\in[14,16]$ TeV jets.

Applying softdrop after reclustering with $R(p_T)$ leads to increased separation between top and QCD jets; however, the separation is reduced between top and W jets, and between W and QCD jets.
The secondary peaks in the $m_{\text{jet}}$ distributions of top and $W$ jets appear when the condition in Equation \ref{eq:SD} removes one of the softer prongs from the multi-pronged jets. This is more evident in 2--4 TeV jets. In 14--16 TeV jets, along with the prongs carrying enough energy to satisfy Equation \ref{eq:SD}, the high boost concentrates the jet constituents with much smaller angular separation. 

\section{Machine learning techniques for jet tagging}
\label{sec:ML}

Our goal in the following sections is to perform multi-class classification using these observables as well as jet images to explore how machine learning distinguishes between the three categories of jets in the $\mathcal{O}(1) - \mathcal{O}(10)$ TeV range. The motivation behind designing a multi-class tagger instead of separate binary classifiers is due to its ability to tackle scenarios involving multiple jet types in the final states. For example, in the decay of a vector-like quark (VLQ) $B'$, both $t$ and $W$ jets are produced, making it essential to simultaneously distinguish between top, $W$, and QCD jets. Additionally, $t\bar{t}$ jets form a substantial background for BSM processes involving $W^+W^-$ final states, just as $W$ jets can become a major background for BSM processes with $t\bar{t}$ in the final state. A multi-class tagger effectively addresses these challenges within a single framework.

We first introduce the machine learning (ML) models used in this work for top tagging and $W$ tagging purposes. We generate top jets, QCD jets, and $W$ jets using the process described in Section \ref{sec:simu}.  We have used two kinds of ML algorithms in our work: the image-based convolutional neural network (CNN) that uses low-level inputs in the form of jet images and the variable-based XGBoost that uses high-level, physics-motivated input features such as jet substructure observables. The choice of these simple models is to build a foundation based on how well these models work at a 100 TeV collider, before moving on to advanced machine learning taggers, which can be a point of further study.

We do not perform $b$ tagging on the top jets in our analysis to ensure a fair comparison of top, W, and QCD tagging via multi-class classification using only jet substructure observables and jet images. However, we discuss the effect of $b$ tagging briefly in Section \ref{sec:Zp_analysis}. We build and train the ML models separately for each $p_T$ range. This is because, as described earlier, the features have a large variation, going from 2 TeV to 14 TeV. We end up with 7 XGBoost and 7 CNN classifiers.
We now briefly discuss the architecture of the algorithms below.

\subsection{Extreme Gradient Boosted Decision Trees}
\label{ssec:xgb}
XGBoost, or eXtreme Gradient Boosting \cite{DBLP:journals/corr/ChenG16}, is an efficient, robust, and scalable implementation of gradient boosting machines. It is extensively utilized in machine learning due to its impressive speed and performance. XGBoost operates by creating an ensemble of decision trees, where each new tree attempts to correct the errors of the previous ones using a gradient descent algorithm, ultimately minimising the loss function. The key parameters of the XGBoost model include \texttt{objective} (loss function), \texttt{num\_round} (boosting iterations), \texttt{learning\_rate} (step size), \texttt{max\_depth} (tree depth), \texttt{min\_child\_weight} (minimum instance weight), and \texttt{gamma} (minimum loss reduction). Regularisation parameters like \texttt{alpha} (L1 regularisation), \texttt{lambda} (L2 regularisation), \texttt{subsample} (training instance subsample ratio), \texttt{colsample\_bytree} (feature subsample ratio) are used to reduce overfitting. 

In our work, we use the XGBoost model for three-class classification between top jets, $W$ jets, and QCD jets for the various $p_T$ ranges. We use the multi-class logarithmic loss as the objective function to minimise. We use a \texttt{max\_depth} of 3 and set the \texttt{learning\ rate}, \texttt{colsample\_bytree}, \texttt{colsample\_bylevel}, \texttt{colsample\_bynode} to 0.1, 0.3, 0.5, 0.3, respectively.

Using these parameters, we build the tree and use the high-level jet observables as input features. The observables $m_{\text{jet}}$ and $\tau_{NN-1}^{\beta}$ have been calculated on four types of candidates: charged particle tracks ($Obs_{trk}$), particle-flow jets with radius $R(p_T)$ ($Obs_{rec.}$), particle-flow jets with radius $R(p_T)$ that have been groomed using soft drop algorithm ($Obs_{SD}$). We have divided them into several sets of input features. These are described as follows:
\begin{itemize}
    \item Set 1: $m_{\text{jet}}$, $m_{\text{jet}}^{SD}$, $m_{\text{jet}}^{\text{trk}}$, N-Subjettiness ratios $\tau_{NN-1}^{\beta}$ where N = 2--6 and $\beta = 0.5$, 1.0, 2.0, calculated on reclustered jets before applying soft drop ($\tau_{NN-1}^{\beta}$), after applying soft drop ($\tau_{NN-1, SD}^{\beta}$), and on tracks ($\tau_{NN-1, trk}^{\beta}$).
    \item Set 2: $m_{\text{jet}}^{\text{trk}}$, $\tau_{NN-1, trk}^{\beta}$ where N = 2--6 and $\beta$ = 0.5, 1.0, 2.0.
    \item Set 3: $m_{\text{jet}}$, $m_{\text{jet}}^{SD}$, $m_{\text{jet}}^{\text{trk}}$, $\tau_{NN-1, trk}^{\beta}$ where N = 2--4 and $\beta$ = 0.5, 1.0, 2.0.
    \item Set 4: $m_{\text{jet}}$,$\tau_{NN-1}^{\beta}$ where N = 2--6 and $\beta$ = 0.5, 1.0, 2.0.
    \item Set 5: $m_{\text{jet}}^{SD}$, $\tau_{NN-1, SD}^{\beta}$ where N = 2-6 and $\beta$ = 0.5, 1.0, 2.0.
\end{itemize}

Set 1 contains all the observables from $Obs_{trk}$, $Obs_{rec.}$, and $Obs_{SD}$. We include higher N-Subjettiness ratios as well as ratios with varying degrees of $\beta$ parameter to form a comprehensive set of features. Set 2, 4, and 5 are composed of only $Obs_{trk}$, $Obs_{rec.}$, and $Obs_{SD}$, respectively. Set 3 is a mixed set of track observables (up to 4-Subjettiness) along with two jet mass observables — $m_{\text{jet}}$ before and after soft drop. 

\begin{figure}[hbt!]
\centering
\begin{subfigure}[c]{0.5\linewidth}
\includegraphics[width=\linewidth]{ROC_top_2_4_sets.pdf} 
\caption{}
\label{fig:12a}
\end{subfigure}\hfill    
\begin{subfigure}[c]{0.5\linewidth}
\includegraphics[width=\linewidth]{ROC_top_14_16_sets.pdf}
\caption{}
\label{fig:12b}
\end{subfigure}
\begin{subfigure}[c]{0.5\linewidth}
\includegraphics[width=\linewidth]{ROC_ww_2_4_sets.pdf} 
\caption{}
\label{fig:12c}
\end{subfigure}\hfill    
\begin{subfigure}[c]{0.5\linewidth}
\includegraphics[width=\linewidth]{ROC_ww_14_16_sets.pdf}
\caption{}
\label{fig:12d}
\end{subfigure}
\caption{Receiver-Operating-Characteristic (ROC) curve and the Area-Under-the Curve (AUC) for multi-class classification. Figures \ref{fig:12a} and \ref{fig:12b} show the ROC for top vs. QCD and $W$ classification for 2--4 TeV and 14--16 TeV jets, respectively. Figures \ref{fig:12c} and \ref{fig:12d} show the ROC for $W$ vs. QCD and top classification for 2--4 TeV and 14--16 TeV jets, respectively. The red dotted lines in Figures \ref{fig:12a} and \ref{fig:12c} show the ROC for top vs. background and $W$ vs. background classification respectively for 2--4 TeV jets generated at the LHC.}
\label{fig:ROC}
\end{figure}

Using these, we train the XGBoost. The Receiver-Operating-Characteristic (ROC) curve and the Area-Under-the Curve (AUC) are presented in Figure \ref{fig:ROC}. AUC is the measure of how well a classifier can identify positives and negatives in data. A perfect classifier shall have AUC = 1. Figure \ref{fig:12a} and \ref{fig:12b} represent how well the classifier can identify top jets against the background of QCD jets and $W$ jets. Figure \ref{fig:12c} and \ref{fig:12d} represent how well the classifier can identify $W$ jets against the background of QCD jets and top jets. Set 1, being the most expansive set of features, performs the best in all the cases. 

We also compare the performance of Set 1 observables using the same XGBoost model for multi-jet classification at the LHC with $\sqrt{s} = 14~\text{TeV}$. The classification performance is better at LHC energies for 2--4 TeV jets compared to that at the FCC-hh. This is possibly due to the difference in admixture of quark-initiated jets and gluon-initiated jets in the QCD background at LHC and FCC-hh, for the same $p_T$ range. While at LHC, 2-4 TeV QCD consists of both quark-initiated and gluon-initiated jets in equal proportions ($(\frac{\sigma_{pp\to q\bar{q}}}{\sigma_{pp\to gg}})_{\text{PYTHIA~8~LO}} \sim 1$), at FCC-hh, 2-4 TeV QCD is mostly dominated by gluon-initiated jets. Gluon-initiated jets are more difficult to distinguish from top jets because of the increased number of splittings, hence the difference in performance.
 
Using solely track-based features (set 2) does not prove to be very advantageous, especially when identifying boosted $W$ jets. Set 2 is outperformed by sets 3, 4, and 5, especially for $W$ jet identification. In the multi-class tagger, the advantages of track-based features for top vs. QCD discrimination are offset by their inability to fully capture $W$ jet information. One example is the $m_{\text{jet,trk}}$ variable. While it provides useful separation between top and QCD jets, there is significant overlap between $m_{\text{jet,trk}}$ of $W$ jets and QCD jets. 

The charged track multiplicity in $W$ jets is low compared to top jets and QCD jets because of its two-prong decay (top decay has three hard prongs, and QCD jets emit soft radiation) and because $W$ boson is less radiative. At high $p_T$ the tracks from the two decay products overlap. The individual tracks carry more energy than tracks from top decay because there are only two partons to share the energy. The collimation makes it harder to resolve the individual tracks. At 14--16 TeV, much of the information is lost due to overlapping hits in the tracker. This causes the track-based mass and other track-based observables to become less correlated with their particle-flow-based counterparts. On the other hand, less correlation leads to better separation from the top and QCD jets because the track-based and particle-flow-based features are more correlated in the latter case.

\subsection{Convolutional Neural Networks (CNN)}
\label{ssec: ML_cnn}

Convolutional neural networks (CNNs) \cite{Fukushima1980NeocognitronAS, FUKUSHIMA2013103,Lecun} are a powerful deep learning method designed to process visual data. The CNNs accept three-dimensional input images comprising height, width, and depth or channels. The core component of a CNN is the convolutional layer, which utilises two-dimensional kernels or filters to scan the input image, generating a feature map that distills the essential information. This process reduces the spatial dimensions of the data while preserving critical information present in the input image. The network then uses a pooling layer to further down-sample the feature maps by keeping only the most relevant information. In a typical CNN architecture, multiple convolutional layers and pooling layers are stacked sequentially, allowing the network to learn increasingly complex and abstract feature representations of the input data. The output is then flattened and subsequently fed into a fully connected neural network for the final classification task.

\subsubsection{Pre-processing of the images}
\label{sssec:ML_cnn_preprocess}
The $p_T, \eta, \phi$ of the constituents of each jet are mapped into a two-dimensional image where each pixel coordinate (x, y) represents a particle's ($\eta$, $\phi$), and the pixel intensity is the $p_T$ of the particle. Before mapping, the jet constituents are subject to some pre-processing steps \cite{Macaluso:2018tck}. First, the jet is centered at the $p_T$ weighted centroid of the constituents. Then, the constituents are rotated in such a way that the principal axis lies in the vertical direction (along the Y-axis). Lastly, the constituents are reflected on both axes such that the hardest constituents (constituents with the highest $p_T$) lie in the third quadrant. The intensity of the whole image is normalised to 1.

\begin{figure}[hbt!]
\centering
\begin{subfigure}[c]{0.3\linewidth}
\includegraphics[width=\linewidth]{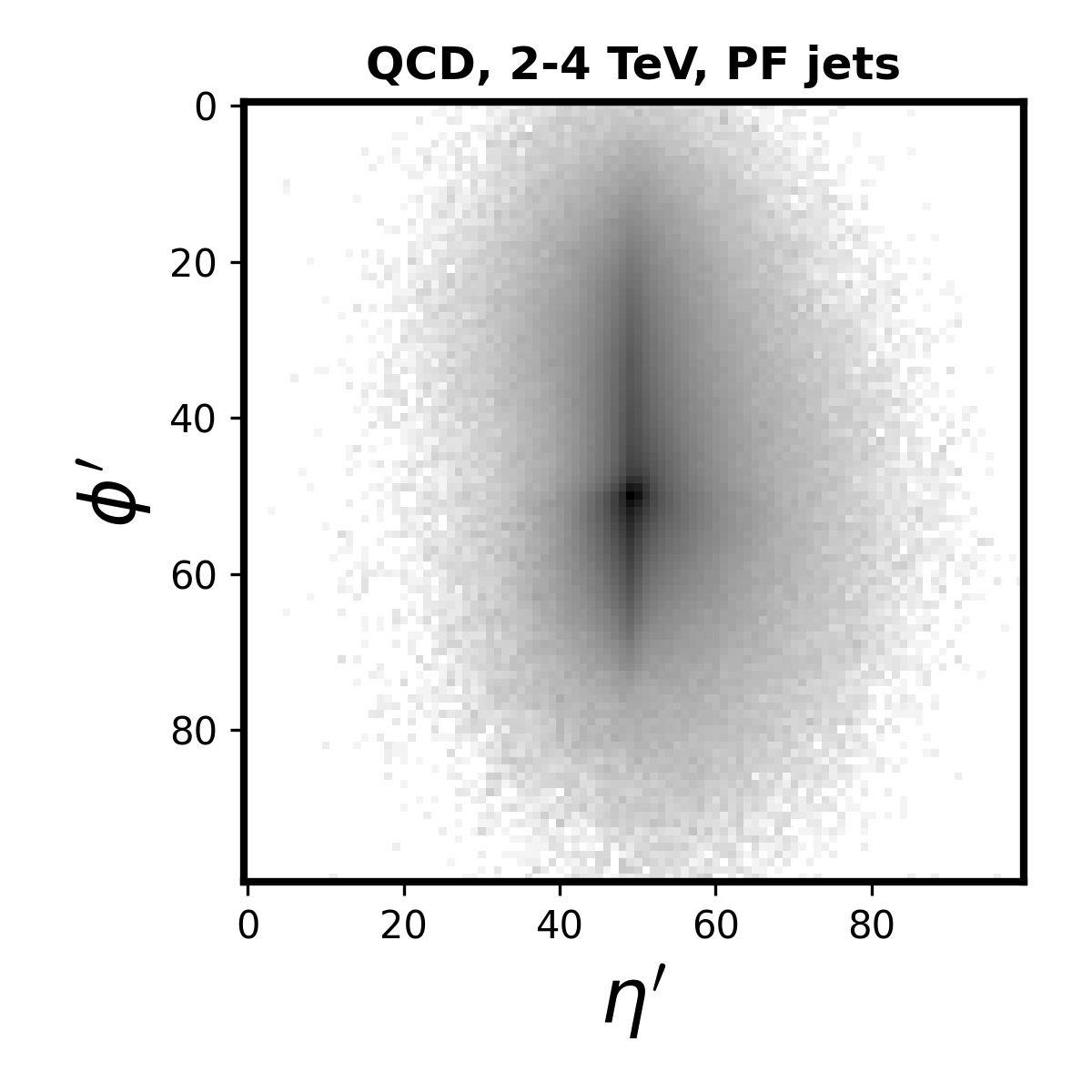} 
\caption{}
\label{fig:13a}
\end{subfigure}\hfill    
\begin{subfigure}[c]{0.3\linewidth}
\includegraphics[width=\linewidth]{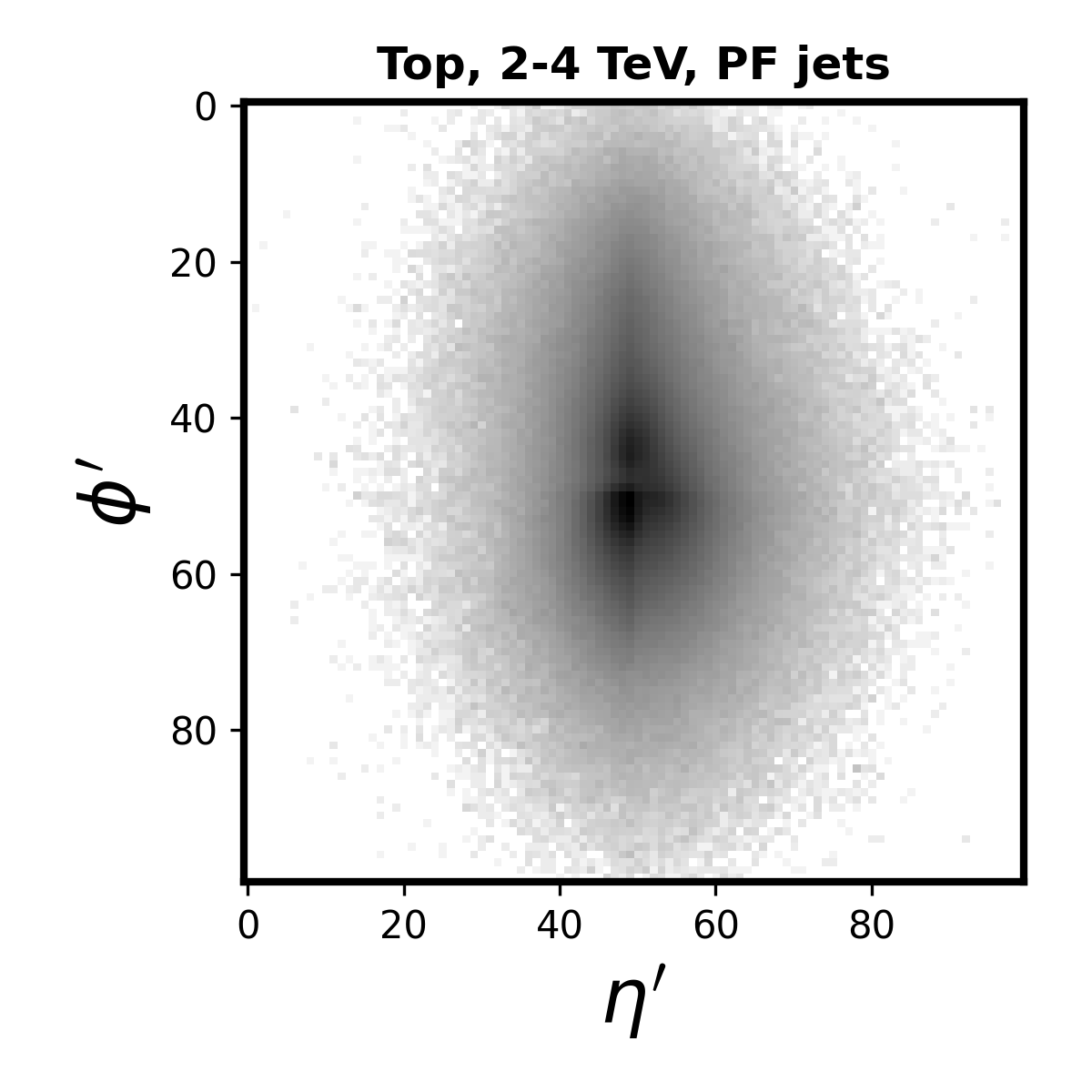}
\caption{}
\label{fig:13b}
\end{subfigure}\hfill
\begin{subfigure}[c]{0.3\linewidth}
\includegraphics[width=\linewidth]{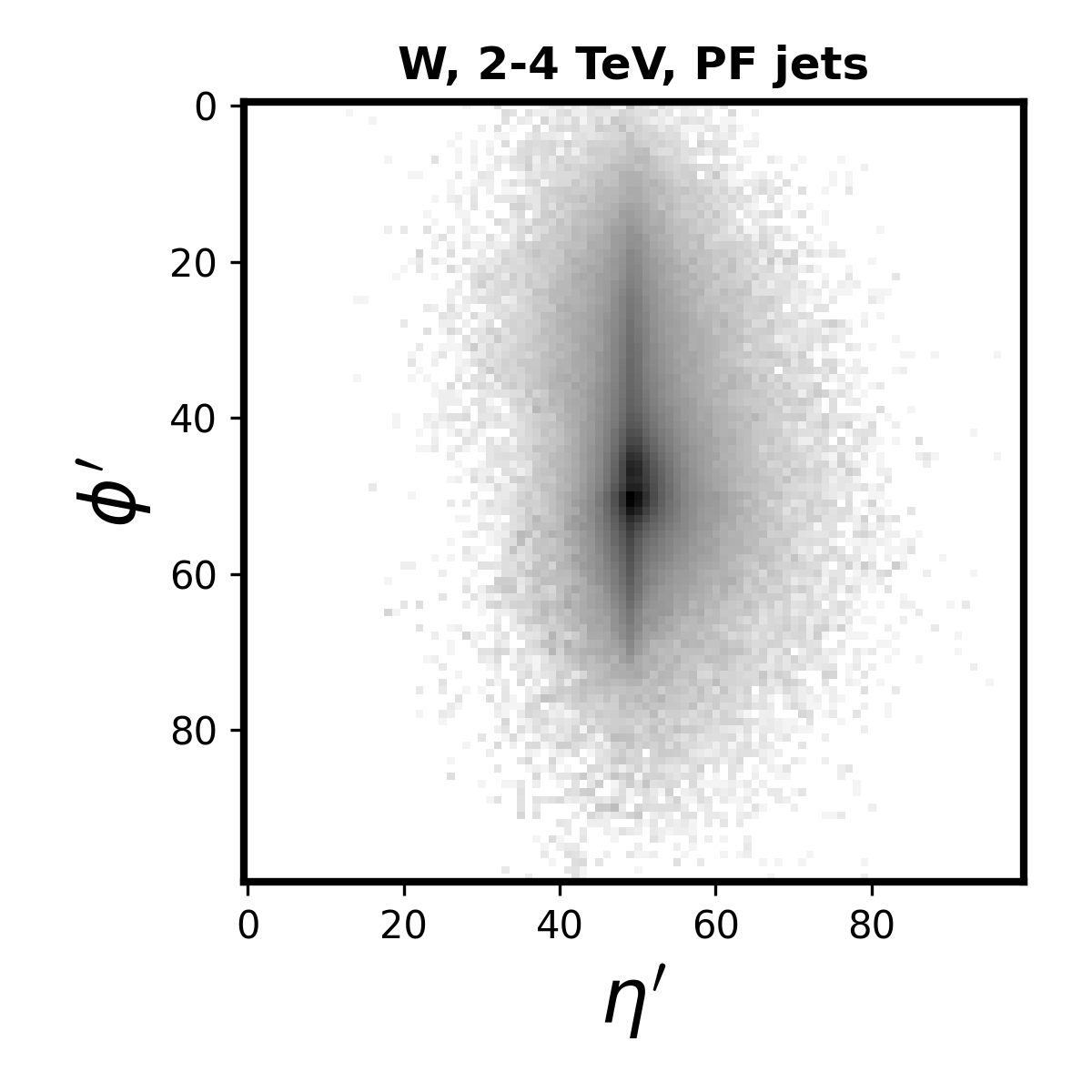}
\caption{}
\label{fig:13c}
\end{subfigure}

\begin{subfigure}[c]{0.3\linewidth}
\includegraphics[width=\linewidth]{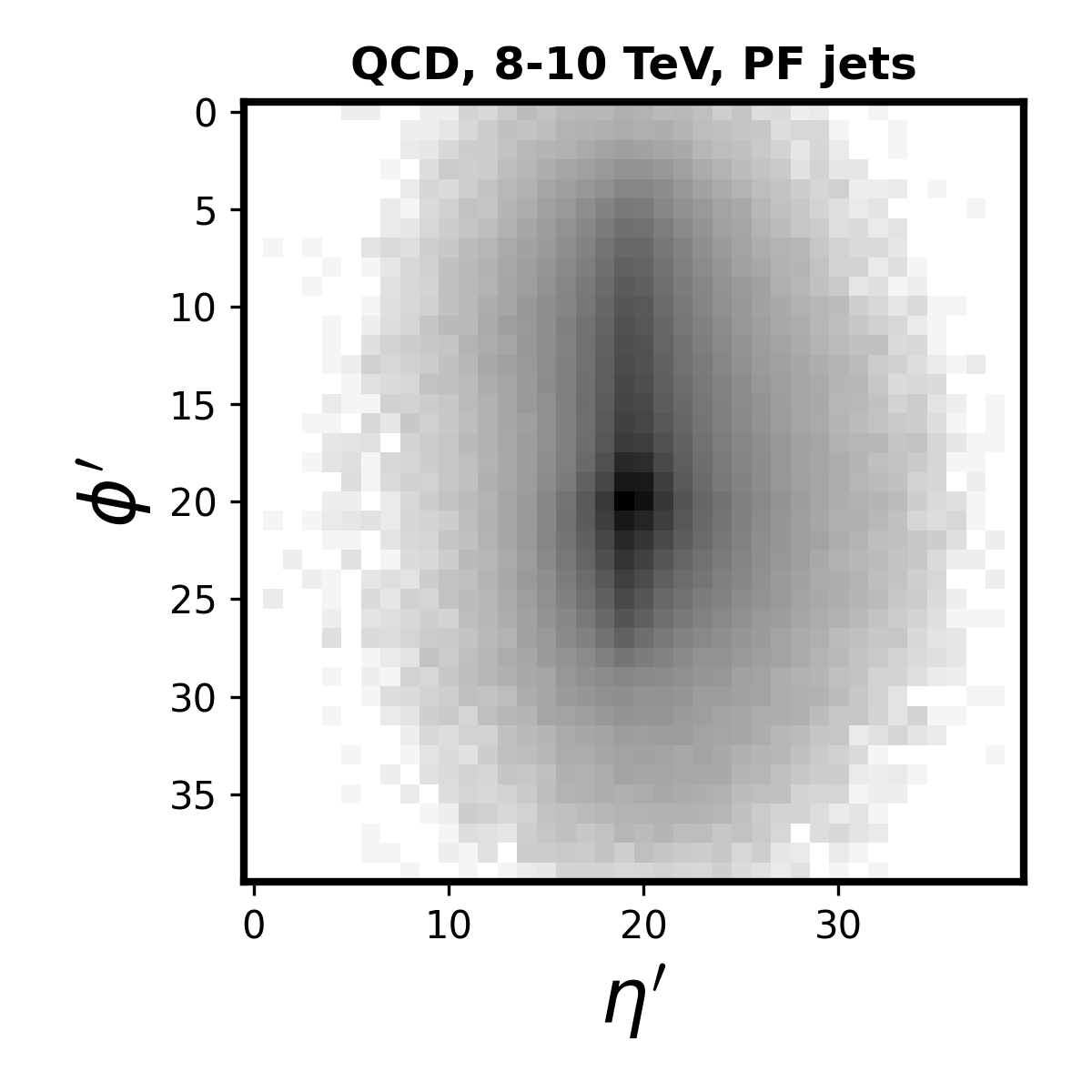} 
\caption{}
\label{fig:13a}
\end{subfigure}\hfill    
\begin{subfigure}[c]{0.3\linewidth}
\includegraphics[width=\linewidth]{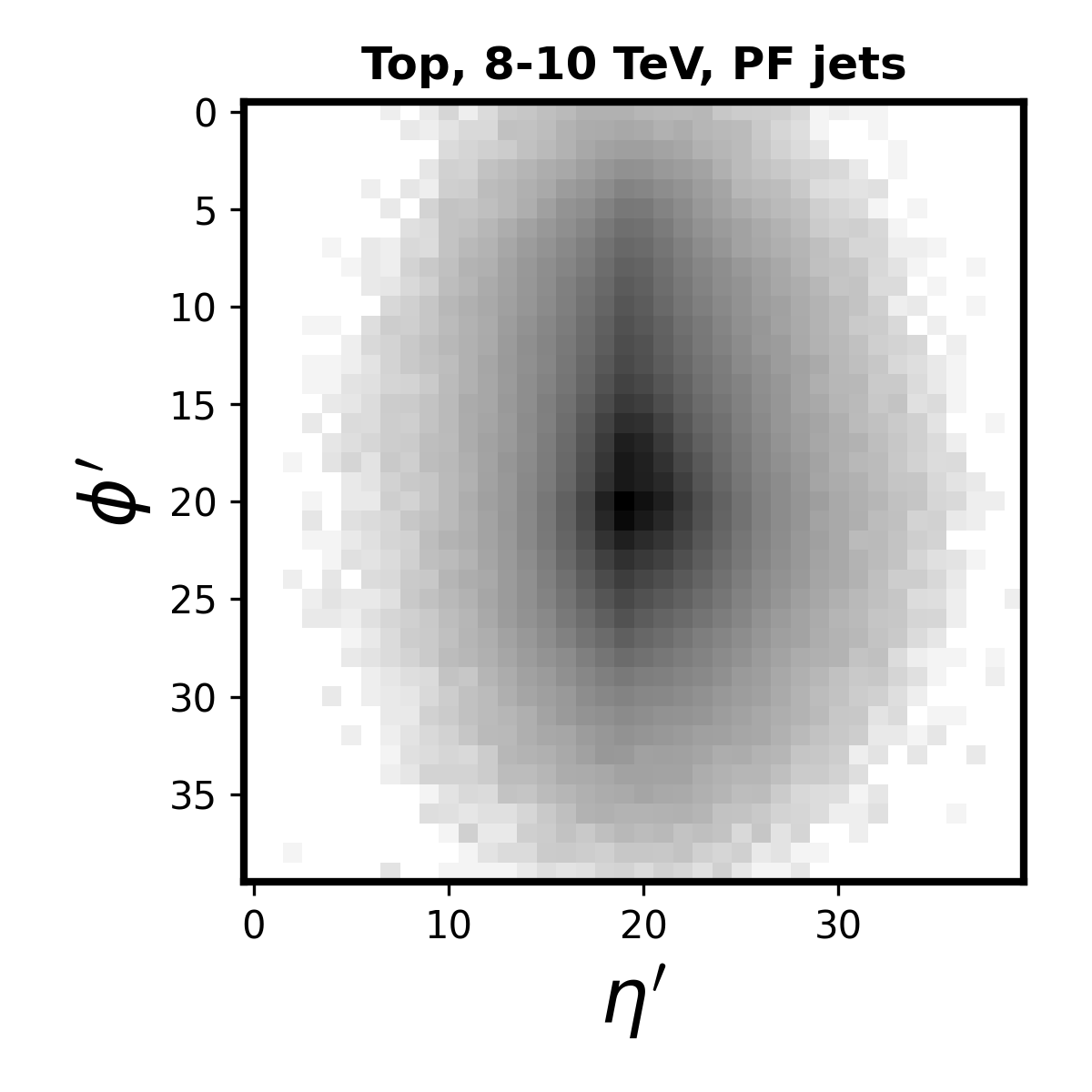}
\caption{}
\label{fig:13b}
\end{subfigure}\hfill
\begin{subfigure}[c]{0.3\linewidth}
\includegraphics[width=\linewidth]{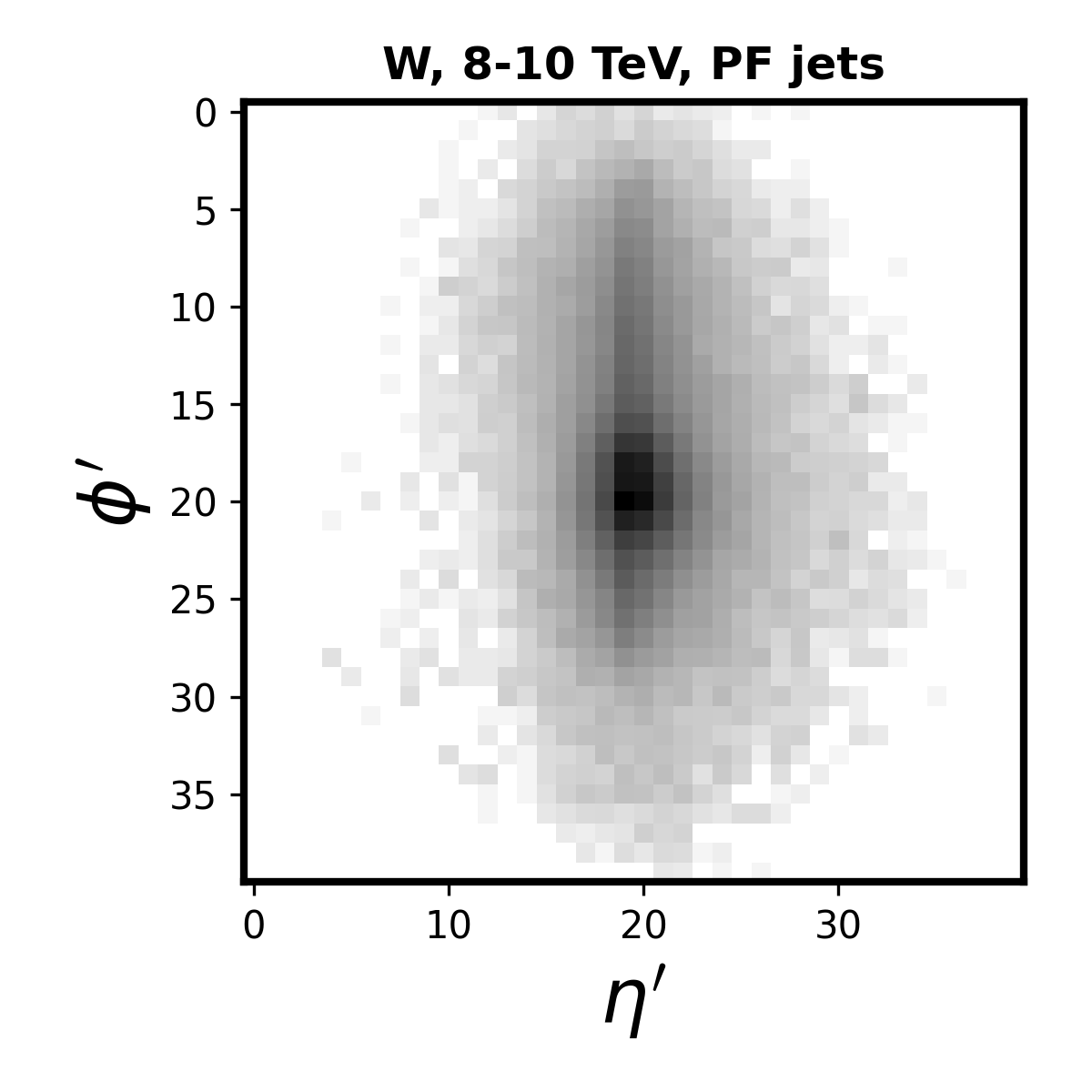}
\caption{}
\label{fig:13c}
\end{subfigure}

\begin{subfigure}[c]{0.3\linewidth}
\includegraphics[width=\linewidth]{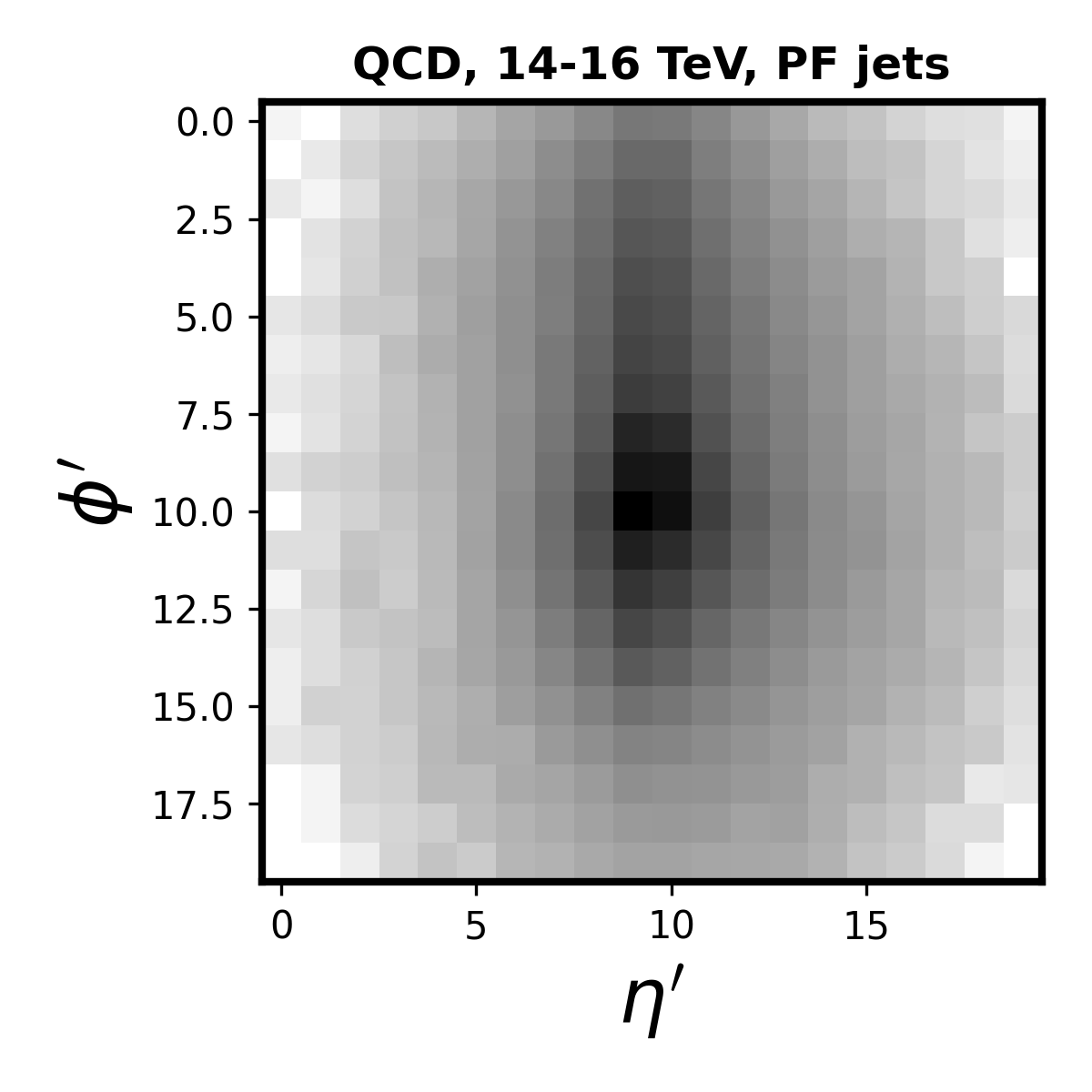} 
\caption{}
\label{fig:13a}
\end{subfigure}\hfill    
\begin{subfigure}[c]{0.3\linewidth}
\includegraphics[width=\linewidth]{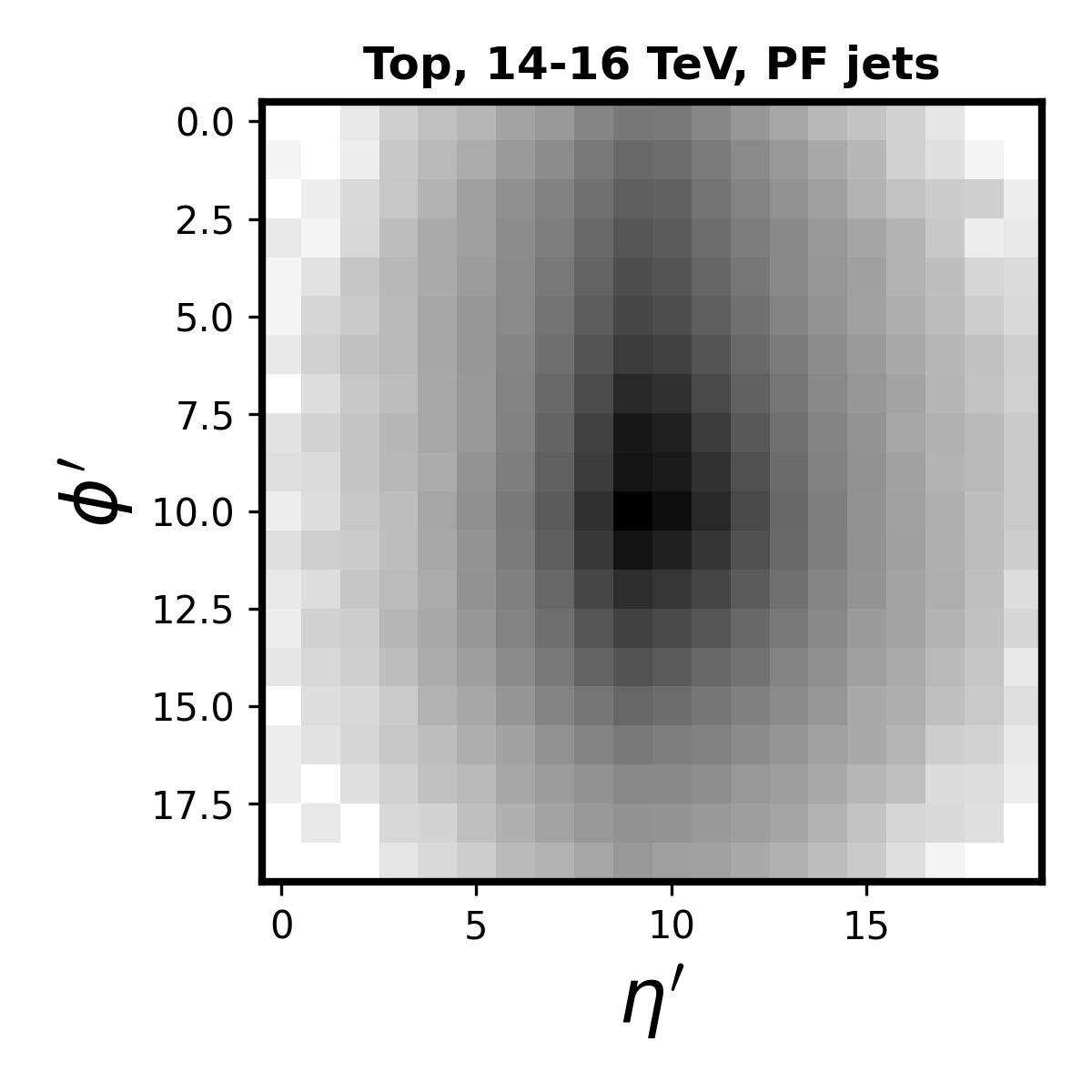}
\caption{}
\label{fig:13b}
\end{subfigure}\hfill
\begin{subfigure}[c]{0.3\linewidth}
\includegraphics[width=\linewidth]{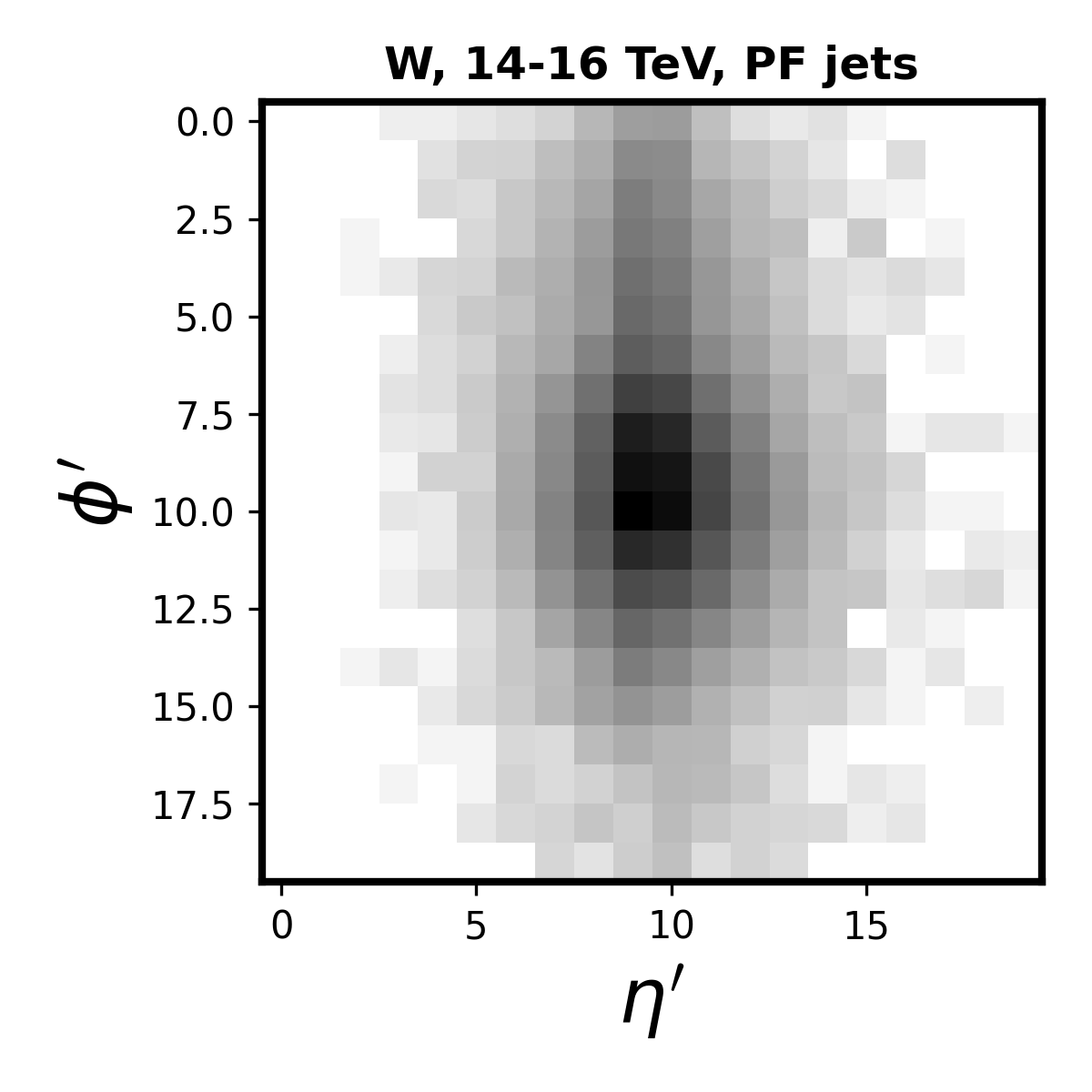}
\caption{}
\label{fig:13c}
\end{subfigure}
\caption{Jet images after averaging over 100,000 events of (a) 2-4 TeV QCD jets, (b) 2-4 TeV top jets, (c) 2-4 TeV W jets, (d) 8-10 TeV QCD jets, (e) 8-10 TeV top jets, (f) 8-10 TeV W jets, (g) 14-16 TeV QCD jets, (h) 14-16 TeV top jets, and (i) 14-16 TeV W jets.}
\label{fig:avg}
\end{figure}

Figure \ref{fig:avg} shows the averaged jet images for QCD jets, top jets, and $W$ jets after pre-processing. As the jet constituents were reclustered into a $p_T$ dependent radius, the dimensions of the images are also dependent on $p_T$. As we go from lower to higher momenta, the three-prong and two-prong features of the core of the top jets and $W$ jets, respectively, shrink and look similar to the QCD jets. The $W$ jets are harder to distinguish from the QCD jets because their two-prong structure appears more similar to that of the QCD jets when highly collimated. However, using the same reclustering radius for all three types of jets leads to some noticeable differences at higher $p_T$, as the constituents of $W$ jets become less spread out compared to those of QCD and top jets for the same image size.

\subsubsection{The network architecture and results}
\label{sssec: ML_cnn_arch}
\begin{figure}[htb!]
    \includegraphics[width=\textwidth]{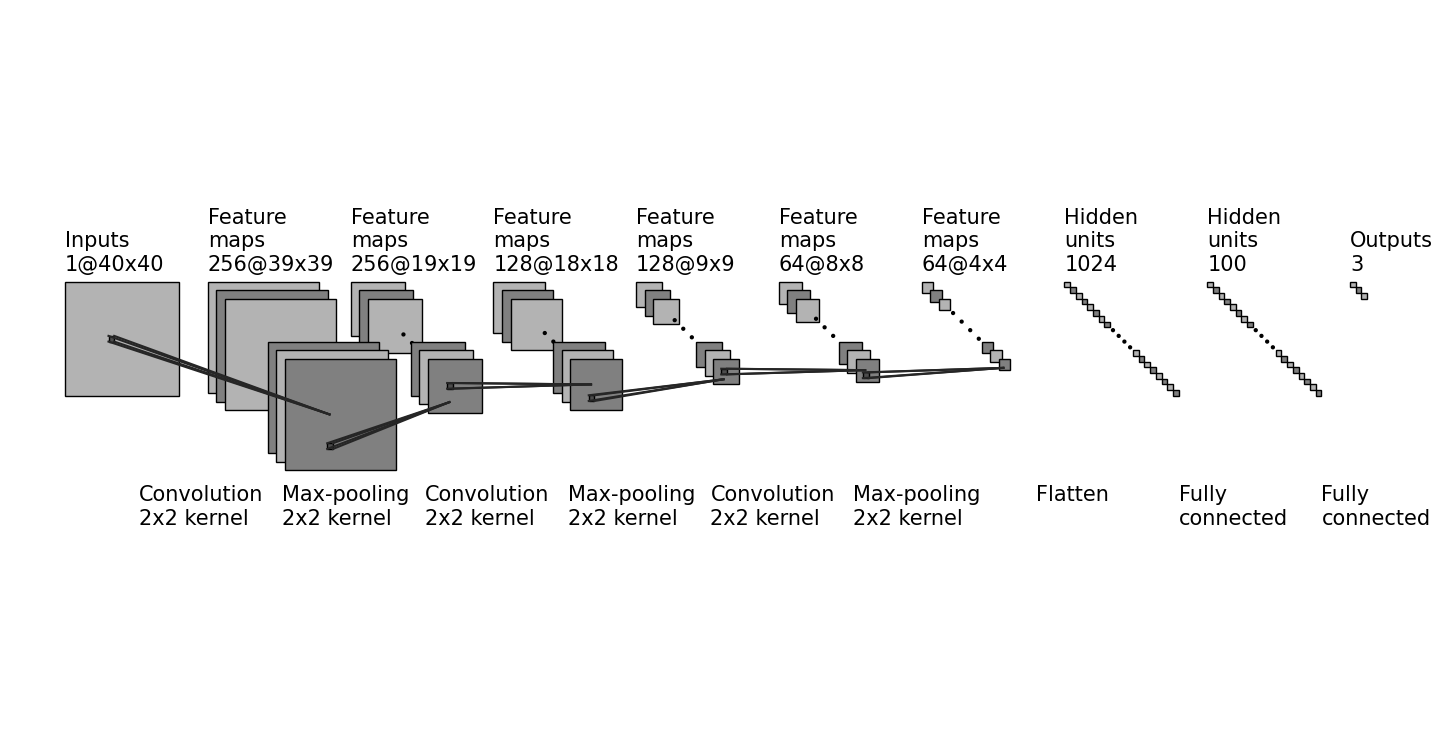}
    \caption{Symbolic representation of the model architecture of CNN used in our analysis. This figure is generated by adapting the code from \href{https://github.com/gwding/draw_convnet}{https://github.com/gwding/draw\_convnet}.}
    \label{fig:arch_cnn}
\end{figure}

We built the CNN model on two 80 GB NVIDIA A100 GPUs using the \texttt{TensorFlow-Keras} framework. Figure \ref{fig:arch_cnn} shows the layer structure in our model. Filters of shape $2\times 2$ were used for every convolution and pooling operation. To regularise the network and reduce overfitting, a dropout of 0.2 was added after the second Max-Pooling Layer, and a dropout of 0.5 was used before the two-dimensional feature maps were flattened. We have also added early-stopping with patience of 5 epochs. The output, in this case, is categorised into three classes, with a probability of a jet being `QCD-like' (0), `top-like' (1), or `W-like' (2). 

With this network, we train the models on $\eta-\phi$ images of particle-flow jets separately for each $p_T$ range. The results for the multi-class classification using CNN are represented by the ROC and AUC in Figure \ref{fig:roc_cnn}. Figure \ref{fig:14a} shows the ROC for the identification of the top jets, while Figure \ref{fig:14b} shows the ROC for the identification of the $W$ jets. The CNN model identifies top jets better than $W$ jets, as indicated by the respective AUC metrics. 

Using the same CNN model for 2--4 TeV jets, we trained and tested on jets generated at $\sqrt{s}=14~\text{TeV}$. We observe a 5\% degradation in the top and $W$ classification performance when moving from $\sqrt{s}=14~\text{TeV}$ to $\sqrt{s}=100~\text{TeV}$. We had previously observed a 1-3\% degradation with XGBoost. This indicates that ML taggers face increased challenges in the high-energy FCC-hh environment. CNN, which is a jet-constituent-based tagger, performs worse than XGBoost, which is a jet-observable-based tagger. The observations in this study call for further investigation in future developments of robust and more complex machine learning taggers. For example, Ref.\,\cite{ATLAS:2022qby} lists the performance of various state-of-the-art taggers, with the ParticleNet~\cite{Qu:2019gqs} achieving an AUC of 0.961. At the LHC, for jets above 2 TeV and signal efficiency ($\epsilon_S$) of 0.8, ParticleNet reaches a background efficiency ($\epsilon_B$) of $\sim$4\%, and $\sim$0.5\% for $\epsilon_S = 0.5$. A detailed evaluation of such taggers in the FCC-hh environment remains an important direction for future work.

\begin{figure}[htb!]
\centering
\begin{subfigure}[c]{0.4\linewidth}
\includegraphics[width=\linewidth]{ROC_cnn_top_tqw.pdf} 
\caption{}
\label{fig:14a}
\end{subfigure}\hfill    
\begin{subfigure}[c]{0.4\linewidth}
\includegraphics[width=\linewidth]{ROC_cnn_ww_tqw.pdf}
\caption{}
\label{fig:14b}
\end{subfigure}
\caption{Receiver-Operating-Characteristic (ROC) curve and the Area-Under-the Curve (AUC) for multi-class classification. Figure \ref{fig:14a} shows the ROC for top vs. QCD and $W$ classification for 2--4 TeV and 14--16 TeV jets. Figure \ref{fig:14b} shows the ROC for $W$ vs. QCD and top classification for 2--4 TeV and 14--16 TeV jets. The red dotted lines in the Figures \ref{fig:14a} and \ref{fig:14b} show the ROC of a multi-class CNN for top vs. background and $W$ vs. background classification, respectively.}
\label{fig:roc_cnn}
\end{figure}

\begin{table}[hbt!]
\centering
\begin{tabular}{|ccccc|}
\hline
\multicolumn{1}{|c|}{$p_T$} & \multicolumn{2}{c|}{CNN results} & \multicolumn{2}{c|}{XGBoost results} \\ \hline
\multicolumn{5}{|c|}{$\epsilon^{top}_S$ = $\epsilon^W_S$ = 30\%} \\ \hline
\multicolumn{1}{|c|}{} & $\epsilon^{top}_B$ & \multicolumn{1}{c|}{$\epsilon^W_B$} & $\epsilon^{top}_B$ & $\epsilon^W_B$ \\ \hline
\multicolumn{1}{|c|}{2--4 TeV} & .71\% & \multicolumn{1}{c|}{.73\%} & .24\% & .16\% \\ \hline
\multicolumn{1}{|c|}{14--16 TeV} & 1.5\% & \multicolumn{1}{c|}{1.6\%} & .68\% & .02\% \\ \hline
\multicolumn{5}{|c|}{$\epsilon^{top}_S$ = $\epsilon^W_S$ = 40\%} \\ \hline
\multicolumn{1}{|c|}{2--4 TeV} & 1.2\% & \multicolumn{1}{c|}{1.4\%} & .48\% & .31\% \\ \hline
\multicolumn{1}{|c|}{14--16 TeV} & 2.4\% & \multicolumn{1}{c|}{2.77\%} & 1.2\% & .06\% \\ \hline
\multicolumn{5}{|c|}{$\epsilon^{top}_S$ = $\epsilon^W_S$ = 60\%} \\ \hline
\multicolumn{1}{|c|}{2--4 TeV} & 3.1\% & \multicolumn{1}{c|}{6.2\%} & 1.8\% & 1.8\% \\ \hline
\multicolumn{1}{|c|}{14--16 TeV} & 5.2\% & \multicolumn{1}{c|}{6.3\%} & 3.4\% & .31\% \\ \hline
\end{tabular}
\caption{Background efficiencies of CNN and XGBoost models for fixed signal efficiencies ($\epsilon_S^{top/W}$) of 30\%, 40\%, and 60\%. The mistagging efficiencies for top identification are indicated as $\epsilon_B^{top}$, while the mistagging efficiencies for $W$ identification are indicated as $\epsilon_B^{W}$.}
\label{tab:cnn_xgb}
\end{table}

We also summarise the results for the best-performing XGBoost tagger (trained with $Obs_{trk}$, $Obs_{rec.}$, and $Obs_{SD}$) and the CNN tagger in Table \ref{tab:cnn_xgb}. The feature-based XGBoost tagger outperforms the image-based CNN tagger. Notably, the $W$ jet identification by XGBoost at high $p_T$ is superior to CNN. 
At 2--4 TeV, a 60\% top tagging efficiency corresponds to a background (including QCD and $W$) mistagging rate of $\sim 3\%$. The same tagging efficiency for $W$ jets yields a background (QCD and top) mistagging rate of $\sim 6\%$. At 14--16 TeV, a 60\% W tagging efficiency results in mistagging rates of $\sim 4\%$ for top jets and $6\%$ for $W$ jets. When the $W$ tagging efficiency exceeds 60\%, the background rejection improves for 14--16 TeV jets compared to 2--4 TeV jets. In our analysis of BSM signatures, we fix the working points of the taggers to obtain 60\% top tagging and 60\% $W$ tagging efficiency.

In this study, the CNN was trained on jet images with the resolution of ECAL. Using track-based images could be a possible way to improve CNN's performance. Track-based jet images offer higher resolution because the tracking detector's granularity (0.001 in $\eta$-$\phi$ space) is much finer compared to the calorimeter's (0.01).  However, to maintain the same jet size, the image dimensions increase by a factor of 10, leading to significant computational challenges when used to train the convolutional neural networks (CNNs). Due to hardware limitations, we opted to use calorimeter-based jet images with a more manageable size. A possible future direction would be to explore convolutional neural networks on track-based images and to further optimize the networks.

In the following section, we will apply the developed taggers to explore physics beyond the Standard Model. As previously noted, the example analyses presented herein serve as illustrations, and the analysis strategy can be further optimized to enhance sensitivity. Our primary objective was to provide a foundation for future investigations rather than achieving the most stringent limits.

\section{Analysis of BSM processes with top and $W$ jets in the final state}
This section explores the three BSM benchmarks described earlier (Section \ref{sec:bsm_theory}). We begin with the analysis of a neutral gauge boson $Z'$ that decays to two top quarks, then proceed to study the production of a heavy neutral Higgs boson in the $W^+W^-$ decay channel. Lastly, we discuss the associated production of a $B'$ VLQ with a light jet, where the $B'$ decays to a top quark and a $W$ boson. 

For our analysis, we choose the working point corresponding to a signal efficiency $\epsilon_S = 60\%$. The XGBoost taggers outperform the CNN taggers; however, for completeness, we also present the results using the CNN tagger. Finally, we combine both taggers to enhance signal purity and achieve high background rejection. In each of the three analyses, we select the two highest $p_T$ jets in each event. If both jets are tagged, we proceed to calculate the invariant mass of the two. 
The resonance peaks are shifted to lower mass values. This effect is expected and can be corrected by applying jet energy correction factors, which are not used in this study. Instead, we factor in the effect of mass peak shift and use this information to apply proper mass cuts for each benchmark accordingly. We select two tagged jets and require the invariant mass to fall within 10-20\% of the mass of the resonant particle. The analyses are summarised in Table \ref{tab:Zp_details}.

\begin{table}[hbt!]
\centering
\small
\begin{tabular}{|ll|}
\hline

\multirow{3}{*}{\textbf{Analysis-I}} & $p_T^{j_1/j_2} > 2$ TeV, $|\eta| < 2.5$ \\
 & $j_1$ AND $j_2$ are XGBoost top/W-tagged \\
 & $m_{res.}-\Delta_1< m_{j_1j_2}< m_{res.}+\Delta_2$ \\ \hline
\multirow{3}{*}{\textbf{Analysis-II}} & $p_T^{j_1/j_2} > 2$ TeV, $|\eta| < 2.5$ \\
 & $j_1$ AND $j_2$ are CNN top/W-tagged \\
 & $m_{res.}-\Delta_1< m_{j_1j_2}< m_{res.}+\Delta_2$ \\ \hline
\multirow{3}{*}{\textbf{Analysis-III}} & $p_T^{j_1/j_2} > 2$ TeV, $|\eta| < 2.5$ \\
 & $j_1$ AND $j_2$ are XGBoost AND CNN top/W-tagged \\
 & $m_{res.}-\Delta_1< m_{j_1j_2}< m_{res.}+\Delta_2$ \\ \hline
\end{tabular}
\caption{Three analyses used for all the benchmark scenarios with XGBoost tagger (I), CNN tagger (II), and both taggers (III).}
\label{tab:Zp_details}
\end{table}

We summarise the analysis strategy as follows for \textit{all} three benchmarks:
\begin{itemize}
  \item Preselect events by requiring the two hardest jets to satisfy $p_T>2$~TeV and $|\eta|<2.5$.
  \item Recluster the jets with a dynamic radius $R(p_T)$, followed by grooming using softdrop algorithm. Additionally, form track‐only jet collections.
  \item Compute the variables of Set 1 (Section \ref{ssec:xgb}) with \texttt{FASTJET} and form jet-images per Section \ref{sssec:ML_cnn_preprocess}.
  \item Jet tagging is performed using XGBoost and CNN taggers corresponding to the correct $p_T$ range, selected based on the individual jet's $p_T$. For example, in an event where a $Z'$ boson produces two jets with $p_T$ values of 6.5~TeV and 3~TeV, the 6--8 TeV ML tagger is used for the first jet, while the 2--4 TeV ML tagger is applied to the second jet. Each tagger returns a probability score, which is then used to classify the jet as a top or a $W$ jet. 
  \item Form the invariant mass of the two leading tagged jets and apply a resonance‐dependent mass window (10–20\% around the resonance mass).
  \item Evaluate background yields ($B$) and signal efficiency ($\epsilon_S$), then extract 95\% CL upper limits ($\sigma_{UL}$) from the following significance formula~\cite{Cowan:2010js}:
    \begin{equation}    
    \eta_S = \sqrt{2(S + B) \ln \left( \frac{(S + B)(B + \sigma_b^2)}{B^2 + (S + B)\sigma_b^2} \right) - \frac{B^2}{\sigma_b^2} \ln \left( 1 + \frac{\sigma_b^2 S}{B(B + \sigma_b^2)} \right)}
    \end{equation}
    In this study, we obtain the projected upper limits at $\eta_{S}=2$, representing a $95\%$ confidence interval. $\sigma_b$ is the systematic uncertainty times the background yield.
\end{itemize}

Systematic uncertainties originate from uncertainties on the parton distribution function (PDF), uncertainties on the production cross-section, scale uncertainties in the matrix element and parton shower algorithms, uncertainty on the integrated luminosity, and uncertainties on the object identification efficiencies~\cite{FCC:2019hfo}. We compute upper limits at 95\% CL on the product of cross-section and branching ratio assuming an overall 5\% systematic uncertainty $(\sigma_b=5\%\times B)$.

\subsection{Analysis of $Z'$ boson decaying to boosted top jets}
\label{sec:Zp_analysis}
With the models trained to efficiently distinguish between boosted top jets, QCD jets, and $W$ jets, we proceed with the reconstruction of the BSM gauge boson, $Z'$. We have considered the decay $Z'\to t\bar{t}$ for four different benchmark masses of the $Z'$ boson: $m_{Z'} = 5$~TeV, 10~TeV, 15~TeV, and 20~TeV. 
As the SM backgrounds to the $Z'\to t\bar{t}$ process, in addition to QCD dijet and $t\bar{t}$, we have also considered SM production of $W^+W^-$, $W^\pm Z$, and $ZZ$, collectively denoted as $VV$, SM production of $W^\pm+\text{jet}$, and $Z+\text{jet}$, collectively denoted as $Vj$, and the SM $tW$ process.

\begin{figure}[htb!]
\centering
\includegraphics[width=0.5\linewidth]{inv_mass_sig_Zp.pdf} 
\caption{Normalised invariant mass distribution of two hardest jets from $pp\to Z' \to t\bar{t}$ process.}
\label{fig:inv_mass_bkg_Z}
\end{figure}

Figure \ref{fig:inv_mass_bkg_Z} shows the invariant mass distribution of the two jets coming from the signal process. From the width of the distributions, we set $\Delta_1$ as 20\% of $m_Z'$, and $\Delta_2$ as 10\% of $m_Z'$. 
Table \ref{tab: zprime_cut} shows the background yield (B) and signal efficiency ($\epsilon_S$) for the four benchmarks of the $Z'$ boson at different levels of cuts when the XGBoost tagger is used.

\begin{table}[hbt!]
\centering
\small
\begin{tabular}{|cccccc|}
\hline
\multicolumn{1}{|c|}{\multirow{2}{*}{Selection cuts}} & \multicolumn{4}{c|}{Background yield} & $\epsilon_S$ \\ \cline{2-5}
\multicolumn{1}{|c|}{} & $jj$ & $t\bar{t}$ & $tW$ & \multicolumn{1}{c|}{$VV+Vj$} & $Z' \rightarrow t\bar{t}$ \\ \hline
\multicolumn{6}{|c|}{$m_{Z'}$ = 5 TeV} \\ \hline
\multicolumn{1}{|c|}{$p_T^{j_1/j_2} > 2$ TeV, $|\eta| < 2.5$} & $4.0\times 10^9$ & $8.5\times 10^6$ & $1.1\times 10^6$ & \multicolumn{1}{c|}{$2.2\times 10^7$} & 18\% \\
\multicolumn{1}{|c|}{$j_1,~j_2$ are XGBoost top-tagged} & $3.0\times 10^6$ & $3.2\times 10^6$ & 4508 & \multicolumn{1}{c|}{717} & 7\% \\
\multicolumn{1}{|c|}{4.0 TeV $< m_{j_1j_2} <$ 5.5 TeV} & $1.7\times 10^6$ & $1.6\times 10^6$ & 2460 & \multicolumn{1}{c|}{8} & 6\% \\ \hline
\multicolumn{6}{|c|}{$m_{Z'}$ = 10 TeV} \\ \hline
\multicolumn{1}{|c|}{$p_T^{j_1/j_2} > 2$ TeV, $|\eta| < 2.5$} & $4.0\times 10^9$ & $8.5\times 10^6$ & $1.1\times 10^6$ & \multicolumn{1}{c|}{$2.2\times 10^7$} & 70\% \\
\multicolumn{1}{|c|}{$j_1,~j_2$ are XGBoost top-tagged} & $3.0\times 10^6$ & $3.2\times 10^6$ & 4508 & \multicolumn{1}{c|}{717} & 24\% \\
\multicolumn{1}{|c|}{8.0 TeV $< m_{j_1j_2} <$ 11.0 TeV} & $5.3\times 10^5$ & $2.5\times 10^5$ & 656 & \multicolumn{1}{c|}{19} & 16\% \\ \hline
\multicolumn{6}{|c|}{$m_{Z'}$ = 15 TeV} \\ \hline
\multicolumn{1}{|c|}{$p_T^{j_1/j_2} > $ 2 TeV, $|\eta| < 2.5$} & $4.0\times 10^9$ & $8.5\times 10^6$ & $1.1\times 10^6$ & \multicolumn{1}{c|}{$2.2\times 10^7$} & 82\% \\
\multicolumn{1}{|c|}{$j_1,~j_2$ are XGBoost top-tagged} & $3.0\times 10^6$ & $3.2\times 10^6$ & 4508 & \multicolumn{1}{c|}{717} & 27\% \\
\multicolumn{1}{|c|}{12.0 TeV $< m_{j_1j_2} <$ 16.5 TeV} & $4.4\times 10^4$ & $3.0\times 10^4$ & 22 & \multicolumn{1}{c|}{94} & 17\% \\ \hline
\multicolumn{6}{|c|}{$m_{Z'}$ = 20 TeV} \\ \hline
\multicolumn{1}{|c|}{$p_T^{j_1/j_2} > $ 2 TeV, $|\eta| < 2.5$} & $4.0\times 10^9$ & $8.5\times 10^6$ & $1.1\times 10^6$ & \multicolumn{1}{c|}{$2.2\times 10^7$} & 86\% \\
\multicolumn{1}{|c|}{$j_1,~j_2$ are XGBoost top-tagged} & $3.0\times 10^6$ & $3.2\times 10^6$ & 4508 & \multicolumn{1}{c|}{717} & 29\% \\
\multicolumn{1}{|c|}{16.0 TeV $< m_{j_1j_2} <$ 22.0 TeV} & 6530 & 5221 & 10 & \multicolumn{1}{c|}{3} & 15\% \\ \hline
\end{tabular}
\caption{Background yield and signal efficiency ($\epsilon_S$) for the four benchmark mass-points of $Z'$ boson. The $m_{j_1 j_2}$ cuts are different for different values of $Z'$ boson mass. Here, $V$ denotes $W^\pm$, and $Z$.}
\label{tab: zprime_cut}
\end{table}

As reported in table \ref{tab: zprime_cut}, after applying the cut on $P_{top}^{XGB}$ for $j_1$ and $j_2$, the QCD background reduces to 0.08\% of its original value. The $tW$ reduces to 0.4\% of its original value. However, due to the large cross-section of QCD, it still remains comparable to the $t\bar{t}$ background. The background yields from diboson ($VV$) and boson-plus-jet ($Vj$) processes have been combined. They are not dominant background sources; however, the $Vj$ cross-section is higher than the $t\bar{t}$ cross-section, resulting in the high background yield initially. After applying all the selection cuts, these backgrounds are reduced to negligible levels.
For $m_{Z'}$ = 5 TeV, the signal efficiency is very low due to the jets failing the $p_T > 2$ TeV criterion. Heavier $Z'$ bosons decay into harder jets, hence we see an increase in the signal efficiency with mass. 
Both signal efficiency and background yield drop after using the cut on the invariant mass. The drop is more for higher resonant masses as the distribution becomes wider.

\begin{table}[hbt!]
\centering
\small
\begin{tabular}{|c|c|cccc|c|c|}
\hline
$m_{Z'}$ & \multirow{2}{*}{Analysis} & \multicolumn{4}{c|}{Background yield at 30 $\text{ab}^{-1}$} & Signal & $\sigma(pp \to Z'\to t\bar{t})_{UL}$ \\ \cline{3-6}
 &  & $jj$ & $t\bar{t}$ & $tW$ & $VV+Vj$ & Efficiency & at 95\% CL [fb] (5\% sys.) \\ \hline
5 & I & $1.7\times 10^6$ & $1.6\times 10^6$ & 2460 & 8 & 0.063 & 4.3 (395)\\
 & II & $3.2\times 10^6$ & $1.4\times 10^6$ & 7057 & 6016 & 0.053 & 6.0 (665)\\
 & III & $1.6\times 10^5$ & $9.6\times 10^5$ & 989 & 8 & 0.04 & 4.2 (229)\\ \hline
10 & I & $5.3\times 10^5$ & $2.5\times 10^5$ & 656 & 19 & 0.16 & 0.82 (37)\\
 & II & $1.1\times 10^6$ & $2.6\times 10^5$ & 1208 & 103 & 0.18 & 0.95 (57)\\
 & III & $4.4\times 10^5$ & $1.8\times 10^5$ & 310 & 0.6 & 0.12 & 0.95 (39)\\ \hline
15 & I & 44161 & 30299 & 22 & 94 & 0.17 & 0.24 (3.4)\\
 & II & 88637 & 32083 & 139 & 138 & 0.18 & 0.28 (5.1)\\
 & III & 22557 & 21016 & 9 & 83 & 0.12 & 0.26 (2.8)\\ \hline
20 & I & 6530 & 5221 & 10 & 3 & 0.16 & 0.099 (0.56)\\
 & II & 22041 & 5264 & 32 & 7 & 0.17 & 0.14 (1.24)\\
 & III & 2083 & 3318 & 9 & 0.3 & 0.11 & 0.099 (0.39)\\ \hline
\end{tabular}
\caption{Table showing the number of events obtained for background processes, the signal efficiency after applying all cuts, and the upper limit of $pp\to Z' \to t\bar{t}$ cross-section at 95\% CL for the three different analyses listed on Table \ref{tab:Zp_details}. Here, $V$ denotes $W^\pm,~Z$. The limits at 95\% CL with 5\% systematic uncertainty on the background are listed within parentheses.}
\label{tab:zprime_limits}
\end{table}

Table \ref{tab:zprime_limits} shows the projected 95\% CL upper limits on the cross-section for the process $pp \to Z' \to t\bar{t}$ at the FCC-hh for three different analysis strategies (I, II, III). We also quote the results with 5\% systematic uncertainty on the background in Table \ref{tab:zprime_limits}. 
Among the three strategies, Analysis III consistently provides the best or comparable sensitivity due to improved background suppression. The most stringent limit is obtained for $m_{Z'} = 20$~TeV, where the cross-section times branching can be constrained down to 0.099~fb  at 95\% CL. 
The CNN taggers did not perform well in comparison to the XGBoost taggers previously, resulting in weaker constraints. 
A CMS study in Ref.~\cite{CMS:2018rkg} conducted $t\bar{t}$ resonance search and placed upper limits at 95\% CL on the cross-section times branching of a $\Gamma/m = 10\%~Z'$ boson of mass 5 TeV to be 9.7 fb.

For $m_{Z'}=5~\text{TeV}$, the 5\% systematic uncertainty worsens the limit by a factor of approximately 100. The backgrounds in the low-mass region are large, and thus, the sensitivity in this region is limited by systematics. Choosing a stricter point on the ROC could potentially reduce the QCD background at the cost of a lower signal efficiency; however, the SM $t\bar{t}$ background remains irreducible. The impact of systematics decreases with increasing resonance mass, as we move from a low-$x$ to a high-$x$ region in FCC-hh ($x$ is the momentum fraction), where background cross-sections fall steeply.

\subsubsection{Effect of $b$ tagging}

Before we proceed to the next analysis of the heavy Higgs boson, we would like to briefly mention the effect of $b$ tagging. The top jets contain a $b$ jet as one of the three subjets. We have conducted our analysis without employing any $b$ tagging. In the case of top tagging, if the two jets in an event obtained after applying selection cuts were also required to be $b$ tagged, it would reduce the QCD background further. For FCC-hh, the $b$ tagging efficiency employed in \texttt{DELPHES} follows the functional form $0.82*(1-p_T/15$~TeV) \cite{Selvaggi:2717698} for the jets in our selected phase space ($|\eta| < 2.5$). Table \ref{tab:btag} shows the effect of using $b$ tagging as an additional criterion at the selection level.

\begin{table}[hbt!]
\centering
\small
\begin{tabular}{|ccc|}
\hline
\multicolumn{3}{|c|}{Effect of $b$ tagging on background events, with $b$ tagging efficiency = $0.82*(1.0 - p_T/15$~TeV)} \\ \hline
Selection cuts & $pp\rightarrow jj$ & $pp\rightarrow t\bar{t}$ \\ \hline
$p_T^{j_1/j_2} > $ 2 TeV, $|\eta| < 2.5$ & $4.0\times 10^9$ & $8.5\times 10^6$ \\
$p_T^{j_1/j_2} > $ 2 TeV, $|\eta| < 2.5$, btag($j_1/j_2$) = 1 & $3.7\times 10^7$ & $2.5\times 10^6$ \\ \hline
\end{tabular}

\caption{Effect of $b$ tagging on the two hardest jets at the selection level for QCD and $t\bar{t}$ backgrounds.}
\label{tab:btag}
\end{table}

\begin{figure}[hbt!]
\centering
\begin{subfigure}[c]{0.5\linewidth}
\includegraphics[width=\linewidth]{Bplus_pT.pdf} 
\caption{}

\end{subfigure}\hfill    
\begin{subfigure}[c]{0.5\linewidth}
\includegraphics[width=\linewidth]{Dplus_pT.pdf}
\caption{}

\end{subfigure}
\caption{Normalised distribution of displacement ($dT$) of (a) boosted $B^{\pm}$ hadron and (b) boosted $D^{\pm}$ hadron from the decay of 500~GeV, 2~TeV, 10~TeV, and 14~TeV top quark. The dashed line in black represents the dimension of the inner tracking layer of the FCC-hh detector.}
\label{fig:Bpl_Dpl_dT}
\end{figure}

The QCD background efficiency reduces to 0.95\% of its original value with the use of the $b$ tagging criterion. Since QCD and $t\bar{t}$ are the leading backgrounds, a simple calculation using $b$ tagging shows that the 95\% CL upper limits on $\sigma(pp \to Z'\to t\bar{t})$ with 5\% systematic uncertainty reduce to $\sim 60~\text{fb}$ using Analysis III. However, the question remains how the $b$ tagging performance in the actual FCC-hh experiment will be affected in the case of multi-TeV jets. A study in \cite{btag_2018} demonstrated that in the multi-TeV regime, the reconstruction of highly boosted $B$-hadrons becomes challenging owing to their high displacement and high collimation of decay products. In Figure \ref{fig:Bpl_Dpl_dT}, we have shown the displacement of the $B^{\pm}$ and $D^{\pm}$ in the radial direction $(dT)$, with increasing $p_T$ of the top jet. The line in black shows the dimension of the inner tracking layer, which consists of pixel sensors of 25--33.3 $\mu$m $\times$ 50--400 $\mu$m and extends in the radial direction up to 200 mm \cite{FCC_CDR3}.

Boosted hadrons coming from a 500~GeV top are well contained within the inner tracking layer. In contrast, for top quarks with $p_T = 2$~TeV, 10~TeV, and 14~TeV, approximately 4\%, 26\%, and 32\% of $B^\pm$ decay outside the pixel tracker. A hit multiplicity approach to $b$ tagging was proposed in Ref. \cite{btag_hit} to tackle this challenge for boosted $b$ jets. Using the information from the number of hits in each detector layer improved the $b$ tagging in multi-TeV jets. Further studies could explore the effectiveness of this technique in improving $b$ tagging for high-$p_T$ top jets.

\subsection{Analysis of neutral heavy Higgs boson decaying to boosted $W$ jets}
\label{sec:H_analysis}

We consider the production of a heavy Higgs boson that decays to the SM $W$ bosons: $pp\to H\to W^+W^-$. We consider the hadronic decay mode of the two $W$ bosons. The final state thus consists of two highly boosted $W$ jets. We consider four different masses of the heavy neutral Higgs to be 5~TeV, 10~TeV, 15~TeV, and 20~TeV. Depending on the mass, the $W$ jets will carry a few TeV of energy. The SM background consists of QCD dijet production and the SM production of $W^+W^-$. We have considered other backgrounds, such as SM production of $W^\pm/Z+\text{jet}$, $WZ$, and $ZZ$. Additionally, top jets, which can be misidentified as $W$ jets, are included as part of the background for the $H \to WW$ in the $t\bar{t}$ and $tW$ processes.

\begin{figure}[htb!]
\centering
\includegraphics[width=0.5\linewidth]{inv_mass_sig_H.pdf} 
\caption{Normalised invariant mass distribution of two hardest jets from $pp\to H \to W^+W^-$ process.}
\label{fig:inv_mass_bkg_H}
\end{figure}

\begin{table}[hbt!]
\centering
\small
\begin{tabular}{|cccccc|}
\hline
\multicolumn{1}{|c|}{\multirow{2}{*}{Selection cuts}} & \multicolumn{4}{c|}{Background yield} & $\epsilon_S$ \\ \cline{2-5}
\multicolumn{1}{|c|}{} & $VV$ & $jj$ & $Vj$ & \multicolumn{1}{c|}{$tW+t\bar{t}$} & $H \rightarrow W^+W^-$ \\ \hline
\multicolumn{6}{|c|}{$m_{H}$ = 5 TeV} \\ \hline
\multicolumn{1}{|c|}{$p_T^{j_1/j_2} > 2$ TeV, $|\eta| < 2.5$} & $4.2\times 10^5$ & $4.0\times 10^9$ & $2.2\times 10^7$ & \multicolumn{1}{c|}{$9.6\times 10^6$} & 32\% \\
\multicolumn{1}{|c|}{$j_1,~j_2$ are XGBoost W-tagged} & $1.6\times 10^5$ & $2.7\times 10^6$ & $6.5\times 10^5$ & \multicolumn{1}{c|}{8211} & 17\% \\
\multicolumn{1}{|c|}{4.0 TeV $< m_{j_1j_2} <$ 5.5 TeV} & 46444 & $1.3\times 10^6$ & $3.4\times 10^5$ & \multicolumn{1}{c|}{4151} & 16\% \\ \hline
\multicolumn{6}{|c|}{$m_{H}$ = 10 TeV} \\ \hline
\multicolumn{1}{|c|}{$p_T^{j_1/j_2} > 2$ TeV, $|\eta| < 2.5$} & $4.2\times 10^5$ & $8.5\times 10^6$ & $1.1\times 10^6$ & \multicolumn{1}{c|}{$2.2\times 10^7$} & 74\% \\
\multicolumn{1}{|c|}{$j_1,~j_2$ are XGBoost W-tagged} & $1.6\times 10^5$ & $2.7\times 10^6$ & $6.5\times 10^5$ & \multicolumn{1}{c|}{8211} & 42\% \\
\multicolumn{1}{|c|}{8.0 TeV $< m_{j_1j_2} <$ 11.0 TeV} & 27810 & $5.8\times 10^5$ & 46694 & \multicolumn{1}{c|}{460} & 36\% \\ \hline
\multicolumn{6}{|c|}{$m_{H}$ = 15 TeV} \\ \hline
\multicolumn{1}{|c|}{$p_T^{j_1/j_2} > $ 2 TeV, $|\eta| < 2.5$} & $4.2\times 10^5$ & $8.5\times 10^6$ & $1.1\times 10^6$ & \multicolumn{1}{c|}{$2.2\times 10^7$} & 80\% \\
\multicolumn{1}{|c|}{$j_1,~j_2$ are XGBoost W-tagged} & $1.6\times 10^5$ & $2.7\times 10^6$ & $6.5\times 10^5$ & \multicolumn{1}{c|}{8211} & 41\% \\
\multicolumn{1}{|c|}{12.0 TeV $< m_{j_1j_2} <$ 16.5 TeV} & 9020 & 77324 & 4020 & \multicolumn{1}{c|}{154} & 35\% \\ \hline
\multicolumn{6}{|c|}{$m_{H}$ = 20 TeV} \\ \hline
\multicolumn{1}{|c|}{$p_T^{j_1/j_2} > $ 2 TeV, $|\eta| < 2.5$} & $4.2\times 10^5$ & $8.5\times 10^6$ & $1.1\times 10^6$ & \multicolumn{1}{c|}{$2.2\times 10^7$} & 82\% \\
\multicolumn{1}{|c|}{$j_1,~j_2$ are XGBoost W-tagged} & $1.6\times 10^5$ & $2.7\times 10^6$ & $6.5\times 10^5$ & \multicolumn{1}{c|}{8211} & 38\% \\
\multicolumn{1}{|c|}{16.0 TeV $< m_{j_1j_2} <$ 22.0 TeV} & 3408 & 1218 & 477 & \multicolumn{1}{c|}{31} & 31\% \\ \hline
\end{tabular}
\caption{Background yield and signal efficiency ($\epsilon_S$) for the four benchmark mass-points of $H$ boson. The $m_{j_1 j_2}$ cuts are different for different values of $H$ boson mass. Here, $V$ denotes $W^\pm$, and $Z$.}
\label{tab:H_cut}
\end{table}

Similar to the $Z'$ analysis, we employ multi-class classification using XGBoost and CNN taggers. 
For each event, we require the presence of two $W$-tagged jets. Using this criterion and the analyses listed in Table \ref{tab:Zp_details}, we calculate the signal and background efficiencies. 
The cuts on the invariant mass of the two jets tagged as W are specific to the signal benchmark. The distribution of jet invariant mass for each signal benchmark is shown in Figure \ref{fig:inv_mass_bkg_H}. We use $\Delta_1 = 20\%~\text{of}~m_{H}$ and $\Delta_2 = 10\%~\text{of}~m_{H}$. Table \ref{tab:H_cut} summarises the results for all benchmarks using Analysis I.

The dominant background is QCD due to its large cross-section. The SM $W/Z+j$ background is the next most dominant contribution. Since the SM $WW$ final state is identical to that of the signal, we evaluate the fractional contributions from its different $p_T$ bins to the background in the $H\to WW$ analysis. 
The $t\bar{t}$ and $tW$ backgrounds are greatly reduced by the multi-class taggers. 
Using these numbers, we calculate the upper limit on the cross-section. Table \ref{tab:H_limits} summarises the background yields and signal efficiency after all cuts, and the upper limit on the cross-section for the process $pp \to H \to W^+W^-$ at 95\% CL, for heavy Higgs boson masses of 5~TeV, 10~TeV, 15~TeV, and~20 TeV. The upper limits, considering systematic uncertainty of 5\%, are shown in parentheses. 

\begin{table}[hbt!]
\centering
\small
\begin{tabular}{|c|c|cccc|c|c|}
\hline
$m_{H}$ & \multirow{2}{*}{Analysis} & \multicolumn{4}{c|}{Background yield at 30 $\text{ab}^{-1}$} & Signal & $\sigma(pp \to H\to W^+W^-)_{UL}$ \\ \cline{3-6}
 &  & $VV$ & $jj$ & $Vj$ & $tW+t\bar{t}$ & Efficiency & at 95\% CL [fb] (5\% sys.) \\ \hline
5 & I & 46444 & $1.3\times 10^6$ & $3.4\times 10^5$ & 4152 & 0.16 & 1.15 (78) \\
 & II & 30385 & $4.5\times 10^6$ & $6.1\times 10^5$ & 4949 & 0.11 & 3.0 (355)\\
 & III & 23827 & 72450 & 99008 & 1535 & 0.095 & 0.69 (16) \\ \hline
10 & I & 27810 & $5.8\times 10^5$ & 46694 & 460 & 0.36 & 0.32 (13) \\
 & II & 28505 & $2.6\times 10^6$ & $1.4\times 10^5$ & 2723 & 0.36 & 0.69 (60)\\
 & III & 21460 & 32151 & 27138 & 222 & 0.29 & 0.14 (2.2)\\ \hline
15 & I & 9020 & 77324 & 4019 & 154 & 0.35 & 0.12 (1.95)\\
 & II & 9550 & $9.8\times 10^5$ & 13894 & 368 & 0.39 & 0.37 (19) \\
 & III & 6881 & 52693 & 1766 & 94 & 0.28 & 0.13 (1.65) \\ \hline
20 & I & 3407 & 1218 & 477 & 31 & 0.31 & 0.034 (0.128)\\
 & II & 3579 & $4.8\times 10^5$ & 2825 & 111 & 0.35 & 0.29 (11) \\
 & III & 2380 & 1218 & 380 & 30 & 0.23 & 0.041 (0.141)\\ \hline
\end{tabular}
\caption{Table showing the number of events obtained for background processes, the signal efficiency after applying all cuts, and the upper limit of $pp\to H\to W^+W^-$ cross-section at 95\% CL for the three different analyses listed on Table \ref{tab:Zp_details}. Here, $V$ denotes $W^\pm,~Z$. The limits at 95\% CL with 5\% systematic uncertainty on the background are listed within parentheses.}
\label{tab:H_limits}
\end{table}

The strongest upper bounds on $\sigma(pp\to H\to W^+W^-)$ at 95\% CL are obtained with Analysis-III. At $m_H = 5$ TeV, Analysis-III constrains $\sigma(pp\to H\to W^+W^-)$ at 95\% CL up to 0.69 fb (16 fb with 5\% systematics). As the Higgs mass increases, the limits tighten further, reaching as low as 0.034 fb (0.128 fb) in Analysis I, and 0.041 fb (0.141 fb) in Analysis III, respectively, for $m_H=20~\text{TeV}$.

\subsection{Analysis of vector-like, heavy $B'$ quark decaying to boosted top jet and $W$ jet}
\label{sec:Bp_analysis}

In this subsection, we study the production of a vector-like quark in association with a jet via the process $pp\to B' j$, and $B' \to t W$, for $m_{B'} = 5$~TeV and 10~TeV, where both the top quark and the $W$ boson decay hadronically.

The signal topology consists of a top jet and a $W$ jet. The SM background consists of the $pp\to tW$, $pp\to jj$, $pp\to t\bar{t}$, $pp\to VV$, and $pp\to Vj$ processes, where $V$ represents $W^\pm/Z$. We select the two hardest jets with $p_T >  2$~TeV and $|\eta| < 2.5$ jets and use the ML taggers to classify them. Following the procedure highlighted in the previous sections, we demand that either one of the two jets should be identified as the top jet and the other one should be identified as the $W$ jet with $\epsilon_{t/W}=60\%$. 

\begin{table}[hbt!]
\centering
\small
\begin{tabular}{|c|c|cccc|c|c|}
\hline
$m_{B'}$ & Analysis & \multicolumn{4}{c|}{Background yield at 30 $\text{ab}^{-1}$} & $\epsilon_S$ & $\sigma_{UL}$ at \\ \cline{3-6}
 &  & $tW$ & $jj$ & $Vj$ & $VV+t\bar{t}$ &  & 95\% CL [fb] (5\% sys.) \\ \hline
5 & I & $8.1\times 10^5$ & $3.5\times 10^6$ & $6.6\times 10^5$ & $1.8\times 10^5$ & 0.086 & 4.0 (459) \\
 & II & $6.3\times 10^5$ & $7.3\times 10^6$ & $5.0\times 10^5$ & $1.3\times 10^5$ & 0.070 & 6.1 (921) \\
 & III & $4.9\times 10^5$ & $3.6\times 10^5$ & $2.3\times 10^5$ & 81363 & 0.055 & 3.0 (162) \\ \hline
10 & I & 51989 & $1.6\times 10^6$ & 66347 & 10370 & 0.084 & 2.3 (155) \\
 & II & 51692 & $3.7\times 10^6$ & 79516 & 17366 & 0.090 & 3.2 (303) \\
 & III & 38729 & $2.2\times 10^5$ & 33449 & 963 & 0.064 & 1.2 (36) \\ \hline
\end{tabular}

\caption{Table showing the number of events obtained for background processes, the signal efficiency ($\epsilon_S$) for $pp\to jB'\to jtW$ after all cuts, and the upper limit of $pp\to jB'\to jtW$ cross-section ($\sigma_{UL}$) at 95\% CL,for the three different analyses listed on Table \ref{tab:Zp_details}. Here, $V$ denotes $W^\pm,~Z$. The limits at 95\% CL with 5\% systematic uncertainty on the background are listed within parentheses.}
\label{tab:GammaFullWidth}
\end{table}

Table \ref{tab:GammaFullWidth} presents the upper limits on the cross-section for the process $pp \to jB' \to jtW$ at 95\% CL for $m_{B'} = 5$~TeV and $m_{B'} = 10$~TeV. The values are computed for Analyses I, II, and III. We apply an invariant mass cut on the signal benchmarks: 4.5~TeV $< m_{j_1j_2} <$ 5.5~TeV, and 9~TeV $< m_{j_1j_2} <$ 11~TeV for $m_{B'} = 5$~TeV and 10~TeV, respectively. For $m_{B'} = 5\,\text{TeV}$, Analysis III yields $\sigma_{\text{UL}} = 3.0\,\text{fb}$, compared to $4.0\,\text{fb}$ and $6.1\,\text{fb}$ for Analyses I and II, respectively. For $m_{B'} = 10\,\text{TeV}$, the upper limits improve due to lower backgrounds, with Analysis III achieving $\sigma_{\text{UL}} = 1.2\,\text{fb}$, outperforming Analysis I ($2.3\,\text{fb}$) and II ($3.2\,\text{fb}$).
\\\\
\textbf{Effect of width}

Before we summarise our work, we would like to discuss the effect of the width of the BSM particles on our results. At the 100~TeV collider, particles with masses in the tens of TeV range are expected to have very large widths. This large width makes it difficult to accurately reconstruct the mass of these particles because their decay products are spread over a wider range, leading to less distinct mass peaks. The decay width ($\Gamma$) increases with the mass of the particle, causing the resonance to become broader. So far, we have been using full-width calculations for the signal vs background analysis of a vector-like $B'$ quark~\cite{Barger:1987nn}. 

\begin{figure}[htb!]
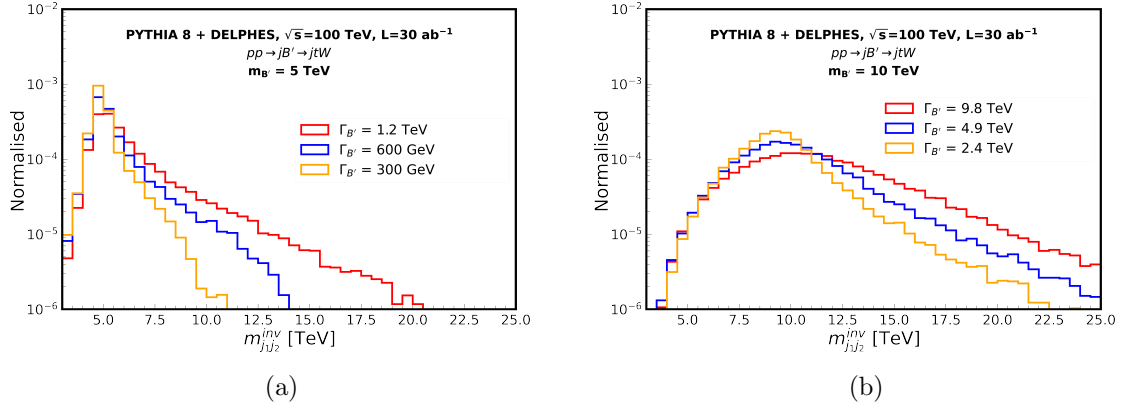

\centering
\begin{subfigure}[c]{0.5\linewidth}
\includegraphics[width=\linewidth]{inv_mass_sig_Bp_5.pdf} 
\caption{}
\label{fig:17a}
\end{subfigure}\hfill    
\begin{subfigure}[c]{0.5\linewidth}
\includegraphics[width=\linewidth]{inv_mass_sig_Bp_10.pdf}
\caption{}
\label{fig:17b}
\end{subfigure}
\caption{Normalised invariant mass distribution of two hardest jets from $pp\to jB' \to jtW$ process for 5~TeV \textit{(left)} and 10~TeV \textit{(right)} $B'$ quark when the width is varied from $\Gamma$ to $\Gamma/4$.}
\label{fig:inv_mass_Bp}
\end{figure}

To demonstrate the effect of width, we show the distributions of $m_{j_1j_2}^{\text{inv}}$ of the signal with the full-width ($\Gamma$), half-width ($\Gamma/2$), and quarter-width ($\Gamma/4$) in Figure \ref{fig:inv_mass_Bp}. The distribution peaks more accurately at lower widths. We use the reduced width to obtain signal efficiency and upper limits on cross-section. The results are listed in Table \ref{tab:GammaHalfQuarterWidth}. The background yields remain the same.

\begin{table}[hbt!]
\small
\centering
\begin{tabular}{|cccccccc|}
\hline
\multicolumn{1}{|c|}{$\Gamma_B'$} & \multicolumn{1}{c|}{Cut on $m_{j_1j_2}$} & \multicolumn{4}{c|}{Background yield at 30 $\text{ab}^{-1}$} & \multicolumn{1}{c|}{$\epsilon_S$} & $\sigma_{UL}$ at 95\% CL \\ \cline{3-6}
\multicolumn{1}{|c|}{[TeV]} & \multicolumn{1}{c|}{[TeV]} & $tW$ & $jj$ & $Vj$ & \multicolumn{1}{c|}{$VV+t\bar{t}$} & \multicolumn{1}{c|}{} & [fb] (5\% sys.) \\ \hline
\multicolumn{8}{|c|}{$m_{B'} = 5~\text{TeV}$} \\ \hline
\multicolumn{1}{|c|}{1.2} & \multicolumn{1}{c|}{4.5 $< m_{j_1j_2} <$ 5.5} & $4.9\times 10^5$ & $3.6\times 10^5$ & $2.3\times 10^5$ & \multicolumn{1}{c|}{81363} & \multicolumn{1}{c|}{0.055} & 3.0 (162) \\
\multicolumn{1}{|c|}{0.6} & \multicolumn{1}{c|}{4.6 $< m_{j_1j_2} <$ 5.4} & $4.0\times 10^5$ & $3.6\times 10^5$ & $1.8\times 10^5$ & \multicolumn{1}{c|}{72086} & \multicolumn{1}{c|}{0.058} & 2.5 (128) \\
\multicolumn{1}{|c|}{0.3} & \multicolumn{1}{c|}{4.6 $< m_{j_1j_2} <$ 5.4} & $4.0\times 10^5$ & $3.6\times 10^5$ & $1.8\times 10^5$ & \multicolumn{1}{c|}{72086} & \multicolumn{1}{c|}{0.066} & 2.2 (113) \\ \hline
\multicolumn{8}{|c|}{$m_{B'} = 10~\text{TeV}$} \\ \hline
\multicolumn{1}{|c|}{9.8} & \multicolumn{1}{c|}{9.0 $< m_{j_1j_2} <$ 11.0} & 38729 & $2.2\times 10^5$ & 33449 & \multicolumn{1}{c|}{5140} & \multicolumn{1}{c|}{0.064} & 1.2 (36) \\
\multicolumn{1}{|c|}{4.9} & \multicolumn{1}{c|}{9.2 $< m_{j_1j_2} <$ 10.8} & 28358 & $1.6\times 10^5$ & 23663 & \multicolumn{1}{c|}{3927} & \multicolumn{1}{c|}{0.073} & 0.93 (22) \\
\multicolumn{1}{|c|}{2.4} & \multicolumn{1}{c|}{9.0 $< m_{j_1j_2} <$ 10.5} & 32477 & $1.9\times 10^5$ & 27813 & \multicolumn{1}{c|}{4330} & \multicolumn{1}{c|}{0.086} & 0.86 (22) \\ \hline
\end{tabular}
\caption{Comparison of signal efficiency, and $\sigma(pp \to jB'\to jtW)_{UL}$ at 95\% CL for $m_{B'} = 5$~TeV and $m_{B'} = 10$~TeV with $\Gamma' = \Gamma$, $\Gamma' = \Gamma/2$ and $\Gamma' = \Gamma/4$ with Analysis III.}
\label{tab:GammaHalfQuarterWidth}
\end{table}

At smaller widths, such as half ($\Gamma/2$) or one-fourth ($\Gamma/4$) of the full width, we obtain better results. This is because reducing the width helps reduce the overlap of the signal with the background and the spread in the particle's mass, making it easier to identify and reconstruct the mass of heavy particles. We adapt the invariant mass selection for each width scenario to improve sensitivity to lower widths from increased signal efficiency and lower background yields.

\section {Conclusion}
\label{sec:conclusion}
The FCC-hh will provide a unique opportunity to explore the high-energy frontier, enabling direct searches for physics beyond the Standard Model at an unprecedented energy scale of a few tens of TeV. However, this comes with its own set of challenges, like how well the top quark tagging and $W$ boson tagging will work for particles with extremely high transverse momenta. In this paper, we have investigated the identification of boosted $W$ boson jets and boosted top quark jets in the context of three BSM scenarios: heavy neutral gauge boson ($Z'$), heavy neutral Higgs boson ($H$), and heavy vector-like quark ($B'$). Jets emanating from the decay of such high-mass particles are highly energetic up to tens of TeV and, thus, are highly boosted. 


The kinematic properties, such as jet mass, are strongly influenced by the $p_T$ of the jets. For ultra-boosted jets, jet mass and other variables will be significantly modified by the presence of underlying events, particularly initial and final state radiations, proportional to the jet radius. We have demonstrated the importance of using a dynamic jet radius, dependent on the transverse momentum of jets, to improve the reconstruction of top and $W$ jets. Our results show that this approach can significantly reduce jet contamination and correct various jet properties to some extent. Additionally, we have shown the extent to which improved calorimeter resolution can affect jets with high transverse momentum. Track-based information is more powerful in terms of providing an accurate description of the jets owing to their higher resolution. We perform a brief hypothetical study showing that increasing the angular resolution from 1 mrad to 5 mrad increases the overlap between signal and background distributions of track-based observables, and the effect is more pronounced for $\mathcal{O}(10)$~TeV jets.

Furthermore, we have explored the application of machine learning techniques, including Extreme Gradient Boosted Decision Trees (XGBoost) and convolutional neural networks (CNNs), to distinguish top jets and $W$ jets from QCD jets. 
The variable-based tagging performs the best when both tower and track-based observables are used as input features. In the case of image-based classification, we have used particle-flow (PF) jet images with ECAL tower resolution. Our findings during the comparative study of 14 TeV and 100 TeV jets indicate that the low-level jet constituent-based CNN, which outperforms observable-based XGBoost at 14 TeV, performs poorly at 100 TeV. Our focus in this study has been exploratory. This study excludes the simulation of pile-up. To extract the maximum sensitivity, the next step would be to use the ML taggers from this study and fit the resultant invariant mass spectrum with detailed signal and background models involving full pile-up simulation, and performing a full statistical analysis, as is usually done in LHC resonance searches. 

We have tested the three BSM scenarios with the trained models and calculated the upper limits on cross-sections at 95\% CL. 
Assuming only statistical uncertainty, the 95\% CL upper limits on the cross-section from Analysis III are as follows. For the $Z' \to t\bar{t}$ channel, the limits are 4.2 fb, 0.95 fb, 0.26 fb, and 0.099 fb for $m_{Z'} = 5$, 10, 15, and 20 TeV, respectively. For the heavy scalar $H \to W^+W^-$ channel, the limits are 0.69 fb at $m_H = 5$ TeV, 0.14 fb at 10 TeV, 0.13 fb at 15 TeV, and 0.041 fb at 20 TeV. For the vector-like quark process $pp \to jB' \to jtW$, the corresponding limits are 3.0 fb at $m_{B'} = 5$ TeV and 1.2 fb at 10 TeV. 

The variation in the upper limits with 5\% systematics also presents an interesting implication. The resonant masses that lie at the extreme reach of the LHC, and are only marginally accessible at the LHC, shall also be the most challenging to probe at the FCC, due to the dominance of systematic effects in the regime. Overcoming this challenge shall require a more detailed analysis of the background. Finally, we show that the results also depend on the width of the BSM particle. As the width increases with the mass, reconstructing the particle at the FCC-hh will pose a challenge for multi-TeV resonances.

This study advances our understanding of jets in the exceptionally high-energy regime envisioned for future particle physics experiments. The results presented in this paper demonstrate the feasibility of identifying boosted top and $W$ jets at the FCC-hh and highlight the potential for discovering new physics at a very high mass scale.
By developing robust and efficient tagging techniques for hadronically decaying top quarks and $W$ bosons, our work contributes to the ongoing efforts to demonstrate the discovery potential of the FCC-hh. We plan on further optimization of the top/$W$ jet tagging performance, using more advanced ML techniques in future studies.

\section{Acknowledgments}
The work of SM is supported by an initiation grant ($\text{IITK}/\text{PHY}/2023282$) received from IIT Kanpur. The works of BB, SB, and DC are supported by the Core Research Grants CRG/2021/007579 and CRG/2022/001922 of the Science and Engineering Research Board (SERB), Government of India. BB also acknowledges the MATRICS Grant (MTR/2022/000264) of the Science and Engineering Research Board (SERB), Government of India. DC also acknowledges funding from an initiation grant IITK/PHY/2019413 at IIT Kanpur. CB and BB are grateful to the Center for High Energy Physics, Indian Institute of Science, for the cluster facility.

\clearpage
\begin{appendices}

\section{Jet energy response and resolution}
\label{app:mig_jet}
As stated earlier, the top jets and the QCD have been generated in fixed $p_T$ bins of 2 TeV. However, boosted objects with high $p_T$ might lose energy due to radiation and smearing within the detector and lie in a $p_T$ bin with lower energy. On the other hand, calorimeters and tracking systems introduce smearing effects that can slightly alter the reconstructed $p_T$ of the jet compared to the original $p_T$, causing the reconstructed jet $p_T$ to acquire higher values than the generated jet $p_T$. 

Generally, a typical general-purpose detector cannot measure the energy (and $p_T$) of jets perfectly. The measured energy of a jet is usually corrected to match the energy of generated particle jets. The correction factors are generally functions of momentum and pseudorapidity, and they help to bring the $p_T$ of the reconstructed jet in line with the generated jet. We study the extent to which the $p_T$ of the reconstructed jets ($p_T^{(jet)}$) differ from the $p_T$ of the generated jets ($p_T^{(gen)}$) at FCC-hh.

\begin{figure}[hbt!]
\centering
\includegraphics[width=0.5\textwidth]{pt_top.pdf}
\caption{$p_T^{(jet)}$ distributions of top jets for each $p_T^{(gen)}$ bin.}
\label{fig:mig_pT}
\end{figure}

Figure \ref{fig:mig_pT} illustrates the jet energy response and resolution for top jets at the FCC-hh. As the $p_T^{(gen)}$ increases, the reconstructed $p_T^{(jet)}$ shows an increasingly broad distribution, reflecting the deterioration of jet resolution at higher energies. The response is biased towards lower reconstructed values, though a small fraction of jets are also smeared upward. Similar behavior is observed for QCD and $W$ jets.

We study the impact of using a poorer energy resolution and calorimeter segmentation on the jet mass observable. Firstly, we increase the $S$ and $C$ co-efficients of Equation~\ref{eq:energy_res} to 30\%. Additionally, we set the ECAL granularity to HCAL granularity. Simulating the FCC-hh detector conditions using these parameters, we construct PF jets with the $p_T$ dependent radius. Figure~\ref{fig:worse_mjet} shows the variation in the $m_{jet}$ distribution for the default and changed scenarios. 

\begin{figure}[hbt!]
    \centering
    \includegraphics[width=0.5\linewidth]{rec_mjet_2_4TeV_sigma_30.pdf}~
    \includegraphics[width=0.5\linewidth]{rec_mjet_2_4TeV_gran_hcal.pdf}\\
    \includegraphics[width=0.5\linewidth]{rec_mjet_14_16TeV_sigma_30.pdf}~
    \includegraphics[width=0.5\linewidth]{rec_mjet_14_16TeV_gran_hcal.pdf}
    \caption{Normalised $m_{\text{jet}}$ distribution of 2--4~TeV ((a), (b)) and 14--16~TeV ((c), (d)) top jets and QCD jets using different calorimeter configurations. Subfigures (a), (c) show the change in $m_{\text{jet}}$ when the energy resolution is varied. Subfigures (b), (d) show the change in $m_{\text{jet}}$ when the angular resolution is varied.}
    \label{fig:worse_mjet}
\end{figure}

We observe that the jet mass resolution worsens due to the increased smearing and poor reconstruction in the calorimeter, and the impact is greater for 14--16 TeV jets compared to 2--4 TeV jets.

\clearpage
\section{Efficiencies used by DELPHES}
\label{app:delphes}

In the following table, we list the identification efficiencies and tracking efficiency for charged hadrons used by the FCC-hh card of \texttt{DELPHES}.
The charged hadrons and electron tracking efficiencies are approximated to be similar in \texttt{DELPHES}.

\begin{table}[h]
\centering
\small
\begin{tabular}{|l|c|c|c|c|}
\hline
\textbf{Type} & \textbf{$|\eta| < 2.5$} & \textbf{$2.5 < |\eta| < 4.0$} & \textbf{$4.0 < |\eta| < 6.0$} & \textbf{$|\eta| > 6.0$} \\
\hline
\multicolumn{5}{|c|}{Tracking Efficiency (Charged Hadrons)} \\
\hline
$p_T < 0.5$ GeV     & 0\%  & 0\%  & 0\%  & 0\% \\
$0.5 < p_T < 1.0$ GeV & 90\% & 85\% & 80\% & 0\% \\
$p_T > 1.0$ GeV      & 95\% & 90\% & 85\% & 0\% \\
\hline
\multicolumn{5}{|c|}{Electron Identification Efficiency} \\
\hline
$p_T < 4$ GeV        & 0\%  & 0\%  & 0\%  & 0\% \\
$p_T > 4$ GeV        & 95\% & 90\% & 85\% & 0\% \\
\hline
\multicolumn{5}{|c|}{Muon Identification Efficiency} \\
\hline
$p_T < 4$ GeV        & 0\%  & 0\%  & 0\%  & 0\% \\
$p_T > 4$ GeV        & 99\% & 99\% & 99\% & 0\% \\
\hline
\multicolumn{5}{|c|}{Photon Identification Efficiency} \\
\hline
$1 < p_T < 5$ GeV    & 70\% & 60\% & 50\% & 50\% \\
$5 < p_T < 10$ GeV   & 85\% & 80\% & 70\% & 70\% \\
$p_T > 10$ GeV       & 95\% & 90\% & 80\% & 80\% \\
\hline
\end{tabular}
\caption{Efficiencies (in \%) for tracking and particle identification as a function of $p_T$ and $|\eta|$ for charged hadrons, electrons, muons, and photons~\cite{Selvaggi:2717698}.}
\label{tab:delphes_eff}
\end{table}

\clearpage
\section{Jet after soft drop}
\label{app:soft drop}
For fixed radius jets, the soft drop algorithm is unable to correct the long tail in the jet mass distribution of 14--16 TeV jets. We see this in Figure~\ref{fig:groom}.
\begin{figure}[hbt!]
    \centering
    \includegraphics[width=0.5\linewidth]{mjet_groom.pdf}
    \caption{Jet mass distribution for top, QCD, and W jets with $p_T=14~\text{TeV}$, after applying grooming on fixed R and variable R jets.}
    \label{fig:groom}
\end{figure}

We now present the distributions of the jet mass ($m_{\text{jet}}$) after applying the softdrop algorithm on the reclustered jets. The distributions are shown for two extreme ends of the $p_T$ spectrum, 2--4~TeV jets and 14--16~TeV jets, to highlight the effect of the softdrop with jet $p_T$. We vary the parameters of softdrop $z_{\text{cut}}$ and $\beta$ defined in Equation~\ref{eq:SD}. 

\begin{figure}[hbt!]
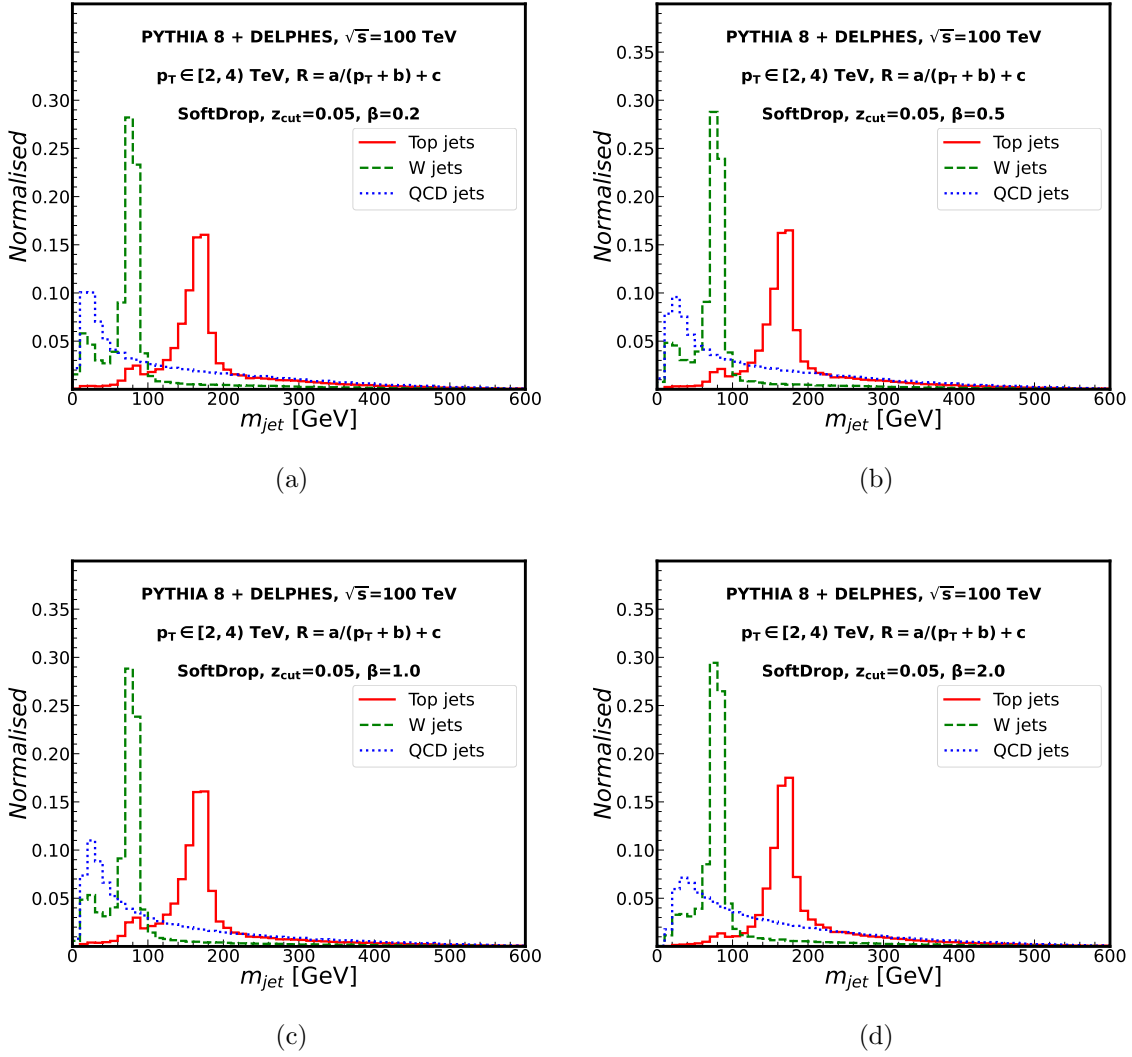

\centering
\begin{subfigure}[c]{0.5\linewidth}
\includegraphics[width=\linewidth]{zcut005_beta02_2_4TeV.pdf} 
\caption{}
\label{fig:A3a}
\end{subfigure}\hfill    
\begin{subfigure}[c]{0.5\linewidth}
\includegraphics[width=\linewidth]{zcut005_beta05_2_4TeV.pdf}
\caption{}
\label{fig:A3b}
\end{subfigure}
\begin{subfigure}[c]{0.5\linewidth}
\includegraphics[width=\linewidth]{zcut005_beta10_2_4TeV.pdf}
\caption{}
\label{fig:A3c}
\end{subfigure}\hfill 
\begin{subfigure}[c]{0.5\linewidth}
\includegraphics[width=\linewidth]{zcut005_beta20_2_4TeV.pdf}
\caption{}
\label{fig:A3d}
\end{subfigure} 
\caption{Normalised $m_{\text{jet}}$ distribution of 2--4~TeV top jets and QCD jets, after applying softdrop grooming with parameters $z = 0.05$ \textit{(fixed)}, and (a) $\beta$ = 0.2, (b) $\beta$ = 0.5, (c) $\beta$ = 1.0, (d) $\beta$ = 2.0.}
\label{fig:mjet_SD_z005_2}
\end{figure}

Figures \ref{fig:mjet_SD_z005_2} and \ref{fig:mjet_SD_z005_14} show the $m_{\text{jet}}$ distributions of 2--4~TeV jets and 14--16~TeV jets respectively. The $z$ parameter is fixed at 0.05 while $\beta$ is varied from 0.2 to 2.0. Similar $m_{\text{jet}}$ distributions for a fixed $z$ = 0.08 are presented in  Figures \ref{fig:mjet_SD_z008_2} and \ref{fig:mjet_SD_z008_14} for 2--4~TeV jets and 14--16~TeV jets respectively. Figures \ref{fig:mjet_SD_z010_2} and \ref{fig:mjet_SD_z010_14} show the distributions for $z$ =0.1, while Figures \ref{fig:mjet_SD_z020_2}  and \ref{fig:mjet_SD_z020_14} show the distributions for $z=0.2$.

\begin{figure}[hbt!]
\centering
\begin{subfigure}[c]{0.5\linewidth}
\includegraphics[width=\linewidth]{zcut005_beta02_14_16TeV.pdf} 
\caption{}
\label{fig:A3e}
\end{subfigure}\hfill    
\begin{subfigure}[c]{0.5\linewidth}
\includegraphics[width=\linewidth]{zcut005_beta05_14_16TeV.pdf}
\caption{}
\label{fig:A3f}
\end{subfigure}
\begin{subfigure}[c]{0.5\linewidth}
\includegraphics[width=\linewidth]{zcut005_beta10_14_16TeV.pdf}
\caption{}
\label{fig:A3g}
\end{subfigure}\hfill 
\begin{subfigure}[c]{0.5\linewidth}
\includegraphics[width=\linewidth]{zcut005_beta20_14_16TeV.pdf}
\caption{}
\label{fig:A3h}
\end{subfigure}
\caption{Normalised $m_{\text{jet}}$ distribution of 14--16~TeV top jets and QCD jets, after applying softdrop grooming with parameters $z = 0.05$ \textit{(fixed)}, and (a) $\beta$ = 0.2, (b) $\beta$ = 0.5, (c) $\beta$ = 1.0, (d) $\beta$ = 2.0.}
\label{fig:mjet_SD_z005_14}
\end{figure}

As $z$ increases, the soft components of the jets are removed more aggressively, hence we see an increase in small peaks at lower $m_{\text{jet}}$ values, in top and $W$ jet distributions. Increasing $\beta$ for a fixed $z$ results in less grooming, since a large $\beta$ indicates that only wide-angle radiations are removed.

\begin{figure}[hbt!]
\centering
\begin{subfigure}[c]{0.5\linewidth}
\includegraphics[width=\linewidth]{zcut008_beta02_2_4TeV.pdf} 
\caption{}
\end{subfigure}\hfill    
\begin{subfigure}[c]{0.5\linewidth}
\includegraphics[width=\linewidth]{zcut008_beta05_2_4TeV.pdf}
\caption{}
\end{subfigure}
\begin{subfigure}[c]{0.5\linewidth}
\includegraphics[width=\linewidth]{zcut008_beta10_2_4TeV.pdf}
\caption{}
\end{subfigure}\hfill 
\begin{subfigure}[c]{0.5\linewidth}
\includegraphics[width=\linewidth]{zcut008_beta20_2_4TeV.pdf}
\caption{}
\end{subfigure} 
\caption{Normalised $m_{\text{jet}}$ distribution of 2--4~TeV top jets and QCD jets, after applying softdrop grooming with parameters $z = 0.08$ \textit{(fixed)}, and (a) $\beta$ = 0.2, (b) $\beta$ = 0.5, (c) $\beta$ = 1.0, (d) $\beta$ = 2.0.}
\label{fig:mjet_SD_z008_2}
\end{figure}

\begin{figure}[hbt!]
\centering
\begin{subfigure}[c]{0.5\linewidth}
\includegraphics[width=\linewidth]{zcut008_beta02_14_16TeV.pdf} 
\caption{}
\end{subfigure}\hfill    
\begin{subfigure}[c]{0.5\linewidth}
\includegraphics[width=\linewidth]{zcut008_beta05_14_16TeV.pdf}
\caption{}
\end{subfigure}
\begin{subfigure}[c]{0.5\linewidth}
\includegraphics[width=\linewidth]{zcut008_beta10_14_16TeV.pdf}
\caption{}
\end{subfigure}\hfill 
\begin{subfigure}[c]{0.5\linewidth}
\includegraphics[width=\linewidth]{zcut008_beta20_14_16TeV.pdf}
\caption{}
\end{subfigure}
\caption{Normalised $m_{\text{jet}}$ distribution of 14--16~TeV top jets and QCD jets, after applying softdrop grooming with parameters $z = 0.08$ \textit{(fixed)}, and (a) $\beta$ = 0.2, (b) $\beta$ = 0.5, (c) $\beta$ = 1.0, (d) $\beta$ = 2.0.}
\label{fig:mjet_SD_z008_14}
\end{figure}

\begin{figure}[hbt!]
\centering
\begin{subfigure}[c]{0.5\linewidth}
\includegraphics[width=\linewidth]{zcut010_beta02_2_4TeV.pdf} 
\caption{}
\end{subfigure}\hfill    
\begin{subfigure}[c]{0.5\linewidth}
\includegraphics[width=\linewidth]{zcut010_beta05_2_4TeV.pdf}
\caption{}
\end{subfigure}
\begin{subfigure}[c]{0.5\linewidth}
\includegraphics[width=\linewidth]{zcut010_beta10_2_4TeV.pdf}
\caption{}
\end{subfigure}\hfill 
\begin{subfigure}[c]{0.5\linewidth}
\includegraphics[width=\linewidth]{zcut010_beta20_2_4TeV.pdf}
\caption{}
\end{subfigure} 
\caption{Normalised $m_{\text{jet}}$ distribution of 2--4~TeV top jets and QCD jets, after applying softdrop grooming with parameters $z = 0.1$ \textit{(fixed)}, and (a) $\beta$ = 0.2, (b) $\beta$ = 0.5, (c) $\beta$ = 1.0, (d) $\beta$ = 2.0.}
\label{fig:mjet_SD_z010_2}
\end{figure}

\begin{figure}[hbt!]
\centering
\begin{subfigure}[c]{0.5\linewidth}
\includegraphics[width=\linewidth]{zcut010_beta02_14_16TeV.pdf} 
\caption{}
\end{subfigure}\hfill    
\begin{subfigure}[c]{0.5\linewidth}
\includegraphics[width=\linewidth]{zcut010_beta05_14_16TeV.pdf}
\caption{}
\end{subfigure}
\begin{subfigure}[c]{0.5\linewidth}
\includegraphics[width=\linewidth]{zcut010_beta10_14_16TeV.pdf}
\caption{}
\end{subfigure}\hfill 
\begin{subfigure}[c]{0.5\linewidth}
\includegraphics[width=\linewidth]{zcut010_beta20_14_16TeV.pdf}
\caption{}
\end{subfigure}
\caption{Normalised $m_{\text{jet}}$ distribution of 14--16~TeV top jets and QCD jets, after applying softdrop grooming with parameters $z = 0.1$ \textit{(fixed)}, and (a) $\beta$ = 0.2, (b) $\beta$ = 0.5, (c) $\beta$ = 1.0, (d) $\beta$ = 2.0.}
\label{fig:mjet_SD_z010_14}
\end{figure}

\begin{figure}[hbt!]
\centering
\begin{subfigure}[c]{0.5\linewidth}
\includegraphics[width=\linewidth]{zcut020_beta02_2_4TeV.pdf} 
\caption{}
\end{subfigure}\hfill    
\begin{subfigure}[c]{0.5\linewidth}
\includegraphics[width=\linewidth]{zcut020_beta05_2_4TeV.pdf}
\caption{}
\end{subfigure}
\begin{subfigure}[c]{0.5\linewidth}
\includegraphics[width=\linewidth]{zcut020_beta10_2_4TeV.pdf}
\caption{}
\end{subfigure}\hfill 
\begin{subfigure}[c]{0.5\linewidth}
\includegraphics[width=\linewidth]{zcut020_beta20_2_4TeV.pdf}
\caption{}
\end{subfigure} 
\caption{Normalised $m_{\text{jet}}$ distribution of 2--4~TeV top jets and QCD jets, after applying softdrop grooming with parameters $z = 0.2$ \textit{(fixed)}, and (a) $\beta$ = 0.2, (b) $\beta$ = 0.5, (c) $\beta$ = 1.0, (d) $\beta$ = 2.0.}
\label{fig:mjet_SD_z020_2}
\end{figure}

\begin{figure}[hbt!]
\centering
\begin{subfigure}[c]{0.5\linewidth}
\includegraphics[width=\linewidth]{zcut020_beta02_14_16TeV.pdf} 
\caption{}
\end{subfigure}\hfill    
\begin{subfigure}[c]{0.5\linewidth}
\includegraphics[width=\linewidth]{zcut020_beta05_14_16TeV.pdf}
\caption{}
\end{subfigure}
\begin{subfigure}[c]{0.5\linewidth}
\includegraphics[width=\linewidth]{zcut020_beta10_14_16TeV.pdf}
\caption{}
\end{subfigure}\hfill 
\begin{subfigure}[c]{0.5\linewidth}
\includegraphics[width=\linewidth]{zcut020_beta20_14_16TeV.pdf}
\caption{}
\end{subfigure}
\caption{Normalised $m_{\text{jet}}$ distribution of 14--16~TeV top jets and QCD jets, after applying softdrop grooming with parameters $z = 0.2$ \textit{(fixed)}, and (a) $\beta$ = 0.2, (b) $\beta$ = 0.5, (c) $\beta$ = 1.0, (d) $\beta$ = 2.0.}
\label{fig:mjet_SD_z020_14}
\end{figure}

\clearpage
\bibliographystyle{JHEP}
\bibliography{refs}
\end{appendices}

\end{document}